\documentclass[journal]{IEEEtran}

\usepackage{cite}

\usepackage{amsmath}
\usepackage{amssymb}

\usepackage{amsfonts}

\usepackage{graphicx}
\usepackage{amsmath,amssymb,amsfonts}
\usepackage{graphicx}
\usepackage{textcomp}
\usepackage{booktabs}
\usepackage{float}
\usepackage{xurl}
\usepackage{placeins}
\usepackage{soul}
\usepackage{tikz}
\usepackage{pgfplots}
\pgfplotsset{compat=1.18}
\usepackage[colorlinks=true, linkcolor=blue, citecolor=blue, urlcolor=blue]{hyperref}
\usepackage{array}
\usepackage{booktabs}
\usepackage{tabularx}
\newcolumntype{C}{>{\centering\arraybackslash}X}
\usepackage{hyperxmp}
\usepackage{enumitem}
\usepackage{xcolor}
\usepackage{pifont}

\usepackage{url}

\begin{document}

\title{Agentic and Generative AI for Open-Source Intelligence and Cyber Investigations: Taxonomy, Evaluation, Challenges, and Future Directions}

\author{Eduardo~Almeida~Palmieri,
        Mohamed~Chahine~Ghanem *,
        Dipo~Dunsin,
        Zubair~Baig,
        Ed~de~Quincey,
        and~Kim-Kwang~Raymond~Choo
\thanks{E. A. Palmieri and E. de Quincey are with the School of Computer Science and Mathematics, Keele University, Newcastle-under-Lyme ST5 5AA, U.K. (e-mail: e.a.palmieri@keele.ac.uk; e.de.quincey@keele.ac.uk).}%
\thanks{M. C. Ghanem is with the School of Computer Science and Mathematics, Keele University, Newcastle-under-Lyme ST5 5AA, U.K., and also with the Cybersecurity Institute, University of Liverpool, Liverpool L69 3BX, U.K. (corresponding author, e-mail: mohamed.chahine.ghanem@liverpool.ac.uk).}%
\thanks{D. Dunsin is with the Department of Applied Computing IICL, University of Wales Trinity Saint David, London E14 4HA, U.K. (e-mail: d.dunsin@uwtsd.ac.uk).}%
\thanks{Z. Baig is with the Deakin Cyber Research and Innovation Hub, Deakin University, Waurn Ponds, VIC 3216, Australia (e-mail: zubair.baig@deakin.edu.au).}%
\thanks{K.-K. R. Choo is with the Department of Information Systems and Cyber Security, The University of Texas at San Antonio, San Antonio, TX 78249, USA (e-mail: raymond.choo@utsa.edu).}%
\thanks{Manuscript received 02 July 2026.}}

\markboth{IEEE Communications Surveys \& Tutorials,~Vol.~XX, No.~X, 2026}%
{Palmieri \MakeLowercase{\textit{et al.}}: Agentic AI for Open-Source Intelligence and Cyber Investigation}

\maketitle

\begin{abstract}
The rapid growth of publicly available digital information has rendered manual open-source intelligence (OSINT) analysis insufficient for contemporary intelligence, cybersecurity, and Cyber Investigation requirements. Large language models (LLMs) and \emph{agentic} AI systems, which select tools, perform multi-step reasoning, and iteratively produce intelligence, have emerged as promising responses, yet published capability demonstrations have substantially outpaced the evaluation infrastructure required to validate operational deployment. This survey systematically reviews 74 unique studies on the application of agentic AI, generative AI, and LLMs to OSINT, cyber threat intelligence (CTI), and Cyber Investigation. Its contribution is fourfold. \emph{First}, it treats agentic AI as a distinct analytical category rather than a variant of LLM prompting, organising the literature through an 11-category taxonomy spanning LLM foundations, agentic architectures, retrieval-augmented generation (RAG), knowledge graphs, prompt engineering, domain adaptation, evaluation benchmarks, and risk. \emph{Second}, it establishes the \emph{hallucination--validation gap} as a corpus-level finding: although hallucination is named a reliability concern in more than twenty studies, end-to-end hallucination is empirically measured in only one OSINT-specific system, a RAG-augmented architecture reporting a 4\% rate under favourable, non-reproducible conditions; the reasoning-error and factual-correction results reported elsewhere are measured in general-domain question answering, not on OSINT hallucination, and do not close this gap. \emph{Third}, it maps the corpus onto the OSINT workflow lifecycle, showing that collection and analysis are well served while verification, reporting, dissemination, and decision support remain systematically underexplored. \emph{Fourth}, it derives a ten-point research agenda, covering evaluation, benchmarking, hallucination measurement, adversarial robustness, dark-web coverage, multimodal processing, and governance, directly from the gaps the corpus exposes. The review further finds that no standardised, open, community-adopted benchmark exists for cross-study comparison of OSINT AI systems, and that agentic systems are evaluated exclusively under benign conditions despite documented adversarial threats. It concludes that a structured human--AI co-pilot model, in which LLMs support collection and triage while analysts retain responsibility for verification, reporting, and decision support, is the most defensible near-term deployment architecture under the current evidence base.
\end{abstract}

\begin{IEEEkeywords}
Open-source intelligence, OSINT, large language models, agentic AI, generative AI, cyber threat intelligence, retrieval-augmented generation, knowledge graphs, hallucination, Cyber Investigation, systematic literature review.
\end{IEEEkeywords}

\section{Introduction}
\label{sec:1-introduction}

\IEEEPARstart{T}{he} amount of digital information available to the public has grown so quickly that traditional manual OSINT analysis is no longer sufficient at operational scale. Intelligence professionals across government, law enforcement, and the private security sector now work in an environment where relevant signals are distributed across millions of surface-web pages, social media platforms, encrypted messaging channels, and dark-web repositories that are continuously updated in multiple languages. The challenge is not access to information, which has never been greater, but the capacity to process, verify, and synthesize it at the speed and scale required by modern threat environments \cite{hwang2022osint,cerny2024implications,chen2026cyberthreateval}.

The emergence of large language models and, more recently, agentic AI systems capable of autonomous tool selection, iterative reasoning, and structured output generation has introduced a qualitative shift in what OSINT analysis can feasibly accomplish. Whereas previous automation approaches could accelerate discrete pipeline stages, such as structured web scraping, keyword filtering, and named entity recognition using fixed classifiers, modern LLMs enable flexible multi-step intelligence tasks: formulating targeted collection queries, extracting and contextualizing entities from unstructured text, synthesizing intelligence across heterogeneous sources, and drafting preliminary analytical reports \cite{almeidapalmieri2025framework,shen2024llmosint,allam2025cybervision}. These are not marginal improvements; they represent a fundamental expansion of the analytical  frontier \cite{allam2025cybervision}.

However, the academic literature documenting this expansion is characterized by a structural imbalance: capability demonstrations have substantially outpaced evaluation practice. Systems are proposed and their outputs are described, but the methodological infrastructure required for systematic validation, standardized benchmarks, reproducible datasets, adversarial test conditions, hallucination validation protocols, and human analyst evaluation remains largely absent. The consequences are significant. When a single empirical measurement of hallucination in an OSINT-specific system---a 4\% rate in one RAG-augmented study \cite{allam2025cybervision}---constitutes the available quantitative evidence base for hallucination rates across the reviewed corpus, confidence in autonomous operational deployment cannot be justified by the current literature. This survey addresses this imbalance. It does not argue against the use of LLMs and agentic AI in OSINT; the evidence of their practical value is substantial and growing. Instead, it argues that capability without validated reliability, provenance tracking, ethical safeguards, and structured human oversight is insufficient for responsible operational deployment, and that the field's evaluation infrastructure must develop alongside its architectural ambition.

\subsection{Motivation and Related Work}
\label{sec:1-1-motivation-and-related-work}

The intelligence and cybersecurity research communities have produced several surveys of LLM applications relevant to OSINT and Cyber Investigation. Broad overviews of LLMs in cybersecurity contexts have catalogued applications in threat detection, vulnerability analysis, and malware classification \cite{hassanin2024llms,yigit2024review}. Surveys of OSINT methodologies have documented the evolution of collection and analysis frameworks from pre-LLM automated pipelines to early generative AI integrations \cite{rajendran2024comprehensive,nagra2024kia,hwang2022osint}. A notable survey of LLM applications to digital forensics introduces the MetaAID ethics framework and acknowledges the dual-use risks of publishing offensive capabilities \cite{yigit2024review}. This survey builds on these contributions while addressing several limitations in the existing review literature.

However, the existing literature exhibits three significant limitations that motivate an independent systematic review. First, prior surveys describe LLM applications for OSINT-adjacent tasks without comparatively evaluating whether those applications are reliable, adversarially robust, or ethically sound under operational conditions. Describing what a system does is not equivalent to assessing whether it can be trusted to perform that task consistently. Second, no prior survey treats agentic AI architectures, including systems with autonomous tool selection, multi-step reasoning loops, and self-correction mechanisms, as a distinct technical category requiring evaluation standards separate from simpler LLM-as-tool approaches. Third, no prior survey identifies the hallucination-validation gap as a corpus-level finding. This gap is the systematic absence of empirical hallucination measurement in a literature that widely acknowledges hallucination as a primary reliability concern. It is not incidental to individual papers but reflects a structural weakness in the field's evaluation culture.

The motivation for this review is therefore both constructive and critical: constructive in synthesizing what the literature has achieved, and critical in making explicit what the literature has failed to measure.

\textbf{The OSINT transformation and LLM enablement.} Open-source intelligence refers to intelligence derived from publicly or commercially available information in support of specific intelligence priorities, requirements, or gaps. The OSINT lifecycle includes collection from diverse publicly accessible sources, processing into structured formats, enrichment through entity extraction and contextualization, analysis to produce intelligence assessments, verification against alternative sources, reporting in structured intelligence products, and dissemination to decision-makers \cite{hwang2022osint}. Prior to the LLM era, each of these stages was addressed by purpose-built tools: web scrapers, keyword classifiers, and named entity recognisers trained on fixed corpora whose performance was predictable but whose flexibility was limited. The introduction of GPT-3 \cite{brown2020fewshot} and the chain-of-thought prompting paradigm \cite{wei2022cot} fundamentally changed the flexibility-performance trade-off: a sufficiently large model could adapt to diverse OSINT tasks through natural-language specification alone, without additional training. This shift from task-specific to task-agnostic intelligence automation provides the technical foundation for the subsequent OSINT AI work reviewed in this corpus.

\textbf{Agentic AI and the OSINT frontier.} The subsequent development of agentic AI architectures, in which LLMs autonomously plan, select external tools, execute multi-step workflows, observe results, and iterate, has extended this flexibility into complete intelligence pipelines. The ReAct (Reason + Act) paradigm \cite{almeidapalmieri2025framework,shen2024llmosint}, in which an agent alternates between reasoning steps and tool actions within a continuous observation loop, enables OSINT agents to orchestrate Shodan, Censys, Maltego, VirusTotal, theHarvester, and Tor-based dark web crawlers within a single coherent workflow \cite{almeidapalmieri2025framework,shen2024llmosint}. This architecture produces intelligence outputs that would previously have required specialized analyst expertise for each individual tool. However, the absence of standardized evaluation frameworks for agentic OSINT means that the capability claims of these systems, although credible in isolation, cannot be compared across studies or generalized beyond the specific conditions under which they were evaluated.

\textbf{The cloud-versus-on-premise tension.} A substantive architectural disagreement runs through the corpus concerning the appropriate deployment model for OSINT AI. Early demonstrations of LLM-based OSINT automation relied on commercial cloud APIs, such as GPT-3 via OpenAI's API, for their underlying language model \cite{radoi2023ai}. This approach is technically accessible, cost-effective, and rapidly deployable, but it introduces an operational-security risk that several studies identify as unacceptable for sensitive intelligence contexts: all OSINT queries and the intelligence derived from them pass through third-party infrastructure \cite{yurtalan2025redefining}. Yurtalan \cite{yurtalan2025redefining} presents the most direct argument for on-premise LLM deployment in law enforcement OSINT contexts, demonstrating that capability-adequate open-weight models can be deployed locally without cloud API exposure. This position is independently supported by evidence from military intelligence contexts \cite{nila2023llms} and receives additional legal grounding from the Italian Data Protection Authority's ruling that cloud-based LLM data processing lacked a valid basis under GDPR for certain personal data uses \cite{golda2024privacy}. The corpus does not resolve this tension but provides the evidence required to frame it as context-dependent: cloud deployment is appropriate where data sensitivity and legal constraints permit; on-premise deployment is required where they do not.

\textbf{The benchmark-versus-workflow tension.} A second consequential tension concerns evaluation realism. The CyberMetric benchmark \cite{tihanyi2024cybermetric} compares 25 LLMs against 30 human cybersecurity experts across multiple-choice questions, yielding the striking finding that GPT-4o achieves 91.25\% accuracy versus a human expert mean of 72.24\%. Taken alone, this finding suggests a substantial LLM advantage on cybersecurity knowledge questions. CyberThreat-Eval \cite{chen2026cyberthreateval}, however, demonstrates that this conclusion is an artifact of evaluation design. Security analysts do not attribute breaches or characterize threat actors by selecting from multiple-choice options; they triage streams of unstructured OSINT content, conduct contextual deep searches, and draft intelligence reports with sufficient operational specificity to guide incident response. When LLMs are evaluated against this realistic three-stage CTI workflow derived from a real CTI analyst workflow, none achieves analyst-acceptable performance on complex threat actor characterisation or report quality \cite{chen2026cyberthreateval}. This contrast between \cite{tihanyi2024cybermetric} and \cite{chen2026cyberthreateval} is one of the strongest methodological arguments in the corpus: evaluation design choices shape performance conclusions, and model-centric lexical metrics can overestimate performance on operationally realistic tasks.

\textbf{Why this review is needed.} The preceding tensions among capability and validation, cloud and on-premise deployment, and benchmark and workflow realism are not resolved by any single paper in the existing literature. They require synthesis across the full corpus. This survey provides that synthesis, covering 74 items across twelve thematic folders plus one excluded/peripheral category, with an explicit critical appraisal of evaluation quality, adversarial robustness, ethical rigour, and workflow coverage. It advances beyond prior surveys by: (a) treating agentic AI as a distinct architectural category requiring dedicated evaluation standards; (b) identifying the hallucination-validation gap as a corpus-level finding rather than an incidental limitation of individual papers; (c) providing a cross-study OSINT workflow coverage analysis that reveals systematic blind spots in the literature; and (d) proposing a structured research agenda derived from the evidence rather than speculating from capability trends.

\subsection{Contributions and Review Organisation}
\label{sec:1-2-contributions-and-review-organisation}

This survey makes the following primary contributions to the literature:

\begin{itemize}

  \item \textbf{A systematic 74-item corpus} (see Table~\ref{tab:table-2} and the supplementary Tables~S1--S2), with a cross-batch evidence matrix comprising 21 data fields per primary study and a folder-to-theme mapping documenting the relationship between the pre-organized thematic corpus structure and the 11-category taxonomy developed through analysis.

  \item \textbf{An 11-category taxonomy} of agentic and generative AI in OSINT and Cyber Investigation (Table~\ref{tab:table-3}, Section~3), spanning LLM foundations, LLM-based OSINT workflows, agentic AI and tool orchestration, APIs and toolkits, RAG and knowledge graph architectures, cyber threat intelligence, prompt engineering and chain-of-thought reasoning, fine-tuning and domain adaptation, evaluation and benchmarking, risks and ethics, and dark web and specialized sources.

  \item \textbf{Identification of the hallucination-validation gap} as the central evaluation failure of the field: hallucination is identified as a reliability concern in more than twenty studies but empirically measured---as an end-to-end rate on an OSINT system evaluated on OSINT data---in only one system \cite{allam2025cybervision}, establishing a critical quantitative finding of this review and an urgent priority for future research. (Reasoning-error and factual-correction signals in adjacent domains, such as \cite{zhao2023verifyandedit}, are measured more widely but are distinct from an OSINT hallucination rate; see Section~\ref{sec:3-9-evaluation-benchmarks-and-performance}.)

  \item \textbf{A cross-study OSINT workflow coverage analysis} (Table~\ref{tab:table-6}, Section~4) demonstrating that collection and analysis are the most extensively studied workflow stages, whereas verification, reporting, dissemination, and decision support remain substantially underexplored, creating a structural imbalance with direct consequences for operational intelligence reliability.

  \item \textbf{An adversarial risk synthesis} covering CTI knowledge graph poisoning \cite{ranade2021fake}, multimodal geo-privacy attacks \cite{yang2024geolocator}, and OSINT community misinformation contamination \cite{niu2024bullshint}, showing that each major agentic OSINT capability identified in the corpus has a corresponding adversarial dual-use risk that no system paper addresses in its own evaluation design.

  \item \textbf{A structured ten-point research agenda} (Section~6) derived from the corpus evidence, addressing hallucination measurement, evaluation standardisation, adversarial robustness, dark web methodology, multimodal and multilingual OSINT, privacy and legal compliance, human-AI teaming, and reproducibility.

\end{itemize}

The remainder of this paper is organised as follows. Section II presents the systematic methodology. Section III introduces the 11-category taxonomy with critical analysis of each technical category. Section IV maps taxonomy findings to OSINT application domains across the intelligence lifecycle. Section V presents four case studies drawn from the strongest primary contributions. Section VI synthesizes research challenges and future directions. Section VII concludes. The supplementary material contains the complete evidence matrix and folder-to-theme mapping

\section{Methodology}
\label{sec:2-methodology}

This review follows a systematic, structured approach to corpus construction, data extraction, and synthesis, informed by PRISMA-style systematic review principles. The methodology was designed to balance the breadth required to cover a rapidly expanding interdisciplinary field spanning AI, cybersecurity, intelligence studies, and information science with the analytical depth needed to evaluate evidence quality rather than aggregate findings indiscriminately.

\subsection{Study Selection Criteria}
\label{sec:2-1-study-selection-criteria}

The corpus was constructed to include studies addressing the application of large language models, generative AI, or agentic AI to OSINT workflows, cyber threat intelligence, Cyber Investigation, or social media intelligence (SOCMINT), as well as studies making direct methodological contributions through prompt engineering, fine-tuning, evaluation framework design, or retrieval-augmented architectures with demonstrable applicability to OSINT contexts. A further tier of background and contextual literature was included to establish the foundational capabilities of LLM architecture, few-shot learning, chain-of-thought reasoning, and scaling dynamics on which all primary contributions depend.

\textbf{Inclusion criteria} were defined as follows. Studies were included if they: (i) primarily investigate the application of an LLM, generative model, or agentic AI system to an OSINT, CTI, Cyber Investigation, or SOCMINT task; (ii) make a direct methodological contribution in prompt engineering, fine-tuning, evaluation design, or RAG architecture whose transferability to OSINT contexts is clear and explicitly argued; (iii) provide foundational technical context for LLM capabilities that is directly cited by primary OSINT studies in the corpus; or (iv) represent a directly applicable risk, adversarial, or ethics study relevant to AI-augmented intelligence collection and analysis. Studies were included regardless of whether their findings were positive, negative, or mixed with respect to LLM performance, because the review's purpose is critical synthesis rather than advocacy.

\textbf{Exclusion criteria} were defined as follows. Studies were excluded from the main analytical argument if they: (i) address non-AI OSINT methodologies without any LLM or generative AI component; (ii) apply LLMs to domains entirely unrelated to OSINT, such as quantum computing, robotic surgery, or materials science with no discernible methodological transferability; (iii) focus solely on educational technology applications with no AI or intelligence analysis content; or (iv) provide insufficient recoverable content to support meaningful analysis, as with the CID-encoded PDF (\cite{zhou2024application}).

Two studies assigned study identifiers within the corpus: a gamification application for OSINT skill-building \cite{siddiq2024gamification} and a general OSINT reconnaissance overview predating the LLM era \cite{ram2023exploring} are catalogued in the corpus and appear in the supplementary Evidence Matrix (Table~S1) but are excluded from the main analytical argument on the grounds that neither contains LLM, generative AI, or agentic AI content relevant to the review's research questions. Two further studies, a national security policy paper \cite{gauthier2025osint} and a conflict forecasting preprint using OSINT-derived data \cite{teagan2025forecasting} are retained but classified as peripheral, as peripheral and cited only where their specific contextual contribution is directly relevant to a claim\cite{ranade2021fake}.

Studies with complete bibliographic details, institutional affiliation, and sufficient recoverable content were prioritized for detailed analysis. One paper, a Chinese-language CID-encoded PDF (\cite{zhou2024application}) was flagged as `\texttt{insufficient detail}` in the evidence matrix; because its content was not sufficiently recoverable, it cannot serve as primary evidence in the review.

The inclusion and exclusion criteria applied in this review are summarized in Table I.

\begin{table*}[htbp]
  \centering
  \caption{Study Selection Criteria Summary. Study identifiers are defined in the supplementary Evidence Matrix (Table~S1).}
  \label{tab:table-1}
  \footnotesize
  \setlength{\tabcolsep}{4pt}\renewcommand{\arraystretch}{1.15}
\begin{tabularx}{\textwidth}{@{}p{1.0cm} p{0.7cm} >{\hsize=.9\hsize\raggedright\arraybackslash}X >{\hsize=1.5\hsize\raggedright\arraybackslash}X >{\hsize=.6\hsize\raggedright\arraybackslash}X@{}}
\toprule
\textbf{Type} & \textbf{Label} & \textbf{Description} & \textbf{Studies Affected} & \textbf{Notes} \\
\midrule
Inclusion & (i) & Primary application of LLM, generative, or agentic AI to OSINT, CTI, Cyber Investigation, or SOCMINT & \cite{rajendran2024comprehensive}, \cite{karakikes2025aiosint}, \cite{pastorgalindo2020osint}, \cite{su2024aiagent}, \cite{yuan2024empowering}, \cite{almeidapalmieri2025framework}, \cite{tseng2025disarm}, \cite{mukhopadhyay2024osintclinic}, \cite{radoi2023ai}, \cite{dhyani2024automated}, \cite{shamunesh2023cybercheck}, \cite{nagra2024kia}, \cite{sermakani2024open}, \cite{agrahari2025osintct2}, \cite{pais2014osint}, \cite{yurtalan2025redefining}, \cite{rheault2024active}, \cite{shen2024llmosint}, \cite{su2025opensource}, \cite{allam2025cybervision}, \cite{zhao2023verifyandedit}, \cite{hassanin2024llms}, \cite{andreoni2024enhancing}, \cite{saddi2024genai}, \cite{nila2023llms}, \cite{yigit2024review}, \cite{dekens2023practical}, \cite{kobayashi2024subjective}, \cite{ranade2021fake}, \cite{yang2024geolocator}, \cite{niu2024bullshint}, \cite{mouthami2025political}, \cite{golda2024privacy}, \cite{pervez2023ease}, \cite{hwang2022osint} and \cite{amin2025darklens} & Core evidential tier (37 studies); primary argument across \S3--\S4 \\
\addlinespace[2pt]
Inclusion & (ii) & Methodological contribution in prompting, fine-tuning, evaluation, or RAG with OSINT transferability & \cite{park2024performance}, \cite{cerny2024implications}, \cite{sun2025decision}, \cite{riad2024finetuning}, \cite{yang2024threatmodeling}, \cite{anas2024sentiments}, \cite{tihanyi2024cybermetric}, \cite{chen2026cyberthreateval}, \cite{sikand2024greencode}, \cite{srikanth2024usability} and\cite{shafee2024evaluation} & Methodological support (11); \cite{anas2024sentiments}, \cite{sikand2024greencode} cited at subsection level only \\
\addlinespace[2pt]
Inclusion & (iii) & Foundational LLM-capability context cited by primary OSINT studies & \cite{naveed2024overview}, \cite{bengesi2024advancements}, \cite{wei2022cot}, \cite{kaddour2023challenges}, \cite{rani2023comparative}, \cite{wei2022emergent}, \cite{brown2020fewshot} and \cite{hoffmann2022training} & Background (8); cited sparingly in \S3.1 only \\
\addlinespace[2pt]
Inclusion & (iv) & Risk, adversarial, or ethics study relevant to AI-augmented collection or analysis & \cite{rheault2024active}, \cite{yigit2024review,dekens2023practical,kobayashi2024subjective,ranade2021fake,yang2024geolocator,niu2024bullshint}, \cite{golda2024privacy} and \cite{pervez2023ease} & Within Primary/Core tier; cross-cited in \S3.10, \S6 \\
\addlinespace[2pt]
Exclusion & (i) & Non-AI OSINT methodology with no generative component & \cite{ram2023exploring} & Pre-LLM reconnaissance primer \\
\addlinespace[2pt]
Exclusion & (ii) & LLMs applied to non-OSINT domains with no transferability & \cite{trummer2024llmprinciples}, \cite{ferrag2025reasoning}, \cite{hagos2024recentadvances}, \cite{zhuang2024research}, \cite{chen2024trainstrategy}, \cite{wang2024grovergpt}, \cite{wang2024bim}, \cite{bilgram2023accelerating}, \cite{ayemowa2024recommender}, \cite{shukhman2024intent}, \cite{celik2024dawn}, \cite{majumder2024computervision}, \cite{arulmohan2023extracting} and \cite{guan2024integrated} & 14 studies (quantum, surgery, drug discovery, space) \\
\addlinespace[2pt]
Exclusion & (iii) & Educational technology with no AI or analysis content & \cite{siddiq2024gamification} & OSINT-training gamification app \\
\addlinespace[2pt]
Note & --- & Included despite limited content recovery & \cite{zhou2024application} & CID-encoded PDF; partial extraction only; retained in Primary/Core with caveats \\
\addlinespace[2pt]
Conditional & --- & Peripheral context only; not cited as primary evidence & \cite{gauthier2025osint}, \cite{teagan2025forecasting} & \cite{gauthier2025osint} IC policy paper; \cite{teagan2025forecasting} conflict-forecasting preprint \\
\addlinespace[2pt]
Note & --- & Identifiers assigned but excluded due to folder misplacement & \cite{dhyani2024automated}, \cite{andreoni2024enhancing} & \cite{dhyani2024automated} API-doc generator; \cite{andreoni2024enhancing} vehicle security \\
\bottomrule
\end{tabularx}
\end{table*}

\subsection{Search Strategy and Corpus Construction}
\label{sec:2-2-search-strategy-and-corpus-construction}

The corpus was assembled from a curated collection of research papers organized into twelve thematic folders corresponding to the review's primary research domains before systematic analysis. This pre-organized structure was then validated through explicit reassignment of misplaced papers (\cite{dhyani2024automated}, \cite{andreoni2024enhancing}), verification of relevance ratings, and cross-batch consistency checking throughout the eight-phase analysis process.

The thematic scope of corpus construction was guided by the following Scopus-style conceptual search string, which reflects the principal terminology used in the target literature:

\begin{quote}
(`\texttt{open-source intelligence}\texttt{ OR }\texttt{OSINT}\texttt{ OR }\texttt{open source intelligence}\texttt{ OR }\texttt{SOCMINT}\texttt{ OR }\texttt{social media intelligence}\texttt{ OR }\texttt{cyber threat intelligence}\texttt{ OR }\texttt{CTI}\texttt{ OR }\texttt{Cyber Investigation}\texttt{) AND (}\texttt{large language model}\texttt{ OR }\texttt{LLM}\texttt{ OR }\texttt{generative AI}\texttt{ OR }\texttt{generative artificial intelligence}\texttt{ OR }\texttt{GPT}\texttt{ OR }\texttt{transformer}\texttt{ OR }\texttt{agentic AI}\texttt{ OR }\texttt{agent}\texttt{ OR }\texttt{retrieval-augmented generation}\texttt{ OR }\texttt{RAG}\texttt{ OR }\texttt{knowledge graph}\texttt{) AND (}\texttt{intelligence analysis}\texttt{ OR }\texttt{information extraction}\texttt{ OR }\texttt{named entity recognition}\texttt{ OR }\texttt{NER}\texttt{ OR }\texttt{hallucination}\texttt{ OR }\texttt{misinformation}\texttt{ OR }\texttt{prompt engineering}\texttt{ OR }\texttt{fine-tuning}\texttt{ OR }\texttt{evaluation}\texttt{ OR }\texttt{benchmark}\texttt{ OR }\texttt{dark web}\texttt{ OR }\texttt{privacy}\texttt{ OR }\texttt{ethics}`)
\end{quote}

Papers were sourced from the ACM Digital Library, IEEE Xplore, arXiv, Google Scholar, and institutional repositories. Supplementary papers referenced within high-relevance primary studies were assessed for inclusion through forward and backward citation analysis. The search string above characterizes the conceptual scope applied to candidate-paper identification. The corpus as presented reflects expert curation within this scope rather than a fully reproducible automated database export, and independent replication using the same search string may yield a different but overlapping candidate set depending on the databases and date ranges applied.

\begin{figure}[htbp]  
    \includegraphics[width=1.03\columnwidth]{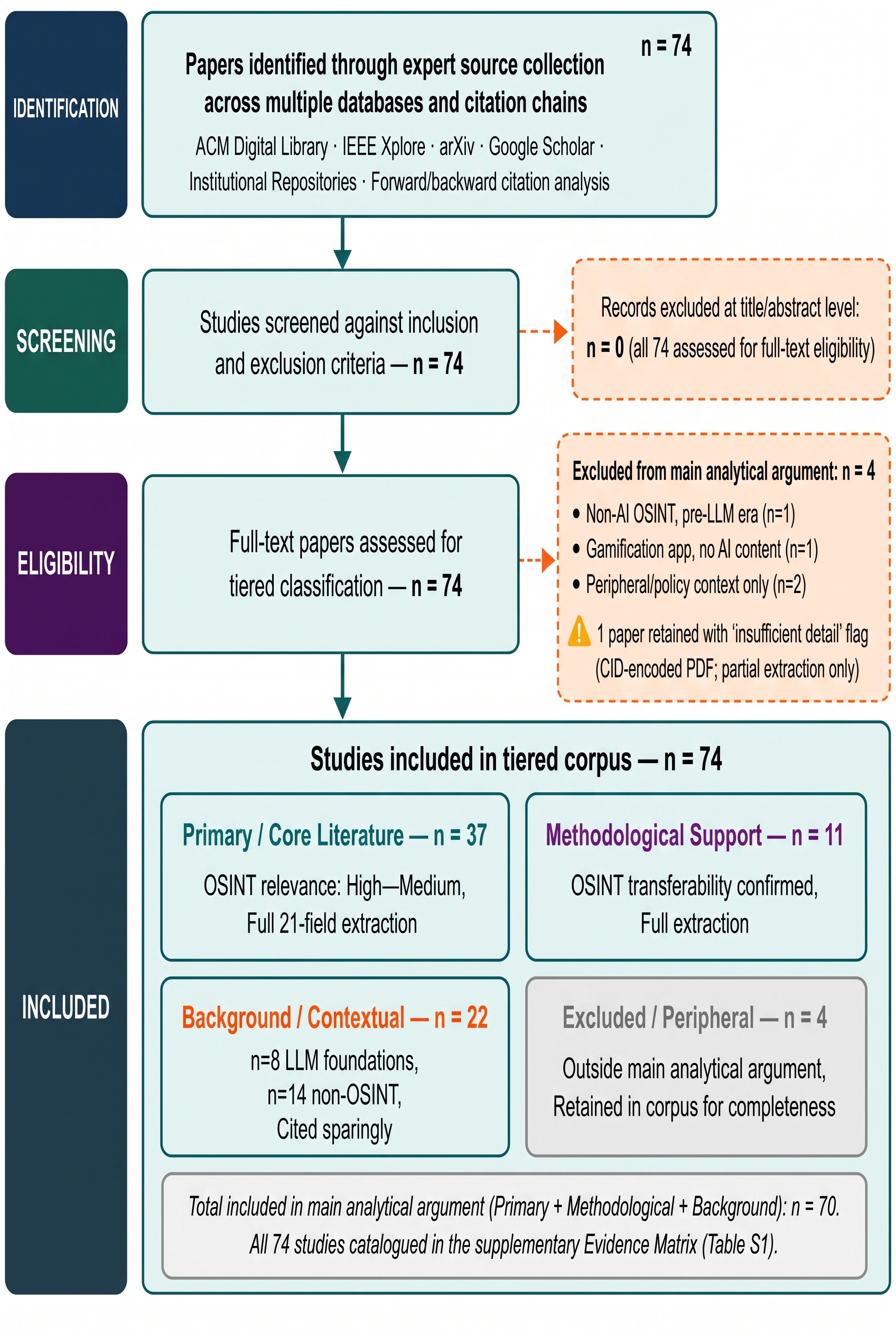}  
  \caption{PRISMA-style corpus selection and classification flow. Source counts are verified from corpus analysis; the flow begins at project source collection and excludes automated database-export counts.}  
  \label{fig:prisma_flow}  
\end{figure}

\textbf{Search window and search log.} Database and repository searches, together with the forward and backward citation snowballing described above, were conducted during the first half of 2026 (the precise search dates should be substituted here if a reviewer requires an exact, replayable log). Because the corpus was assembled by expert curation within the conceptual scope above rather than by a single automated Boolean export, this review does not report per-database hit counts of the form \emph{Scopus}~$n$, \emph{IEEE Xplore}~$n$, and so on that a fully automated query protocol would yield; presenting such figures retrospectively would misrepresent how the corpus was actually identified. The screening that \emph{was} performed---records identified, duplicates removed, records screened, full texts assessed for eligibility, and studies included---is documented in the PRISMA-style flow of Figure~\ref{fig:prisma_flow}, and the final disposition of every item (including the fourteen non-OSINT exclusions and the four peripheral retentions) is enumerated in Table~\ref{tab:table-2} and the supplementary Evidence Matrix (Table~S1). The included corpus spans publications from 2014 \cite{pais2014osint} to 2026 \cite{chen2026cyberthreateval}, with the temporal distribution detailed below; the reproducibility limitation that expert curation entails is stated explicitly in Section~\ref{sec:7-limitations}.

Following identification, all candidate papers were assessed against the inclusion and exclusion criteria in Section 2.1. Each included paper was assigned a stable study identifier (see Table~\ref{tab:table-2} and the supplementary Folder-to-Theme Mapping, Table~S2) within its thematic folder. Two duplicate file pairs were identified because the same papers had been stored under variant filenames within the collection. These were resolved by retaining the primary file and subsuming the duplicate under the same study identifier. The complete folder-to-study mapping is provided in the supplementary Folder-to-Theme Mapping (Table~S2).

The corpus construction process is summarised in Figure 1.

\begin{table*}[htbp]
  \centering
  \caption{Corpus classification by thematic folder, analytical role, and OSINT relevance tier. Study identifiers (S01--S74) index the supplementary Evidence Matrix (Table~S1); ${\dagger}$ marks a study cross-listed across folders.}
  \label{tab:table-2}
  \footnotesize
  \setlength{\tabcolsep}{4pt}\renewcommand{\arraystretch}{1.15}
\begin{tabularx}{\textwidth}{@{}>{\raggedright\arraybackslash}p{0.235\linewidth} >{\raggedright\arraybackslash}X c >{\raggedright\arraybackslash}p{0.165\linewidth} l c@{}}
\toprule
\textbf{Folder} & \textbf{Study IDs} & \textbf{\#} & \textbf{Analytical Role} & \textbf{Relevance} & \textbf{\S} \\
\midrule
01 Foundations of LLMs & \cite{naveed2024overview}, \cite{bengesi2024advancements}, \cite{wei2022cot}, \cite{kaddour2023challenges}, \cite{rani2023comparative}, \cite{wei2022emergent}, \cite{brown2020fewshot}, \cite{hoffmann2022training} & 8 & Background & Background & 3.1 \\
02 LLMs for OSINT Overview & \cite{rajendran2024comprehensive}, \cite{karakikes2025aiosint}, \cite{zhou2024application}, \cite{pastorgalindo2020osint}, \cite{tseng2025disarm} & 5 & Primary/Core & Med--High & 3.2 \\
03 OSINT Agents and Tool Use & \cite{su2024aiagent}, \cite{yuan2024empowering}, \cite{almeidapalmieri2025framework}, \cite{mukhopadhyay2024osintclinic} & 4 & Primary/Core & Low--High & 3.3 \\
04 APIs, Toolkits, Automation & \cite{radoi2023ai}, \cite{dhyani2024automated}, \cite{shamunesh2023cybercheck}, \cite{nagra2024kia}, \cite{sermakani2024open}, \cite{agrahari2025osintct2}, \cite{pais2014osint}, \cite{yurtalan2025redefining} & 8 & Primary/Core (\cite{dhyani2024automated} excl.) & Med--High & 3.4 \\
05 Knowledge Graphs, RAG, Memory & \cite{rheault2024active}, \cite{shen2024llmosint}, \cite{su2025opensource}, \cite{allam2025cybervision}, \cite{zhao2023verifyandedit} & 5 & Primary/Core & Low--High & 3.5 \\
06 Cyber Threat Intelligence & \cite{hassanin2024llms}, \cite{andreoni2024enhancing}, \cite{saddi2024genai}, \cite{nila2023llms}, \cite{yigit2024review}, \cite{ranade2021fake}$^{\dagger}$ & 6 & Primary/Core (\cite{andreoni2024enhancing,saddi2024genai} min.) & Low--High & 3.6 \\
07 Prompt Engineering and CoT & \cite{park2024performance}, \cite{cerny2024implications}, \cite{sun2025decision} & 3 & Methodological & Med--High & 3.7 \\
08 Fine-Tuning, Domain Adaptation & \cite{riad2024finetuning}, \cite{yang2024threatmodeling}, \cite{anas2024sentiments} & 3 & Methodological & Low--Med & 3.8 \\
09 Evaluation and Benchmarks & \cite{tihanyi2024cybermetric}, \cite{chen2026cyberthreateval}, \cite{sikand2024greencode}, \cite{srikanth2024usability}, \cite{shafee2024evaluation} & 5 & Methodological (\cite{sikand2024greencode} excl.) & Low--High & 3.9 \\
10 Risks, Ethics, Hallucination & \cite{dekens2023practical}, \cite{kobayashi2024subjective}, \cite{yang2024geolocator}, \cite{niu2024bullshint}, \cite{mouthami2025political}, \cite{golda2024privacy}, \cite{pervez2023ease}, \cite{hwang2022osint}, \cite{ranade2021fake}$^{\dagger}$ & 9 & Primary/Core (\cite{hwang2022osint} pre-LLM era; cited as context only) & Low--High & 3.10 \\
11 Dark Web, Specialized Sources & \cite{amin2025darklens} & 1 & Primary/Core & Medium & 3.11 \\
12 Other (Non-OSINT) & \cite{trummer2024llmprinciples}, \cite{ferrag2025reasoning}, \cite{hagos2024recentadvances}, \cite{zhuang2024research}, \cite{chen2024trainstrategy}, \cite{wang2024grovergpt}, \cite{wang2024bim}, \cite{bilgram2023accelerating}, \cite{ayemowa2024recommender}, \cite{shukhman2024intent}, \cite{celik2024dawn}, \cite{majumder2024computervision}, \cite{arulmohan2023extracting}, \cite{guan2024integrated} & 14 & Background (excluded) & Background & --- \\
99 Excluded or Peripheral & \cite{gauthier2025osint}, \cite{siddiq2024gamification}, \cite{ram2023exploring}, \cite{teagan2025forecasting} & 4 & Excluded/Peripheral & Excluded & --- \\
\bottomrule
\end{tabularx}
\end{table*}

\smallskip\noindent\textit{Note on counts: the per-folder values in the ``\#'' column sum to 75 \emph{listings}, not 74 studies. The discrepancy is a single cross-listing: \cite{ranade2021fake} (Ranade et al.) appears in both Folder~06 (primary assignment) and Folder~10 (thematic citation) and is marked $^{\dagger}$ in both rows, but is counted once toward the corpus total. The corpus therefore comprises 74 unique studies, consistent with the analytical-tier totals in Section~2.3 (37 primary/core $+$ 11 methodological $+$ 22 background $+$ 4 excluded/peripheral $=$ 74). \cite{dekens2023practical,kobayashi2024subjective} were physically filed in Folder~06 during source collection but are primarily assigned to Folder~10 for this review and are counted only once, under Folder~10. The complete folder-to-theme mapping is given in the supplementary Folder-to-Theme Mapping (Table~S2).}\smallskip
The temporal coverage of the corpus reflects the rapid acceleration of LLM research following the emergence of large-scale generative models from 2020 onwards and the subsequent wave of public deployment from 2022. Although a small number of pre-LLM-era OSINT framework studies date back to 2014 \cite{pais2014osint} and the foundational generative-model papers begin in 2020 \cite{brown2020fewshot}, most corpus papers were published between 2023 and 2025, and the most recent CTI-evaluation contribution is dated 2026 \cite{chen2026cyberthreateval}. Approximately eleven papers in the corpus do not carry recoverable publication dates and are noted as undated in the evidence matrix; the findings of these papers are treated with corresponding caution where temporal context is relevant.

\subsection{Screening and Categorisation}
\label{sec:2-3-screening-and-categorisation}

\begin{figure*}[!t]
  \centering
  \includegraphics[width=0.85\linewidth]{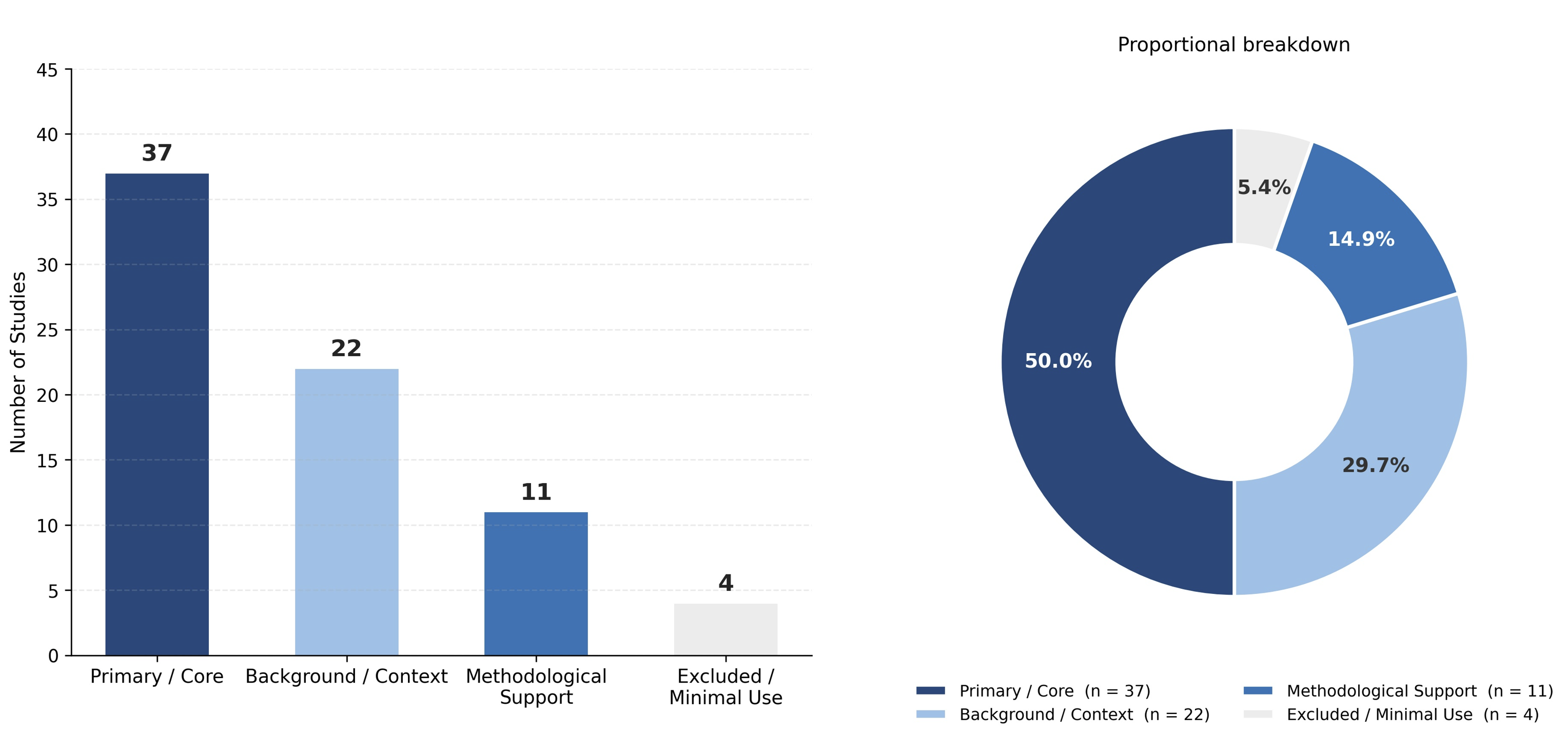}
  \caption{Number of Studies by Analytical Category}
  \label{fig:chart1}
\end{figure*}

\begin{figure}[htbp]
  \centering
  \includegraphics[width=1.05\columnwidth]{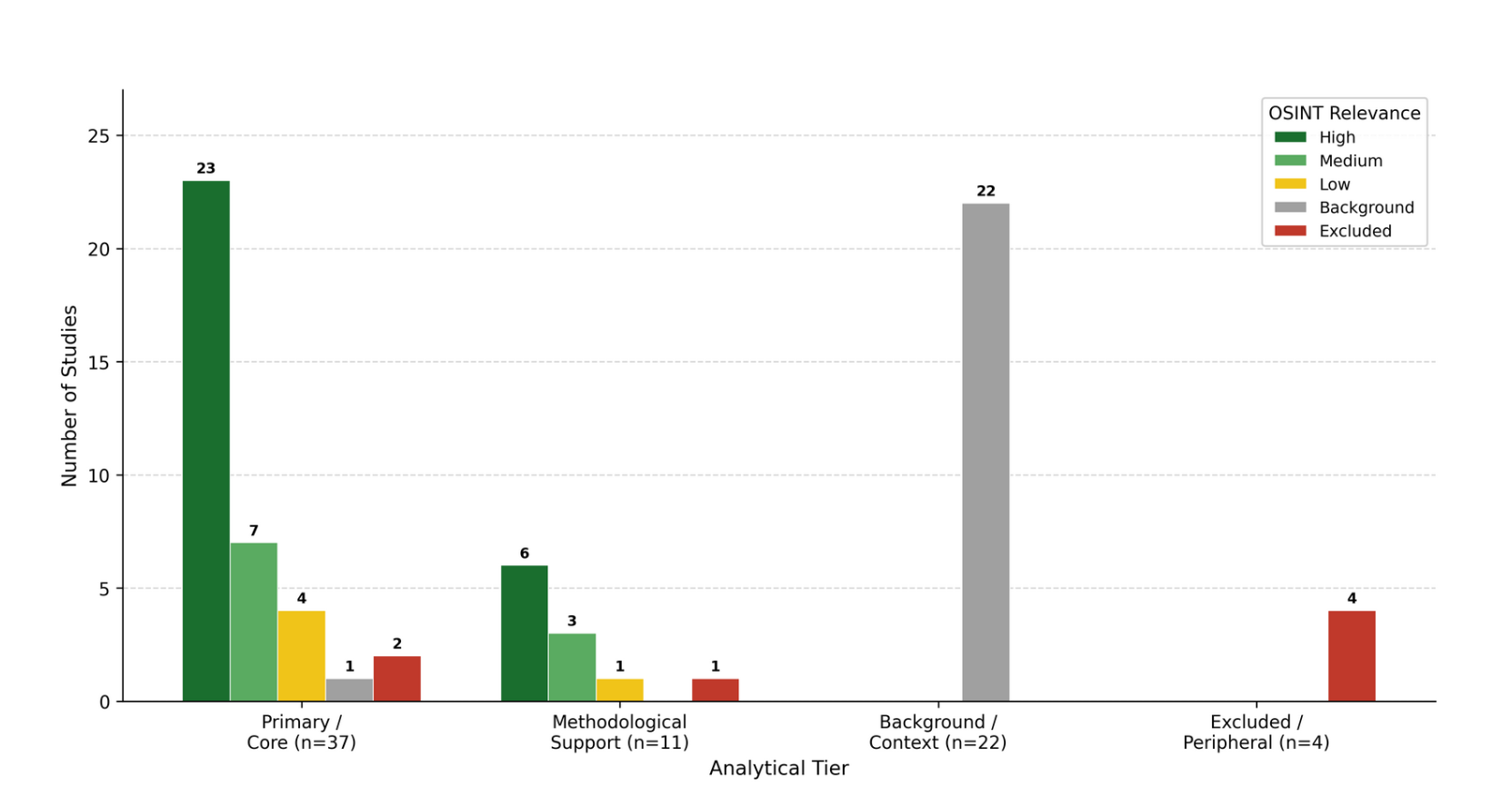}
  \caption{OSINT Relevance Rating Distribution by Analytical Tier}
  \label{fig:chart2}
\end{figure}

All 74 studies were classified into one of four analytical tiers based on their role in the review and their proximity to the review's primary research questions. This tiered classification determines how prominently each study appears in the synthesis and how much weight is assigned to its findings. The tier distribution is summarised in Figure~\ref{fig:chart1} and Figure~\ref{fig:chart2}.

\textbf{Primary and Core Literature (37 studies; see Table~\ref{tab:table-2})} encompasses papers whose primary subject matter is the application of LLMs, generative AI, or agentic AI to OSINT, CTI, Cyber Investigation, or SOCMINT tasks. These papers constitute the evidential core of the review and are analyzed using the full 21-field evidence matrix described in Section~2.4. They include agentic OSINT frameworks \cite{almeidapalmieri2025framework,mukhopadhyay2024osintclinic,shen2024llmosint}, RAG-augmented intelligence systems \cite{allam2025cybervision,zhao2023verifyandedit}, adversarial risk demonstrations \cite{ranade2021fake,yang2024geolocator,niu2024bullshint}, and ethical and governance analyses \cite{rheault2024active,yigit2024review,dekens2023practical,kobayashi2024subjective,golda2024privacy}.

\textbf{Methodological Support Literature (11 studies; see Table~\ref{tab:table-2})} comprises papers that make direct methodological contributions in prompt engineering, domain-specific fine-tuning, or evaluation benchmarking, with findings transferable to OSINT contexts even when the primary application domain is not OSINT-specific. This category includes the CyberThreat-Eval evaluation framework derived from real analyst workflows \cite{chen2026cyberthreateval}, the LASIGE LLM chatbot evaluation on OSINT-derived Twitter CTI data \cite{shafee2024evaluation}, the LLM CTI usability study with practising security professionals \cite{srikanth2024usability}, and the prompt engineering framework for OSINT collection \cite{cerny2024implications}.

\textbf{Background and Contextual Literature (22 studies; see Table~\ref{tab:table-2})} consists of foundational LLM papers and non-OSINT application studies whose role is to establish the technological context for the primary literature rather than to provide OSINT-specific evidence. Background papers are cited sparingly and only where they establish foundational capabilities, such as few-shot learning \cite{brown2020fewshot}, chain-of-thought reasoning \cite{wei2022cot}, emergent abilities \cite{wei2022emergent}, and LLM challenge taxonomies \cite{kaddour2023challenges} that cannot be adequately established by citing a primary OSINT paper directly. This tier must not dominate the review's argument; its purpose is contextualisation, not corroboration.

\textbf{Excluded and Minimal-Use Literature (4 studies; see Table~\ref{tab:table-2})} comprises papers outside the OSINT AI scope or conditionally useful for peripheral contextualisation only. A gamification application for OSINT skill training \cite{siddiq2024gamification} and a general OSINT reconnaissance overview predating the LLM era \cite{ram2023exploring} are excluded from the main analytical argument and are not cited as evidence in Sections~3 through~7. A national security policy paper \cite{gauthier2025osint} and a conflict forecasting preprint \cite{teagan2025forecasting} are cited only conditionally where their specific contextual value is directly relevant.

Several specific categorisation decisions warrant explicit justification. Two papers were found to be misplaced in their assigned thematic folders: an API documentation generator \cite{dhyani2024automated} had been filed within the OSINT APIs and Toolkits folder, and a study of autonomous vehicle and robotic security systems \cite{andreoni2024enhancing} had been filed within the Cyber Threat Intelligence folder. Neither paper contains substantive OSINT or intelligence analysis content, and both are excluded from the main narrative. Their misplacement is recorded in the supplementary Folder-to-Theme Mapping (Table~S2). Two further papers by the same sole author, from the same institutional affiliation and publication venue \cite{su2024aiagent,su2025opensource}, are IEEE IAEAC submissions proposing LLM-based OSINT systems for military contexts without empirical evaluation. They are retained in the corpus but cited only as illustrative background, not as primary evidence of system performance.

\subsection{Data Extraction and Coding}
\label{sec:2-4-data-extraction-and-coding}

A structured evidence matrix was applied to all primary and methodological support papers (48 studies; see Table~\ref{tab:table-2}) to ensure systematic, comparable, and transparent extraction of the information required to address the review's research questions. The complete evidence matrix is provided in the supplementary Evidence Matrix (Table~S1).

For primary and core papers, a 21-field extraction schema was applied, encompassing the following fields: study ID; filename; thematic folder; full title; authors; year; venue or source; main aim; research type; main technology; OSINT relevance rating (1--5); OSINT workflow stage or stages covered; data sources used; models, tools, APIs, or frameworks; evaluation method and metrics; hallucination measured (Yes / No / Rate); key findings; key strengths; key weaknesses; methodological limitations; risks and ethical issues; relationship to other studies in the corpus; contribution to the SLR; and suggested SLR section or sections.

The explicit inclusion of a hallucination measurement field, recorded as a binary yes/no indicator and, where an empirical rate was reported, as the reported value, was a deliberate analytical choice reflecting the centrality of hallucination to the review’s research questions. Its consistent completion across all 48 primary and methodological support papers makes the corpus-level finding transparent and directly verifiable from the supplementary Evidence Matrix (Table~S1) rather than dependent on the reviewer's interpretation: of 48 papers analyzed using the full 21-field schema, only one \cite{allam2025cybervision} records an empirical hallucination rate for an OSINT-specific system.

For background and contextual papers (22 studies; see Table~\ref{tab:table-2}), a condensed eight-field schema was applied, covering: study ID; title; year; main topic; reason for background classification; concept supported in the SLR; citation recommendation; and useful figures, tables, or metrics relevant to the review's argument.

Where a study was CID-encoded and text extraction was incomplete, specifically \cite{zhou2024application} (a Chinese-language document), this is recorded in the evidence matrix as `\texttt{Insufficient detail}'. Claims derived from this paper are not presented as primary evidence in the review. \cite{pervez2023ease} (Pervez et al.), whilst partially CID-encoded, yielded sufficient recoverable content for the substantive citations made at Sections 3.4, 3.10, and 6.6.

Extraction was conducted across eight sequential batches corresponding to the corpus's thematic folder structure, allowing cross-batch analytical patterns, including the hallucination-validation gap and the workflow coverage imbalance, to emerge progressively rather than be imposed retrospectively.

\subsection{Synthesis Method}
\label{sec:2-5-synthesis-method}

The synthesis approach is thematic and comparative, organised around the 11-category taxonomy described in Section~3 and the OSINT workflow coverage analysis detailed in Table~\ref{tab:table-6}. Quantitative meta-analysis is inappropriate for this heterogeneous corpus; narrative synthesis is the more appropriate alternative. Four types of synthetic statement are distinguished: \textbf{consensus findings} supported consistently across multiple independent studies; \textbf{tensions} reflecting genuine unresolved disagreements; \textbf{corpus-level observations} that emerge from examining the corpus as a whole --- the OSINT workflow coverage imbalance is grounded in all 74 studies, whilst the hallucination-validation gap is derived from systematic measurement across the 48 primary and methodological support papers subjected to full 21-field extraction (Section~2.4); and \textbf{gaps} absent from the corpus despite operational relevance, used to motivate the research agenda in Section~6.

\section{Taxonomy of Agentic and Generative AI in OSINT and Cyber Investigation}
\label{sec:3-taxonomy-of-agentic-and-generative-ai-in-osint-and-digital}

The 74 studies in this corpus do not constitute a uniform body of research. They differ substantially in technological scope, from foundational language-model architectures to fully agentic pipeline systems, as well as in the OSINT workflow stages they address, their empirical depth, and the extent to which their findings can be generalized beyond the specific conditions under which they were obtained. A taxonomy is therefore a necessary organizing tool rather than a merely descriptive convenience: it establishes the conceptual architecture of the field, defines what distinguishes technically distinct approaches, and provides the analytical framework for the comparative assessments in Sections 4 through 6.

The taxonomy proposed in this review comprises eleven categories, organised into five conceptual groups: foundational capabilities (Section 3.1); OSINT-facing applications of those capabilities at increasing levels of architectural complexity (Sections 3.2 through 3.4); knowledge infrastructure for grounding and reliability (Section 3.5); the domain of cyber threat intelligence as the most mature application area (Section 3.6); methodological enablers in prompting and adaptation (Sections 3.7 and 3.8); evaluation frameworks (Section 3.9); risk and governance dimensions (Section 3.10); and specialized source environments (Section 3.11). Table 3 summarises the taxonomy, while Figure 4 illustrates it hierarchically.

No category in the taxonomy operates effectively in isolation: agentic architectures (Section~3.3) depend on chain-of-thought reasoning (Section~3.7) and tool infrastructure (Section~3.4); RAG and knowledge graph systems (Section~3.5) address the hallucination risks that evaluation frameworks (Section~3.9) reveal; risk and governance considerations (Section~3.10) apply to every category simultaneously. The central claim of this review applies across all eleven categories: agentic and generative AI can improve OSINT, but only when combined with knowledge grounding, provenance mechanisms, careful evaluation, and structured human oversight.

\begin{table*}[htbp]
  \centering
  \caption{Taxonomy of agentic and generative AI in OSINT and Cyber Investigation. Study identifiers index the supplementary Evidence Matrix (Table~S1).}
  \label{tab:table-3}
  \footnotesize
  \setlength{\tabcolsep}{4pt}\renewcommand{\arraystretch}{1.15}
\begin{tabularx}{\textwidth}{@{}>{\raggedright\arraybackslash}p{0.13\linewidth} c >{\raggedright\arraybackslash}X >{\raggedright\arraybackslash}p{0.12\linewidth} >{\raggedright\arraybackslash}X >{\raggedright\arraybackslash}X@{}}
\toprule
\textbf{Category} & \textbf{\S} & \textbf{Definition} & \textbf{Rep.\ Studies} & \textbf{Key Strengths} & \textbf{Key Weaknesses} \\
\midrule
Foundations of LLMs & 3.1 & Pre-trained transformers with emergent few-shot, chain-of-thought, and scale-dependent capabilities enabling downstream OSINT use & \cite{wei2022cot}, \cite{kaddour2023challenges}, \cite{wei2022emergent}, \cite{brown2020fewshot}, \cite{hoffmann2022training} & Establishes scale and architectural constraints; motivates large-model requirement & Foundational only; no OSINT evaluation evidence \\
\addlinespace[2pt]
LLMs for OSINT Workflows & 3.2 & Instruction-tuned LLMs applied to discrete pipeline tasks: query formulation, entity extraction, summarisation & \cite{rajendran2024comprehensive}, \cite{pastorgalindo2020osint}, \cite{tseng2025disarm}, \cite{radoi2023ai}, \cite{shafee2024evaluation} & Broadest application scope; cross-platform generality & Mostly survey-level/single-task; inconsistent evaluation \\
\addlinespace[2pt]
Agentic AI and Tool Use & 3.3 & ReAct or multi-step loops in which an LLM autonomously selects and invokes tools for multi-stage OSINT & \cite{yuan2024empowering}, \cite{almeidapalmieri2025framework}, \cite{mukhopadhyay2024osintclinic}, \cite{shen2024llmosint} & Highest capability ceiling; pipeline-level automation & Benign-only evaluation; private task sets; oversight unvalidated \\
\addlinespace[2pt]
APIs, Toolkits, Orchestration & 3.4 & LLMs integrated with OSINT APIs and orchestration frameworks for automated collection and enrichment & \cite{radoi2023ai}, \cite{shamunesh2023cybercheck,nagra2024kia,sermakani2024open,agrahari2025osintct2,pais2014osint,yurtalan2025redefining}, \cite{pervez2023ease} & Operationally realistic; connects capability to deployed tools & Cloud deployment raises GDPR risk; infrastructure exposure shown; on-premise constrains scale \\
\addlinespace[2pt]
RAG, KGs, Memory & 3.5 & Outputs grounded in external knowledge via retrieval, knowledge-graph enrichment, or persistent memory & \cite{shen2024llmosint,su2025opensource,allam2025cybervision,zhao2023verifyandedit}, \cite{sun2025decision} & Directly addresses hallucination; only OSINT hallucination rate (4\%, \cite{allam2025cybervision}) & Private knowledge bases; grounding depends on source verification \\
\addlinespace[2pt]
CTI and Security OSINT & 3.6 & LLMs and agents for threat-report generation, vulnerability classification, attribution, and analyst support & \cite{hassanin2024llms}, \cite{nila2023llms}, \cite{yigit2024review}, \cite{ranade2021fake}, \cite{tihanyi2024cybermetric}, \cite{chen2026cyberthreateval}, \cite{srikanth2024usability}, \cite{shafee2024evaluation} & Most mature domain; two evaluation frameworks; usability evidence & No LLM adequate by most valid benchmark (\cite{chen2026cyberthreateval}); adversarial CTI shown (\cite{ranade2021fake}) \\
\addlinespace[2pt]
Prompt Engineering and CoT & 3.7 & Structured prompting (zero/few-shot, CoT, decomposition, self-reflection) improving OSINT reasoning & \cite{park2024performance}, \cite{cerny2024implications}, \cite{sun2025decision}, \cite{wei2022cot} & Low-cost gains; transferable to analyst practice; bias reduction & No OSINT-specific outcome evidence; cross-language generality uncertain \\
\addlinespace[2pt]
Fine-Tuning, Domain Adaptation & 3.8 & Supervised or parameter-efficient adaptation on domain OSINT/security datasets & \cite{tseng2025disarm}, \cite{hassanin2024llms}, \cite{riad2024finetuning}, \cite{yang2024threatmodeling} & Outperforms zero-shot on domain tasks; PEFT enables on-premise & Private datasets; single-sector focus; cost and version stability \\
\addlinespace[2pt]
Evaluation and Benchmarks & 3.9 & Frameworks from classification F1 to analyst-centric workflow assessment & \cite{yuan2024empowering}, \cite{almeidapalmieri2025framework}, \cite{tihanyi2024cybermetric}, \cite{chen2026cyberthreateval}, \cite{srikanth2024usability}, \cite{shafee2024evaluation} & CyberThreat-Eval most valid; \cite{shafee2024evaluation} highest replicability & No shared open OSINT-AI benchmark; hallucination measured once (4\%, \cite{allam2025cybervision}) \\
\addlinespace[2pt]
Risks, Ethics, Hallucination & 3.10 & Adversarial, reliability, privacy, and governance risks of deployment & \cite{rheault2024active}, \cite{dekens2023practical,kobayashi2024subjective,ranade2021fake,yang2024geolocator,niu2024bullshint}, \cite{golda2024privacy}, \cite{pervez2023ease} & Risk landscape comprehensively documented with empirical evidence & Identification exceeds mitigation; no paper spans more than one risk category \\
\addlinespace[2pt]
Dark Web, Specialized Sources & 3.11 & Collection from Tor, encrypted platforms, and restricted environments & \cite{shen2024llmosint}, \cite{amin2025darklens} & \cite{shen2024llmosint} working integration; \cite{amin2025darklens} six-module design taxonomy & Only two studies; \cite{amin2025darklens} conceptual; legal and chain-of-custody risks unaddressed \\
\bottomrule
\end{tabularx}
\end{table*}

\begin{figure*}[!t]  
  \centering  
  \includegraphics[width=0.75\linewidth]{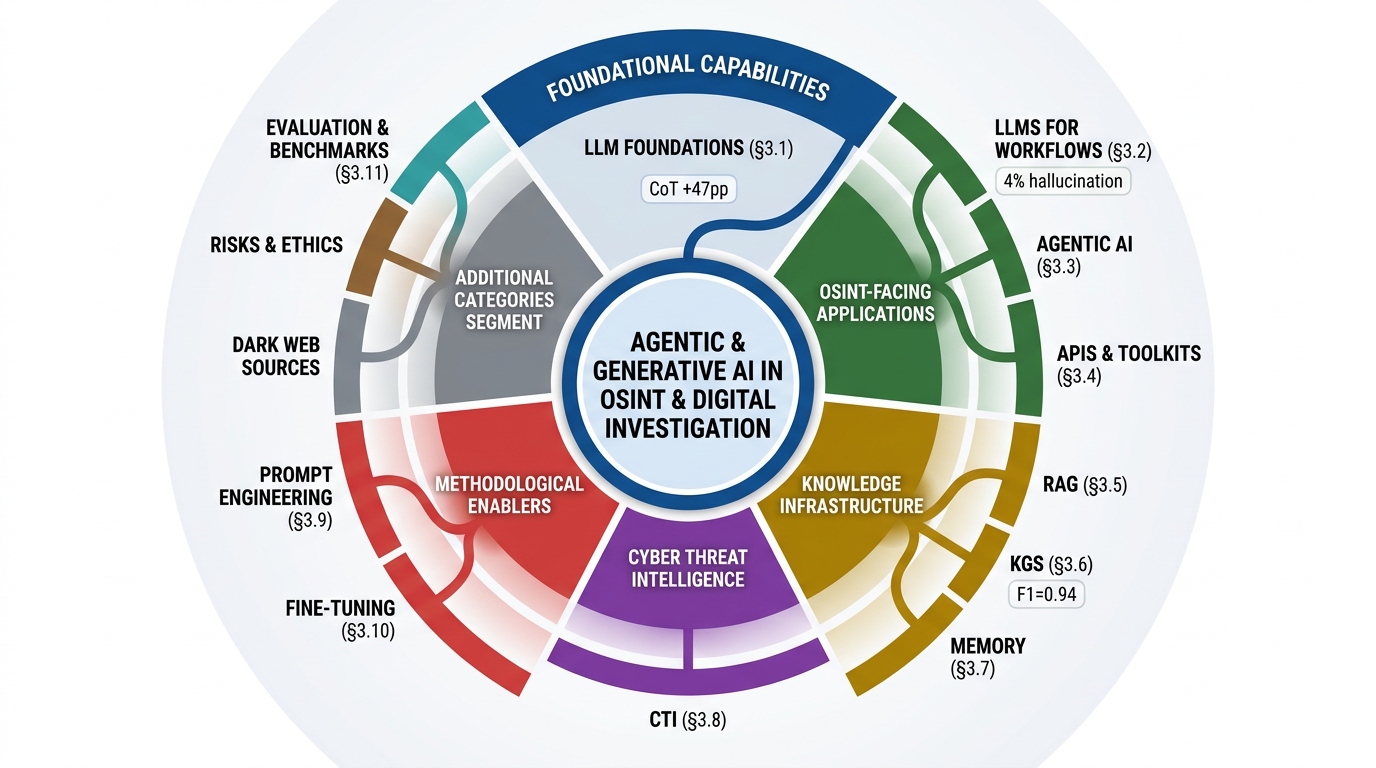}  
  \caption{Hierarchical Taxonomy Map --- Agentic and Generative AI in OSINT and Cyber Investigation. The taxonomy comprises eleven categories organised into five conceptual groups. Section references and representative study IDs are shown at the leaf level.}  
  \label{fig:taxonomy}  
\end{figure*}

\subsection{Foundations of LLMs and Transformer Models}
\label{sec:3-1-foundations-of-llms-and-transformer-models}

The capabilities that make LLMs useful for OSINT analysis, including few-shot task adaptation, chain-of-thought reasoning, and emergent multi-step problem-solving, are not arbitrary properties of modern neural architectures. They emerge from specific conditions of scale, training data volume, and architectural design that only became practically available in 2020 and were only understood theoretically shortly thereafter. For OSINT system designers, these foundational conditions are not merely background knowledge; they define practical constraints on model scale, deployment environment, and prompting strategy.

The paradigm shift underpinning all subsequent LLM-based OSINT work was established by Brown et al. \cite{brown2020fewshot}, who demonstrated that scaling language models to 175 billion parameters (GPT-3) produces strong few-shot performance across diverse natural language tasks without any gradient updates or task-specific fine-tuning. This in-context learning capability, whereby a model adapts its behavior to a new task through examples provided in the input prompt, is the direct precursor to OSINT prompt-engineering approaches in the corpus \cite{cerny2024implications,sun2025decision} and the foundation of the instruction-following behavior on which agentic OSINT systems rely\cite{almeidapalmieri2025framework,shen2024llmosint}. The significance of \cite{brown2020fewshot} for OSINT lies not in the specific tasks demonstrated but in the paradigm it establishes: intelligence tasks can be specified through natural language rather than through labelled training data, dramatically lowering the barrier to OSINT AI deployment for organisations without machine learning infrastructure.

Chain-of-thought prompting, introduced by Wei et al. \cite{wei2022cot}, extends this paradigm to multi-step reasoning tasks tasks by eliciting intermediate reasoning steps before a final answer is produced. On the GSM8K mathematical reasoning benchmark, PaLM 540B with eight chain-of-thought exemplars achieves 80\% accuracy compared to 33\% under standard prompting. This result is important less for its mathematical content than for its demonstration that complex structured reasoning is a prompting-accessible capability in sufficiently large models. For OSINT, this matters because the reasoning steps required for intelligence analysis, including formulating a hypothesis, selecting a collection approach, interpreting ambiguous evidence, and revising an assessment, are precisely the kind of sequential structured reasoning that chain-of-thought prompting enables. Every agentic OSINT architecture in the corpus that employs a thought-action-observation loop \cite{almeidapalmieri2025framework,shen2024llmosint} depends on this capability; without it, the agent reduces to brittle scripting rather than adaptive reasoning.

A critical constraint accompanies this capability: Wei et al. \cite{wei2022cot} show that chain-of-thought prompting is most reliable in very large models, with strong effects reported around the 100-billion-parameter scale Smaller models do not exhibit it predictably, regardless of prompting technique. This finding is directly relevant to the cloud-versus-on-premise architectural debate addressed in Section 3.4: models capable of the agentic reasoning demonstrated in \cite{almeidapalmieri2025framework} and \cite{shen2024llmosint} may may require computational infrastructure that many intelligence organizations cannot provision locally. The emergent abilities framework \cite{wei2022emergent} further establishes that such capabilities cannot be predicted by extrapolating from smaller model performance; they arise discontinuously at scale thresholds, making capability assessment for smaller deployment-feasible models unreliable.

A taxonomy of LLM challenges proposed by Kaddour et al. \cite{kaddour2023challenges} establishes that the problems encountered in the OSINT AI literature are not unique to OSINT but are recognized limitations of the LLM paradigm, spanning design challenges such as latency and context length, behavior challenges such as hallucination, outdated knowledge, and prompt brittleness, and science challenges such as brittle evaluations and poor reproducibility). Each maps directly onto findings in the primary OSINT corpus: hallucination \cite{allam2025cybervision,ranade2021fake}, prompt brittleness \cite{park2024performance,cerny2024implications}, brittle evaluations \cite{tihanyi2024cybermetric,chen2026cyberthreateval}, and poor reproducibility \cite{su2024aiagent,su2025opensource,saddi2024genai}. This contextualisation clarifies which research directions require OSINT-specific investment and which depend on broader advances in the underlying technology.

\subsection{LLMs and Generative AI for OSINT Workflows}
\label{sec:3-2-llms-and-generative-ai-for-osint-workflows}

The first wave of LLM integration into OSINT practice did not take the form of autonomous agentic systems. Instead, LLMs were applied to discrete, well-defined tasks within otherwise conventional intelligence workflows: classification of source documents, extraction of entities from open-source text, summarization of threat reports, and automated formatting of collection outputs. These applications establish the baseline capability level against which more architecturally ambitious approaches should be measured. They also reveal where the capability ceiling of LLM-as-tool approaches lies and why the transition to agentic architectures (Section 3.3) represents an architectural rather than merely incremental advance.

R\u{a}doi \cite{radoi2023ai} provides the earliest practical demonstration in the corpus of LLM-based OSINT automation, integrating GPT-3 via the OpenAI API into a collection pipeline that extracts and structures intelligence from publicly accessible web sources. This approach usefully demonstrates that the few-shot paradigm established in \cite{brown2020fewshot} transfers to OSINT collection tasks without OSINT-specific fine-tuning. However, the study overlooks the operational-security implications of routing intelligence queries through a commercial API, a limitation that subsequent work \cite{yurtalan2025redefining} identifies as a critical barrier to adoption in law-enforcement and classified-intelligence contexts.

The most directly applicable quantitative evidence for LLM performance on OSINT-relevant tasks is provided by Shafee et al. \cite{shafee2024evaluation}, who evaluate seven LLM chatbots on binary classification and named entity recognition (NER) using Twitter-sourced cybersecurity data. This is one of the few studies in the corpus to use real OSINT data for evaluation. For binary classification of cybersecurity-relevant content, GPT-4 achieves an F1 score of 0.94, a level that may be sufficient to support large-scale OSINT triage under human supervision. The open-source GPT4all model achieves F1 = 0.90 on the same task, a finding of considerable practical importance: the performance differential between a proprietary cloud model and an open-weight model deployable on local infrastructure is narrow enough to support on-premise deployment in organizations with operational-security constraints that preclude cloud API use \cite{yurtalan2025redefining}. However, the results change markedly for NER. Across all seven chatbots evaluated, performance on cybersecurity entity extraction falls approximately 25\% below that of specialized pre-trained models in zero-shot and few-shot settings. This deficit is not a minor calibration issue; it indicates a systematic failure to identify the threat indicators, actor names, and technical artifacts that constitute the fundamental currency of OSINT-derived intelligence.

This NER performance gap is analytically significant for two reasons. First, it is precisely measured and robustly evidenced on real OSINT data, making it one of the most reliable quantitative findings in the corpus. Second, it is entirely unaddressed by the fine-tuning literature within the corpus: whilst \cite{riad2024finetuning} fine-tunes for sentiment classification and \cite{yang2024threatmodeling} fine-tunes for banking threat modelling, no paper fine-tunes a model specifically for OSINT-domain cybersecurity entity extraction and measures recovery of the deficit identified by \cite{shafee2024evaluation}. This reveals a nested gap: the corpus identifies the NER problem empirically but does not propose or evaluate a solution for it.

A broader collection of API-based OSINT automation studies \cite{shamunesh2023cybercheck,nagra2024kia,sermakani2024open,agrahari2025osintct2,pais2014osint} demonstrates the feasibility of LLM-augmented collection in constrained, English-language, surface-web contexts, but these studies share a common limitation: none evaluates collection on non-English sources, dark-web content, encrypted platforms, or adversarially manipulated content~\cite{ranade2021fake}.

Almutairi et al. \cite{tseng2025disarm}, applying fine-tuned LLMs to Twitter-sourced data for election-related OSINT using the DISARM disinformation framework, extends the baseline approach to social media intelligence and demonstrates that fine-tuning on domain-specific data consistently outperforms zero-shot and few-shot approaches for targeted classification tasks, a finding consistent with the broader domain adaptation consensus discussed in Section 3.8. The study is noteworthy for its use of an established disinformation ontology (DISARM) as an annotation framework, providing the most structured entity taxonomy for OSINT-relevant disinformation entities in the corpus.

\subsection{Agentic AI, Tool Use, and OSINT Automation}
\label{sec:3-3-agentic-ai-tool-use-and-osint-automation}

The architectural innovation that most directly addresses the limitations of LLM-as-tool approaches is the OSINT agent: a system in which an LLM does not merely process a single query but autonomously plans a multi-step intelligence collection and analysis workflow, selects and invokes external tools, observes the results, updates its reasoning, and iterates until the intelligence objective is satisfied or a stopping condition is met. This architecture is grounded in the ReAct (Reason + Act) paradigm, in which an LLM alternates between explicit reasoning steps (“Thought”) and concrete actions (“Act”) within a continuous observation loop. It transforms the LLM from a document processor into an intelligence-analysis assistant capable of directing its own inquiry \cite{almeidapalmieri2025framework,shen2024llmosint}.

\begin{figure*}[!t]  
  \centering  
  \includegraphics[width=0.8\linewidth]{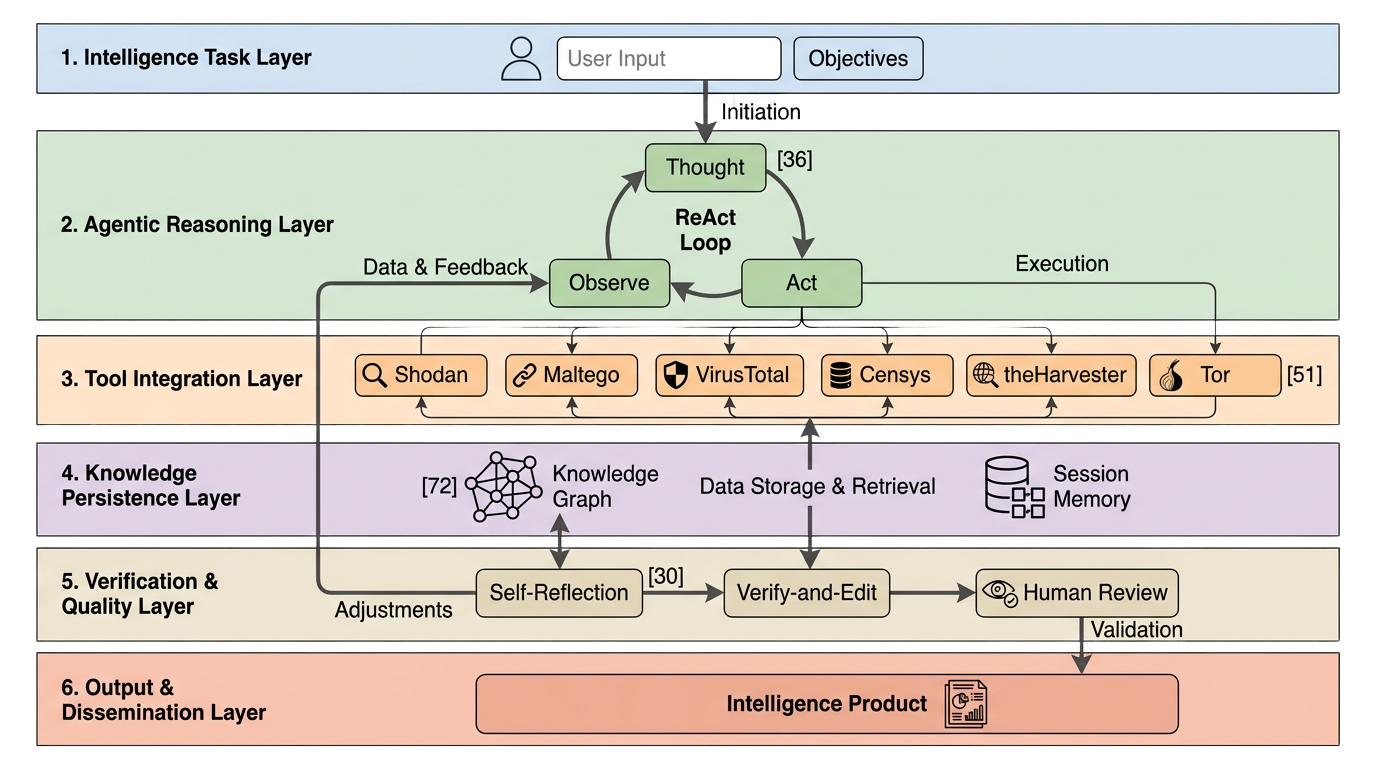}  
  \caption{Composite agentic OSINT reasoning loop, synthesising components described across four studies: the ReAct reasoning loop \cite{almeidapalmieri2025framework}, the human review checkpoint \cite{mukhopadhyay2024osintclinic}, the knowledge graph, session memory, self-reflection, and Tor dark-web integration \cite{shen2024llmosint}, and the Verify-and-Edit post-generation correction mechanism \cite{zhao2023verifyandedit}. No single paper implements all components simultaneously.}  
  \label{fig:reasoning}  
\end{figure*}

Palmieri \cite{almeidapalmieri2025framework} presents one of the most complete agentic OSINT proof-of-concept systems in the corpus. The system orchestrates tool calls across Shodan, Maltego, VirusTotal, Censys, and theHarvester within a ReAct reasoning loop, and the paper reports task-level performance figures across multi-tool intelligence collection tasks. Palmieri presents a unified agentic OSINT framework and proof-of-concept comparison but does not provide a reusable open benchmark or standardised quantitative evaluation protocol comparable across studies. This approach is the most comprehensive agentic OSINT design in the reviewed corpus but constitutes a proof-of-concept comparison rather than a validated operational benchmark independent of the system's specific performance figures. No subsequent study adopts or extends it, however a pattern of non-cumulative evaluation design that is examined critically in Section 3.9.

The most architecturally complete agentic OSINT system in the corpus is presented by Shen et al.\ at Tsinghua University \cite{shen2024llmosint}, which integrates LangChain-based orchestration, Neo4j knowledge graph storage for extracted OSINT entities and their relationships, session memory for multi-step context persistence, a self-reflection mechanism in which the agent evaluates its own outputs against accumulated knowledge graph content before finalising them, and Tor-based dark web crawling for access to .onion domain intelligence sources. This combination is architecturally ambitious: it addresses the tool orchestration requirement (LangChain), the knowledge persistence requirement (Neo4j), the context continuity requirement (session memory), the output quality requirement (self-reflection), and the source coverage requirement (Tor integration) within a single cohesive system. The self-reflection loop in particular represents a convergent architectural development independently proposed in different forms by \cite{zhao2023verifyandedit} (Verify-and-Edit) and \cite{sun2025decision} (iterative OSINT report refinement). This convergence suggests that post-generation verification against accumulated knowledge is emerging as a common design principle for reliable OSINT AI output.

Despite these architectural advances, a critical limitation applies to both \cite{almeidapalmieri2025framework} and \cite{shen2024llmosint} both systems are evaluated exclusively under benign conditions, in which the input data they process is assumed to be genuine. This assumption is directly contradicted by the adversarial risk literature. Ranade et al. \cite{ranade2021fake} demonstrate that fake cyber threat intelligence generated by a comparatively modest GPT-2 model fine-tuned on an OSINT-sourced cybersecurity corpus can deceive professional threat hunters. In the reported human evaluation (n\,=\,10), participants judged generated CTI and authentic CTI as genuine at comparable rates when the content was distributed through the same social media and open-source channels from which agentic OSINT systems collect intelligence. When an agentic system such as \cite{almeidapalmieri2025framework} or \cite{shen2024llmosint} processes content from these channels, as both explicitly do, it operates without any adversarial filtering mechanism. The capability demonstrated under benign evaluation conditions therefore cannot be treated as evidence of reliable performance under the adversarial conditions that characterize real OSINT environments.

Mukhopadhyay and Luther \cite{mukhopadhyay2024osintclinic} provide the strongest case in the corpus for structuring human oversight within agentic OSINT pipelines. The paper's contribution is not primarily technical. Rather than introducing a novel architecture, it makes an epistemological argument: the degree of autonomous decision-making demonstrated in \cite{almeidapalmieri2025framework} and \cite{shen2024llmosint} is inappropriate in intelligence contexts where errors in collection, extraction, or synthesis can have consequential downstream effects on analytical products and operational decisions. The paper's proposed checkpoints, at which human judgement overrides or confirms agent decisions before the pipeline proceeds, operationalize the human-oversight consensus identified across the corpus and are directly relevant to the responsible deployment of the architectures described above. The evidence from \cite{srikanth2024usability}, showing that experienced CTI professionals have a median LLM familiarity of 2 out of 5, complicates this proposal: effective human oversight requires not only architectural oversight mechanisms but also analysts with sufficient LLM literacy to detect and correct agent errors.

\subsection{APIs, Toolkits, and Orchestration Pipelines}
\label{sec:3-4-apis-toolkits-and-orchestration-pipelines}

The practical feasibility of agentic OSINT depends not only on the reasoning capability of the underlying language model but also on the quality, reliability, and accessibility of the external tools the agent invokes. The API and toolkit layer comprising OSINT-specific databases, infrastructure reconnaissance services, threat intelligence platforms, and orchestration frameworks, is a necessary enabling condition for the agentic architectures described in Section 3.3. It is also a layer that introduces operational dependencies, failure modes, and security implications that no study in the corpus evaluates under realistic conditions.

The most frequently integrated tool categories across the corpus are infrastructure intelligence APIs (Shodan, Censys, Maltego), threat intelligence aggregation services (VirusTotal, MISP), open-source collection frameworks (theHarvester, OSINT Framework, Spiderfoot), and agentic orchestration libraries (LangChain in \cite{shen2024llmosint}, implicit orchestration in \cite{almeidapalmieri2025framework}). The prevalence of Shodan and Censys across multiple studies \cite{almeidapalmieri2025framework,shen2024llmosint,nila2023llms,pervez2023ease} reflects the accessibility of these APIs and their usefulness for network reconnaissance, infrastructure mapping, and vulnerability identification. However, this accessibility is itself a security concern: Shodan and Censys provide the same internet-wide scanning data to legitimate OSINT practitioners and adversarial actors conducting offensive reconnaissance. Pervez et al. \cite{pervez2023ease} demonstrate that commercially available OSINT tooling, specifically the combination of Shodan-based network scanning and Maltego-based relationship mapping, can identify unpatched CVEs and operational topology details for critical infrastructure within months of public disclosure. The tools that agentic OSINT agents invoke are, in other words, precisely the tools used for offensive cyber operations, and their integration into automated pipelines without access controls or ethical filtering amplifies the dual-use risk at the system level.

The cloud-versus-on-premise architectural tension, introduced in Section 1.1 as a field-level disagreement, is most directly relevant at the API and toolkit layer. API-based LLM deployment exemplified by \cite{radoi2023ai}'s use of the OpenAI GPT-3 API routes both the intelligence queries formulated by the system and the raw OSINT content processed by the LLM through the infrastructure of a commercial third party. For low-sensitivity, publicly available, non-personal data, such an approach may be operationally acceptable. For intelligence collection involving personal data, classified source references, or legally privileged investigation content, such an approach may be impermissible or operationally unacceptable. Yurtalan \cite{yurtalan2025redefining} addresses this constraint directly by demonstrating that open-weight models, specifically models deployable on local hardware without API dependencies, achieve sufficient performance for law enforcement OSINT tasks while eliminating the data exposure risk associated with commercial API use. The empirical support for this position is strengthened by \cite{shafee2024evaluation}'s finding that the open-source GPT4all model achieves F1 = 0.90 on cybersecurity content classification compared to GPT-4's 0.94. This four-point performance difference may be operationally acceptable in exchange for the security and legal-compliance advantages of on-premise deployment. Corroborating evidence from intelligence and military contexts \cite{nila2023llms} reinforces the position that on-premise deployment is not merely a preference but an operational requirement in high-sensitivity environments.

\subsection{RAG, Knowledge Graphs, and Memory-Augmented OSINT}
\label{sec:3-5-rag-knowledge-graphs-and-memory-augmented-osint}

The most consequential technical challenge in deploying LLMs for OSINT is not generating plausible outputs but generating accurate, verifiable, and traceable outputs from sources that can be validated. LLMs trained on static corpora face three distinct failure modes in knowledge-intensive tasks: hallucination of plausible but false information, reproduction of outdated training data, and conflation of sources without a mechanism for provenance assessment. Retrieval-augmented generation, knowledge graph integration, and session-memory architectures are the primary technical responses in the current corpus. Their effectiveness is empirically supported, but so is their vulnerability to adversarial exploitation.

\begin{figure}[!t]  
  \centering  
  \includegraphics[width=\linewidth]{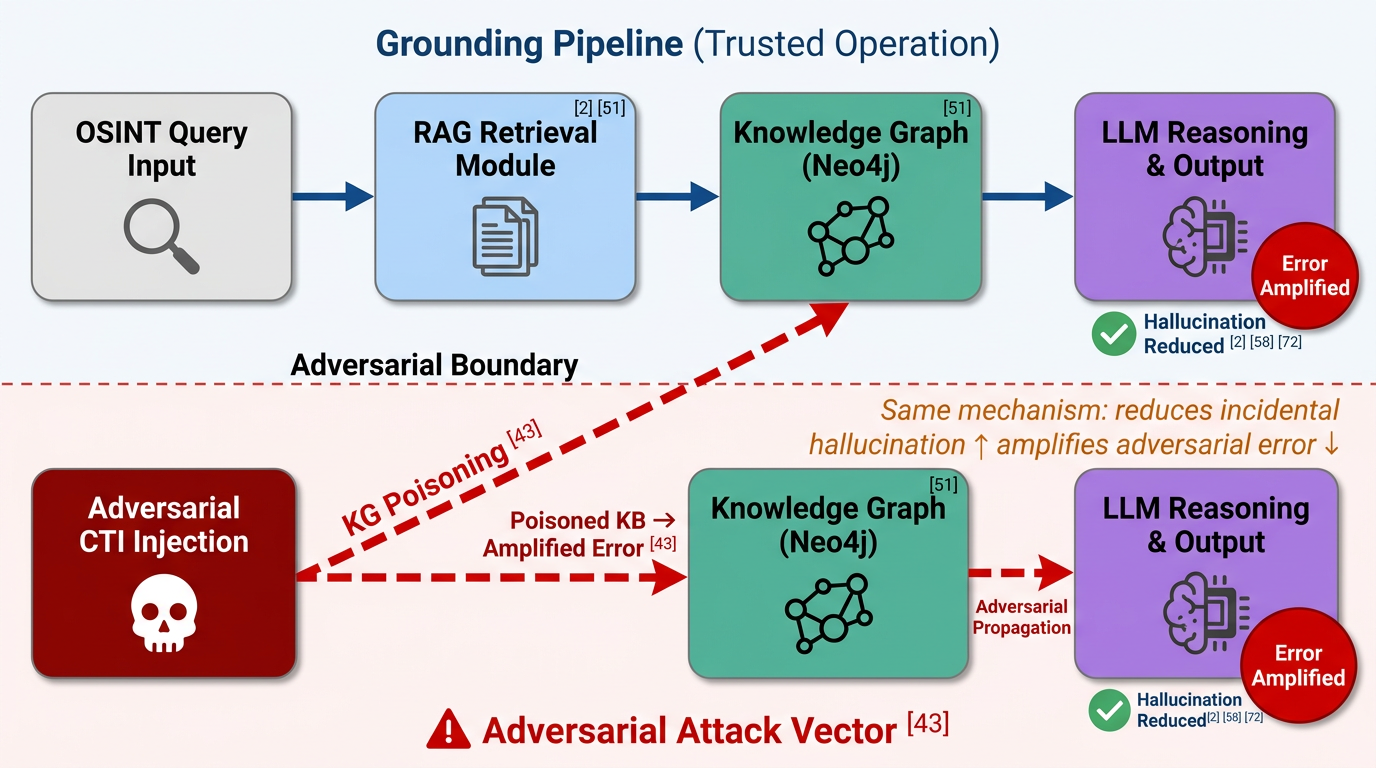}  
  \caption{RAG and Knowledge Graph Enhanced OSINT Architecture. This diagram shows the grounding and knowledge infrastructure pipeline and the adversarial attack vector demonstrated in \cite{ranade2021fake}. The diagram makes explicit that the same mechanisms (RAG retrieval, KG enrichment) that reduce incidental hallucination can amplify adversarial error when the knowledge base is compromised.}  
  \label{fig:ragkg}  
\end{figure}

Allam \cite{allam2025cybervision} provides the only direct measurement of hallucination rate in an OSINT-specific AI system. Using Claude 3.7 Sonnet augmented with a purpose-built RAG knowledge base, the study achieves P@5 = 0.83, MRR = 0.76, a factual accuracy of 92\%, citation accuracy of 96\%, and a hallucination rate of 4\%. The 4\% figure establishes that a well-designed OSINT RAG system can substantially ground LLM outputs in verifiable sources, but it must be interpreted carefully: the evaluation used a private, curated dataset that cannot be independently reproduced; the knowledge base was purpose-built rather than operationally heterogeneous; and the 11\% task-completion failure rate indicates that the system fails to produce a complete output for approximately one in nine intelligence tasks, a non-trivial limitation in time-sensitive contexts. The 4\% hallucination rate is best understood as an empirical floor achieved under favourable conditions, not a transferable benchmark.

The Verify-and-Edit framework \cite{zhao2023verifyandedit} offers a complementary mechanism through post-generation correction rather than pre-generation retrieval. Individual claims in the LLM's reasoning chain are verified against externally retrieved evidence and revised where verification fails, achieving a 4.5\% Exact Match improvement on AdvHotpotQA and 5.9\% on 2WikiMultiHop. The approach directly addresses error propagation across multi-step reasoning chains, a failure mode of particular relevance to agentic OSINT systems. Its convergence with the self-reflection loop in \cite{shen2024llmosint} and the iterative refinement approach in \cite{sun2025decision} suggests that staged verification against accumulated knowledge is emerging as a shared architectural principle across independently conducted research.

Knowledge graph integration \cite{shen2024llmosint,su2025opensource} extends the grounding principle to persistent structured storage. By extracting OSINT entities and their relationships and storing them as structured triples in a graph database, an agentic system can maintain a queryable intelligence picture that accumulates across collection and analysis cycles, directly addressing the temporal currency problem the \texttt{Outdated Knowledge} failure mode identified by \cite{kaddour2023challenges} by storing time-sensitive intelligence in a dynamic external structure rather than frozen model weights.

The critical counterpoint to this grounding optimism is provided by Ranade et al.\ \cite{ranade2021fake}, who demonstrate that knowledge-graph infrastructure is not only a reliability solution but also an attack surface. By fine-tuning GPT-2 on an OSINT-sourced cybersecurity corpus to generate plausible but deliberately false threat intelligence, the authors show that adversarial content distributed through normal OSINT channels can be ingested by NER pipelines without adversarial filtering and incorporated as genuine entities into a cybersecurity knowledge graph. Once poisoned, the graph returns corrupted attribution data, fictitious threat actor profiles, and misleading indicator lists. In a human evaluation study (n,=,10), professional threat hunters judged generated fake CTI and authentic CTI as genuine at comparable rates. The poisoning attack exploits precisely the normalized ingestion pipeline that legitimate knowledge graphs require to function.

The juxtaposition of \cite{allam2025cybervision}, \cite{zhao2023verifyandedit}, and \cite{ranade2021fake} defines the central design tension for OSINT knowledge infrastructure: the same mechanisms that reduce incidental error, including external retrieval, graph-based storage, and entity persistence, can amplify deliberate error when the knowledge base is compromised. No paper in the corpus simultaneously addresses both the grounding benefit and the poisoning vulnerability within a single system design, making this one of the most critical unresolved architectural problems in OSINT knowledge infrastructure.

\subsection{Cyber Threat Intelligence and Security OSINT}
\label{sec:3-6-cyber-threat-intelligence-and-security-osint}

Cyber threat intelligence is the most developed application domain in the corpus, attracting a concentration of primary and methodological support studies \cite{hassanin2024llms,nila2023llms,yigit2024review,yang2024threatmodeling,tihanyi2024cybermetric,chen2026cyberthreateval,srikanth2024usability,shafee2024evaluation}. The domain offers the clearest demonstrations of LLM capability under structured evaluation, the clearest examples of evaluation-design artifacts that can be mistaken for operational findings, and some of the most consequential adversarial risks in the corpus. Its maturity is domain-specific: CTI results depend on structured data, well-defined entity ontologies, and expert annotators that may not be reproducible in general OSINT environments.

Domain adaptation is the most consistent strategy for improving performance in the CTI literature. Hassanin and Moustafa\ \cite{hassanin2024llms} demonstrate that CyberBERT and SecureBERT consistently outperform general-purpose BERT in CTI entity extraction and relationship classification, confirming that the linguistic specificity of cybersecurity text, CVE identifiers, threat-actor conventions, and procedural language constitutes a semantic challenge that general-purpose pre-training cannot address. This study contextualizes the 25\% NER performance gap identified by \cite{shafee2024evaluation} for general-purpose chatbots on cybersecurity Twitter data: the deficit is not an interface limitation but a domain adaptation failure requiring domain-specific pre-training or targeted fine-tuning.

Shafee et al.\ \cite{shafee2024evaluation} establish a practical capability boundary for current LLMs using OSINT-derived CTI data. Binary classification of cybersecurity-relevant tweets achieves F1 = 0.94 (GPT-4) and F1 = 0.90 (GPT4all), levels arguably sufficient for large-scale triage under human supervision. The operationally significant limit lies at named entity extraction: NER performance falls 25\% below specialized models across all seven chatbots evaluated, a deficit that disqualifies unsupervised CTI extraction at operational scale, given that specific threat indicators (IP addresses, CVEs, malware family names) constitute the core analytical output of CTI production.

The deployment architecture question is acute in CTI contexts. \"Oktem \cite{nila2023llms} documents LLM use for OSINT log analysis and threat attribution in in a military-intelligence setting, providing evidence consistent with \cite{yurtalan2025redefining} that operational-security constraints can render cloud API deployment unacceptable in high-sensitivity environments. Crucially, \cite{shafee2024evaluation} finds only a four-point F1 gap between proprietary and open-source alternatives, supporting the position that the cloud-versus-on-premise choice is an operational and legal constraint whose performance penalty is narrowing as open-weight models mature.

Yigit et al.\ \cite{yigit2024review} extend the CTI frame to digital forensics through the MetaAID governance framework, which is the most developed ethical architecture in the CTI sub-corpus. However, the study does not fully address how MetaAID would be operationalized in law-enforcement environments where legal frameworks for AI-assisted investigative evidence remain unsettled, and the framework lacks empirical validation. The usability dimension is captured by Srikanth et al.\ \cite{srikanth2024usability}, whose evaluation with ten cybersecurity professionals establishes a sociotechnical constraint that technical performance metrics miss: practitioners with a median CTI experience of 4 out of 5 report a median LLM familiarity of only 2 out of 5, identifying hallucination and output inconsistency as the primary adoption barriers. The research neglected to consider that a system achieving strong automated performance while remaining opaque to the supervising analyst has not achieved operational readiness.

The field's most consequential evaluation disagreement between CyberMetric \cite{tihanyi2024cybermetric} and CyberThreat-Eval \cite{chen2026cyberthreateval} is examined in Section~3.9 as a corpus-level methodological concern, where the lesson both papers jointly provide applies to OSINT evaluation design at large.

\subsection{Prompt Engineering and Chain-of-Thought Reasoning}
\label{sec:3-7-prompt-engineering-and-chain-of-thought-reasoning}

The analytical potential of LLMs for OSINT is shaped substantially by how models are instructed. Prompt engineering is the most accessible technical intervention available to practitioners who lack the infrastructure for fine-tuning or agentic system development, but the corpus shows that it is both more powerful and more brittle than its accessibility suggests.

The theoretical foundation is established by Wei et al.\ \cite{wei2022cot}, whose demonstration that eliciting explicit intermediate reasoning steps produces 80\% accuracy on GSM8K for PaLM 540B, compared with 33\% for standard prompting, establishes a general principle directly relevant to OSINT: the sequential structured reasoning involved in intelligence analysis, including formulating collection hypotheses, interpreting ambiguous evidence, and revising assessments, is precisely the kind of task that chain-of-thought prompting enables. Every agentic OSINT architecture employing a thought-action-observation loop \cite{almeidapalmieri2025framework,shen2024llmosint} depends on this capability at its core; the ReAct paradigm is, in structural terms, chain-of-thought reasoning applied to tool-using agents in multi-step collection environments. This connection is rarely made explicit in the corpus despite its direct implications for both prompt design and system architecture.

{\v{C}}ern{\'y}\ \cite{cerny2024implications} extend the principle to the OSINT operational context by framing prompt engineering as a professional analyst's competency. Their three-mode taxonomy, targeted, bulk, and monitored collection maps prompt design choices to intelligence collection objectives and position this as a trainable skill for inclusion in professional development curricula. Neither the framework nor the taxonomy has been adopted by any subsequent paper in the corpus, consistent with the broader pattern of non-cumulative development identified in Section~3.9. The QTM decomposition framework \cite{park2024performance} offers a complementary structural approach, achieving an 11.46\% improvement over baseline prompting through the principled decomposition of intelligence questions into constituent task components. Its Korean-language evaluation context limits direct generalisability, but the underlying principle that structured question decomposition outperforms ad hoc prompting is language-agnostic.

Sun et al. \cite{sun2025decision} demonstrate that prompt engineering extends to epistemic quality control: an iterative self-reflection loop in which the model evaluates its output for political balance before finalising it, achieving measurable bias reduction in OSINT report generation. This aligns with \cite{shen2024llmosint}'s self-reflection mechanism and \cite{zhao2023verifyandedit}'s Verify-and-Edit framework, as all three studies independently address different concerns---political bias, architectural completeness, and factual accuracy---while arriving at the same design principle: LLM outputs in OSINT contexts should be treated as drafts subject to structured revision rather than final conclusions.

A critical constraint tempers these gains. The LLM Challenges taxonomy \cite{kaddour2023challenges} classifies prompt brittleness as a core behavioral failure: minor wording changes, synonymous terms, different orderings, and minor paraphrases can produce substantially different and factually incorrect outputs. In intelligence analysis contexts, where errors may include false threat-actor attribution or missed indicators of compromise, such a failure is a safety-critical reliability failure rather than a quality inconvenience. No paper in the corpus systematically evaluates prompt brittleness for OSINT-specific queries, and no system paper includes prompt robustness testing as a validation component. The field acknowledges this failure mode conceptually but has not yet operationalized a response.

\subsection{Fine-Tuning and Domain Adaptation}
\label{sec:3-8-fine-tuning-and-domain-adaptation}

Across the methodological support literature, a single empirical pattern recurs without exception: models adapted to the linguistic, semantic, and ontological characteristics of a specific OSINT or CTI domain consistently outperform general-purpose equivalents on tasks within that domain. This domain adaptation consensus is documented through supervised fine-tuning \cite{tseng2025disarm,hassanin2024llms,yang2024threatmodeling}, parameter-efficient methods \cite{riad2024finetuning}, and retrieval-augmented knowledge specialisation \cite{allam2025cybervision}, and it represents the strongest agreement among the methodological components of the corpus. However, the consensus has a consequential counterargument, which originates not from a limitation of the approach but from one of its most striking demonstrations of success.

Almutairi et al. \cite{tseng2025disarm} provide the clearest OSINT-facing demonstration of fine-tuning advantage. Fine-tuning a transformer model on Twitter-sourced election OSINT, annotated using the DISARM disinformation framework, produces consistently higher classification accuracy on OSINT-relevant disinformation categories than zero-shot or few-shot approaches with the same base model, a finding that establishes domain adaptation as the appropriate technical strategy when labelled in-domain training data are available and the task is sufficiently well-defined for supervised learning. The DISARM annotation framework is a significant secondary contribution: it provides the most structured entity taxonomy for disinformation-relevant OSINT entities in the corpus, offering a reusable annotation schema that subsequent research could build upon. That no subsequent paper does so reflects the absence of a cumulative data infrastructure that characterises the broader field.

Hassanin and Moustafa \cite{hassanin2024llms}'s comparison of CyberBERT and SecureBERT against general-purpose BERT on CTI entity extraction establishes that domain-adaptive pre-training on large unlabelled domain-specific text corpora prior to task-specific fine-tuning provides additional performance gains beyond supervised fine-tuning alone on general-domain pre-trained models. This finding is directly relevant to the NER performance gap documented by \cite{shafee2024evaluation}: if the 25\% deficit relative to specialized models reflects inadequate domain adaptation rather than an inherent LLM chatbot limitation, then domain-adaptive pre-training on cybersecurity OSINT text corpora represents the most promising path to closing it. No paper in the corpus implements or evaluates this approach specifically for cybersecurity OSINT entity extraction. This constitutes a concrete, well-evidenced research gap with a clearly defined solution pathway.

Yang et al. \cite{yang2024threatmodeling} extend the domain adaptation evidence base to financial crime OSINT through banking-sector threat modelling, creating the first purpose-built banking threat annotation dataset in the corpus and demonstrating that fine-tuning on this proprietary domain data achieves performance levels that general-purpose models cannot match. Riad et al. \cite{riad2024finetuning} demonstrate that parameter-efficient fine-tuning via LoRA and PEFT achieves competitive performance relative to full fine-tuning on sentiment classification tasks at substantially lower computational cost. The PEFT finding is particularly relevant for OSINT organisations with constrained GPU infrastructure: the combination of open-weight models \cite{shafee2024evaluation} deployable on local hardware and parameter-efficient adaptation methods \cite{riad2024finetuning} creates a feasible on-premise fine-tuning pathway that avoids both the performance limitations of zero-shot prompting and the infrastructure requirements of full model fine-tuning.

The most significant observation regarding domain adaptation in this corpus is that it has a direct adversarial analogue. Ranade et al. \cite{ranade2021fake} demonstrate that fine-tuning a comparatively modest GPT-2 model on an OSINT-sourced cybersecurity corpus, exactly the domain adaptation approach advocated across the legitimate fine-tuning literature, produces a system capable of generating fake CTI convincing enough for professional threat hunters to be equally likely to judge it as true alongside authentic CTI, with the majority of fake samples labelled true in the reported human evaluation (n\,=\,10). The technique that improves legitimate CTI extraction and the technique that enables the most capable adversarial CTI generation documented in the corpus are methodologically identical. This dual-use equivalence is not addressed by any fine-tuning paper that proposes domain adaptation as a beneficial approach: The corpus focuses mainly on the performance advantages of domain adaptation without examining the adversarial capability implications of making domain-specific fine-tuning methods widely available and documented. A significant governance failure in the methodological support literature is the absence of a responsible development framework that incorporates adversarial evaluation as a mandatory component before publication.

\subsection{Evaluation, Benchmarks, and Performance}
\label{sec:3-9-evaluation-benchmarks-and-performance}

The most important finding about OSINT-AI evaluation is not a performance figure but a structural absence: hallucination, named a primary reliability concern in more than 20 corpus papers, has been empirically measured in only one OSINT-specific system, Allam's 4\% RAG rate \cite{allam2025cybervision}. This claim must be read precisely. It concerns \emph{OSINT-specific, end-to-end hallucination measurement}: the proportion of fabricated or unsupported assertions in the final intelligence output of an OSINT system evaluated on OSINT data. It is deliberately distinguished from \emph{error or factual-correction measurement in adjacent or general domains}. The Verify-and-Edit framework \cite{zhao2023verifyandedit}, for instance, reports Exact-Match accuracy gains of $+4.5\%$ and $+5.9\%$ from revising individual reasoning-chain claims on the open-domain multi-hop question-answering benchmarks AdvHotpotQA and 2WikiMultiHop; these figures quantify the benefit of a correction mechanism on general QA tasks, not a hallucination rate on an OSINT system. Several corpus studies therefore measure related error signals---reasoning-chain correction \cite{zhao2023verifyandedit}, adversarial-content detectability \cite{ranade2021fake}, and political-bias reduction \cite{sun2025decision}---but only \cite{allam2025cybervision} reports an end-to-end hallucination rate for an OSINT pipeline on OSINT data. The remaining papers describing CTI generation, agentic collection, or automated summarisation acknowledge the risk without measuring it against standardised, adversarial, or reproducible protocols. The gap between how often hallucination is named the defining problem and how rarely the field has built the shared benchmarks, adversarial conditions, and reproducible procedures needed to study it is the starkest evidence of evaluation immaturity.

A benchmark-fragmentation problem of equal severity accompanies it, illustrated most sharply by CyberMetric \cite{tihanyi2024cybermetric} and CyberThreat-Eval \cite{chen2026cyberthreateval}. The former evaluates 25 LLMs and 30 experts on a 2{,}000-question multiple-choice benchmark, where GPT-4o reaches 91.25\% against a 72.24\% expert mean, implying super-expert cybersecurity knowledge. The latter, derived from a real CTI workflow (triage of 488 articles, deep search of 55 URLs, expert-judged report drafting), finds no LLM adequate for complex attribution or operational reporting. The results do not contradict: the first measures structured knowledge recall, the second analytical reasoning under operational uncertainty, and CyberThreat-Eval explicitly rejects lexical metrics (ROUGE, BLEU, and BERTScore) as surface measures. The implication for system designers is severe: benchmark performance is no evidence of readiness for open-ended analysis, and the two must not be conflated.

Two further weaknesses recur. Metrics design is non-cumulative: Custom OSINT-agent metrics of \cite{yuan2024empowering}, proof-of-concept figures of \cite{almeidapalmieri2025framework}, six private RAG metrics of \cite{allam2025cybervision}, analyst-centric criteria of \cite{chen2026cyberthreateval}, and the reproducible F1 protocol of \cite{shafee2024evaluation} are each developed independently and adopted by no subsequent group, making cross-study comparison structurally impossible. Human evaluation is systematically underpowered: \cite{ranade2021fake} uses ten evaluators, \cite{yang2024geolocator} three posts, and \cite{srikanth2024usability} ten professionals, none reaching the $n \geq 30$ conventionally required, so the field's most consequential human-centred findings rest on its weakest empirical foundations. Table~\ref{tab:table-4} summarises the comparison.

\begin{table*}[htbp]
  \centering
  \caption{Evaluation and benchmark comparison across key OSINT-AI studies. An end-to-end OSINT hallucination rate was measured empirically in one system only (S21, 4\%, \cite{allam2025cybervision}); other studies measure related error or detection signals rather than an OSINT hallucination rate. Study identifiers index the supplementary Evidence Matrix (Table~S1).}
  \label{tab:table-4}
  \footnotesize
  \setlength{\tabcolsep}{4pt}\renewcommand{\arraystretch}{1.15}
\begin{tabularx}{\textwidth}{@{}l >{\raggedright\arraybackslash}p{0.15\linewidth} >{\raggedright\arraybackslash}X >{\raggedright\arraybackslash}X >{\raggedright\arraybackslash}X l@{}}
\toprule
\textbf{ID} & \textbf{Evaluation Type} & \textbf{Dataset / Scale} & \textbf{Metrics} & \textbf{Key Result} & \textbf{Repl.} \\
\midrule
\cite{yuan2024empowering} & Custom OSINT agent framework & Private agent task set & RA, TUR, TCA, VPR & First formal OSINT-agent evaluation framework; not adopted by later papers & Low \\
\addlinespace[2pt]
\cite{almeidapalmieri2025framework} & Agentic OSINT proof-of-concept & Private multi-tool ReAct set & Task-level figures; no reusable open benchmark & Most complete agentic OSINT PoC; multi-tool ReAct under tested conditions & Low \\
\addlinespace[2pt]
\cite{allam2025cybervision} & Six-metric RAG evaluation & Private OSINT knowledge base & P@5, MRR, factual/citation accuracy, TCR, hallucination & Only OSINT hallucination measurement: 4\% under favourable conditions & Low \\
\addlinespace[2pt]
\cite{zhao2023verifyandedit} & Knowledge-grounded multi-hop QA & Open: AdvHotpotQA, 2WikiMultiHop & Exact-match accuracy & Verify-and-Edit gains +4.5\% / +5.9\% over CoT baseline & High \\
\addlinespace[2pt]
\cite{tihanyi2024cybermetric} & Multiple-choice cyber knowledge & Semi-open MCQ (CyberMetric); 25 LLMs, 30 experts & Accuracy & GPT-4o 91.25\% vs.\ human 72.24\%; largest LLM--human comparison & Mod. \\
\addlinespace[2pt]
\cite{chen2026cyberthreateval} & Analyst-centric CTI workflow & Real-world CTI workflow (public) & Factual accuracy, completeness, actionability & No LLM adequate across all three workflow stages; most valid evaluation & Low \\
\addlinespace[2pt]
\cite{srikanth2024usability} & Usability heuristics + user study & Security professionals; $n=10$ & Adoption barriers; familiarity (1--5) & Median familiarity 2/5; capability--readiness adoption gap & Low \\
\addlinespace[2pt]
\cite{shafee2024evaluation} & F1 classification and NER & Open: Twitter CTI (LASIGE) & F1 (classification), F1 (NER) & GPT-4 F1\,=\,0.94; GPT4all 0.90; 25\% NER gap vs.\ domain models & High \\
\addlinespace[2pt]
\cite{ranade2021fake} & Adversarial human evaluation (fake-CTI credibility) & GPT-2--generated fake CTI; threat hunters $n=10$ & Human true/false credibility judgement & Threat hunters judged fabricated CTI as true at rates comparable to authentic CTI; fluency/citation cues insufficient discriminators & Low \\
\addlinespace[2pt]
\cite{niu2024bullshint} & Empirical OSINT-content reliability analysis & Russo--Ukrainian war OSINT tweets (BullshINT) & Misinformation characterisation and detection & Documents misinformation contamination of OSINT-tagged social content; detection-only, no adversarial or generative evaluation & Low--Mod. \\
\midrule
\multicolumn{6}{@{}>{\raggedright\arraybackslash}p{\textwidth}@{}}{\textit{\textbf{Overall:}} no open, community-adopted benchmark exists for agentic OSINT evaluation; an end-to-end OSINT hallucination rate is measured in one system only (\cite{allam2025cybervision}); private or non-public datasets dominate. \cite{ranade2021fake} and \cite{niu2024bullshint} are included as the adversarial-credibility and misinformation-reliability evaluations that bear directly on OSINT trustworthiness, though both rest on small or non-released samples. The green-code energy study \cite{sikand2024greencode} is \emph{excluded} from this comparison: it evaluates the energy efficiency of generative-AI code generation, not OSINT or CTI system performance, and contributes no OSINT evaluation metric; it is retained in the corpus only as subsection-level context (Section~2.1, criterion ii).} \\
\bottomrule
\end{tabularx}
\end{table*}

\subsection{Risks, Ethics, Hallucination, and Misinformation}
\label{sec:3-10-risks-ethics-hallucination-and-misinformation}

No category in the taxonomy is more empirically grounded in its problems or more inadequate in its solutions. The risk literature \cite{rheault2024active,yigit2024review,dekens2023practical,kobayashi2024subjective,ranade2021fake,yang2024geolocator,niu2024bullshint,golda2024privacy,pervez2023ease} establishes a multi-layered landscape of adversarial attacks on knowledge infrastructure, geo-privacy violations, misinformation contamination, embedded analytical bias, training-data legality, and infrastructure exposure, each documented with evidence sufficient to establish operational concern, yet it produces no coherent mitigation framework, disclosure standard, or compliance pathway addressing them in combination. Table~\ref{tab:table-5} maps the principal risks; the most consequential are summarised here.

The adversarial finding of Ranade et al.\ \cite{ranade2021fake} is the pivot for the review: a GPT-2 model fine-tuned on OSINT-sourced CTI generated fake intelligence that professional threat hunters judged genuine at the authentic rate ($n=10$) and poisoned cybersecurity knowledge graphs through standard NER ingestion. This undermines every agentic architecture in Section~3.3 the ReAct agent \cite{almeidapalmieri2025framework}, the KG-integrated system \cite{shen2024llmosint}, and any system drawing on open feeds operate without adversarial filtering because the assumption of genuine input, never stated but structurally embedded in their evaluations, fails under conditions \cite{ranade2021fake} confirms are routine. The hallucination-validation gap (Section~3.9) is equally a risk: the 4\% rate \cite{allam2025cybervision} is a favourable-condition floor, not a guarantee, and the rate under degraded or adversarially contaminated conditions is unknown, a single data point against a near-universal concern. Two further empirical risks are well evidenced: GeoLocator \cite{yang2024geolocator} infers precise location from a single photograph via a public plugin, and Niu et al.\ \cite{niu2024bullshint} measure a 9\% misinformation base rate across 1.96M tweets from 48 professional OSINT communities, contamination inherited before any LLM processing. Like \cite{yang2024geolocator} and \cite{pervez2023ease}, both \cite{ranade2021fake}, present offensive capabilities without reporting a responsible-disclosure process a norm long standard in cybersecurity.

Two governance risks complete the landscape. Kobayashi and Yamaguchi \cite{kobayashi2024subjective} show LLMs transmit training-corpus political bias into analyses, a validity threat where neutrality is a legal requirement, with self-reflection \cite{sun2025decision} of unproven sufficiency, and the GDPR analysis attributed to Golda et al. \cite{golda2024privacy} establishes that cloud-API processing of personal data carries enforceable EU regulatory risk, elevating the cloud-versus-on-premise question \cite{yurtalan2025redefining,shafee2024evaluation,nila2023llms} from preference to compliance obligation. Against this, Rheault et al.\ \cite{rheault2024active} provide the strongest positive benchmark: an IRB-approved methodology with consent protocols, OSINT-V source validation, and documented ethical rationale, a standard that approximately 60\% of primary papers fail to meet, mostly without acknowledging the omission.

\begin{table*}[htbp]
  \centering
  \caption{Risk, ethics, hallucination, and safeguards in AI-augmented OSINT. ``Mitigation'' is folded into the assessment. Study identifiers index the supplementary Evidence Matrix (Table~S1).}
  \label{tab:table-5}
  \footnotesize
  \setlength{\tabcolsep}{4pt}\renewcommand{\arraystretch}{1.15}
\begin{tabularx}{\textwidth}{@{}>{\raggedright\arraybackslash}X l >{\raggedright\arraybackslash}p{0.12\linewidth} >{\raggedright\arraybackslash}p{0.13\linewidth} >{\raggedright\arraybackslash}X >{\raggedright\arraybackslash}X@{}}
\toprule
\textbf{Risk Category} & \textbf{Studies} & \textbf{Stage} & \textbf{Legal / Ethical} & \textbf{Empirical Evidence} & \textbf{Assessment (incl.\ mitigation)} \\
\midrule
Adversarial CTI generation, KG poisoning & \cite{ranade2021fake} & Analysis, Enrichment & Professional ethics & Strong: threat hunters judged fake CTI as true ($n=10$; not generalisable at scale) & Critical, unmitigated; dual-use fine-tuning; output indistinguishable from authentic CTI \\
\addlinespace[2pt]
LLM hallucination in OSINT output & \cite{allam2025cybervision}, \cite{srikanth2024usability}, \cite{kaddour2023challenges} & All stages & Negligence; evidence law & Weak: one datapoint (4\% RAG, \cite{allam2025cybervision}); qualitative in \cite{srikanth2024usability}, \cite{kaddour2023challenges} & Most critical reliability risk; only partial RAG grounding; no audit framework \\
\addlinespace[2pt]
Geo-privacy attack via multimodal LLM & \cite{yang2024geolocator} & Collection & GDPR; ECHR Art.\,8 & Moderate: geolocation from three social-media images & Responsible-disclosure failure; public capability; no countermeasure \\
\addlinespace[2pt]
Misinformation in OSINT communities & \cite{niu2024bullshint} & Collection, Verification & --- & Strong: 9\% of 1.96M community tweets; GraphSAGE detection & Undermines agentic collection at scale; detection only, no mitigation pipeline \\
\addlinespace[2pt]
LLM subjective political bias & \cite{kobayashi2024subjective} & Analysis, Reporting & Tradecraft standards & Moderate: systematic bias across topics and model families & Validity threat to neutral products; self-reflection (\cite{sun2025decision}) sufficiency unproven \\
\addlinespace[2pt]
Cloud-API GDPR exposure & \cite{golda2024privacy} & Processing, Collection & GDPR; DPA enforcement & Moderate: enforcement precedent documented & EU-constrained for personal data; on-premise (\cite{yurtalan2025redefining}, \cite{nila2023llms}) partial mitigation \\
\addlinespace[2pt]
Critical-infrastructure exposure via tooling & \cite{pervez2023ease} & Collection & CFAA; NIS Directive & Moderate: CVE exposure via commercial OSINT tools & Disclosure failure; dual-use; no confirmed remediation \\
\addlinespace[2pt]
Collection without IRB or consent & corpus-wide; \cite{rheault2024active} & Collection & Privacy law; research ethics & Strong: \char`~60\% of primary papers lack an ethics statement; IRB in one (\cite{rheault2024active}) & Systemic ethics failure; \cite{rheault2024active} OSINT-V benchmark not replicated \\
\addlinespace[2pt]
Source volatility and link rot & \cite{rajendran2024comprehensive,pastorgalindo2020osint} & Collection, Verification & Evidence integrity standards & Moderate: OSINT sources disappear or are altered after collection; no corpus paper preserves source snapshots & Ground truth degrades over time; evaluation datasets become non-replicable; admissibility undermined \\
\addlinespace[2pt]
Attribution uncertainty and chain-of-custody & \cite{allam2025cybervision,almeidapalmieri2025framework,yigit2024review} & Analysis, Reporting & Evidential admissibility; disclosure obligations & Weak: no corpus paper implements provenance tracking sufficient for court-ready intelligence & AI-generated intelligence lacks chain-of-custody; inadmissible in legal proceedings without authenticated source records \\
\addlinespace[2pt]
Analyst-in-the-loop failure under adversarial inputs & \cite{ranade2021fake,srikanth2024usability,mukhopadhyay2024osintclinic} & Human Review, Verification & Professional duty of care & Strong (\cite{ranade2021fake} $n=10$): expert reviewers do not reliably detect adversarial content; median analyst LLM familiarity 2/5 \cite{srikanth2024usability} & Human oversight provides weaker protection than assumed; oversight effectiveness empirically unvalidated in agentic pipelines \\
\midrule
\multicolumn{6}{@{}>{\raggedright\arraybackslash}p{\textwidth}@{}}{\textit{\textbf{Overall:}} the risk literature is strong in problem identification and weak in solution provision; no corpus paper addresses more than one risk category at once; OSINT-specific risks, including source volatility, attribution uncertainty, and analyst-layer failure, are entirely unmitigated.} \\
\bottomrule
\end{tabularx}
\end{table*}

\begin{figure*}[!t]  
  \centering  
  \includegraphics[width=0.85\linewidth]{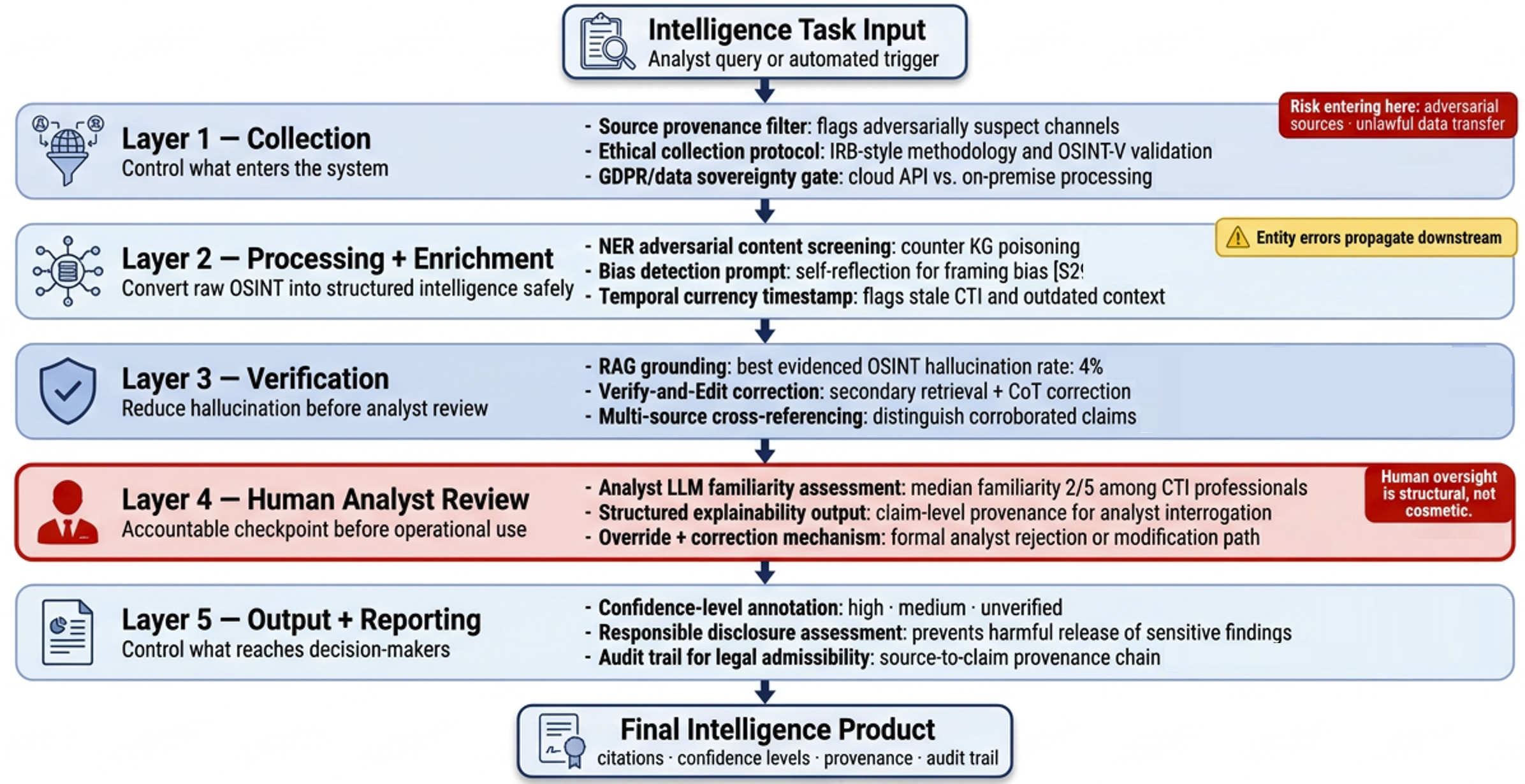}  
  \caption{Responsible Agentic OSINT Safeguard Framework. A five-layer safeguard architecture synthesised from corpus contributions. Each layer is annotated with the studies that motivate it. No existing paper in the corpus implements all five layers simultaneously this framework represents a normative synthesis of scattered contributions rather than a description of any deployed system.}  
  \label{fig:safeguard}  
\end{figure*}

\subsection{Dark Web and Specialized OSINT Sources}
\label{sec:3-11-dark-web-and-specialized-osint-sources}

The dark web is both one of the most intelligence-rich environments accessible to open-source investigators, hosting threat actor communications, criminal marketplaces, leaked data archives, and coordination infrastructure for organised crime, terrorism, and advanced persistent threat groups, and the most systematically neglected source environment in this corpus. Two papers from the 74 studies engage with dark web OSINT substantively: the Tsinghua agentic system \cite{shen2024llmosint}, discussed in Section 3.3 for its architectural completeness, and the DarkLens conceptual framework \cite{amin2025darklens}. This two-paper coverage of a source environment that the threat intelligence literature identifies as a primary vector for advanced threat actor communication is a finding of consequence, not a bibliographic footnote. It indicates that the OSINT-AI research community has concentrated its efforts on the accessible, structured, English-language surface web while leaving the most operationally significant alternative source environment without systematic investigation.

The Tsinghua system \cite{shen2024llmosint} provides the only working dark web OSINT capability in the corpus. Its Tor-based crawling module, incorporating circuit management, IP rotation, and dynamic handling of onion domain URLs, enables the agentic system to access dark web sources within the same Neo4j knowledge graph and LangChain session memory infrastructure used for surface web collection. The architectural significance of this integration is considerable: it demonstrates that dark web intelligence can be ingested, entity-extracted, and stored using the same pipeline as surface web content, without requiring a parallel dark web-specific system. This convergent architecture reduces the implementation barrier for organisations wishing to extend existing surface web OSINT pipelines into dark web collections.

The study did not fully address, however, either the operational security risks of automated dark web crawling or its legal dimensions. Automated access to .onion domains via an agent that can be identified through traffic analysis, timing correlation, or honeypot detection presents operational security risks that are qualitatively different from those of surface web collection. The legal question of whether automated access to dark web resources constitutes unauthorised access under the Computer Fraud and Abuse Act, the UK Computer Misuse Act, or equivalent national statutes receives no analysis. The implications of chain-of-custody for intelligence collected via Tor, as well as the admissibility of such intelligence in criminal prosecution or regulatory proceedings, remain completely unaddressed. These are not peripheral concerns for an operational tool; they are the primary obstacles to dark web OSINT intelligence being used in any context with legal consequences.

DarkLens \cite{amin2025darklens} proposes a conceptual tool architecture for dark web OSINT comprising six feature modules: Search, Alerts, Rules, Profiling, Dashboard, and Severity Classification. The Severity Classification module, which proposes to map dark web content automatically to country-specific legal codes, raises significant due process and jurisdictional concerns that the paper does not engage with: automated legal classification of content without human review and without jurisdiction specification conflates dramatically different legal standards across the many countries in which investigators and subjects may be located. The citation practices of \cite{amin2025darklens}, including "Google Scholar" as a listed reference and multiple entries with placeholder URLs, fall below the standards expected of peer-reviewed research. \cite{amin2025darklens} is useful for establishing the design space for a future dark web OSINT tool and for articulating the requirements that such a tool should meet, but it should not be cited as evidence for any technical or empirical claim.

Beyond these two papers, Telegram and other encrypted messaging platforms referenced in the threat intelligence literature as primary channels for threat actor coordination, criminal market operations, and extremist recruitment receive no dedicated treatment in any paper within the reviewed corpus. The specific NLP challenges posed by encrypted platform OSINT pseudonymous authorship, context-dependent slang, rapidly evolving coded terminology, and multilingual content are as significant as those of dark web text processing and are unaddressed in the corpus. The multilingual dimension of both dark web and encrypted messaging OSINT represents a compounded gap: no paper in the 74-item corpus evaluates LLM-based OSINT on non-English content at scale, and non-English threat actor communications on both the dark web and encrypted platforms constitute a majority of operationally significant traffic in many security contexts.

The most practically valuable contribution that dark web and specialized-source OSINT research could make to operational intelligence practice is not a more sophisticated crawling architecture \cite{shen2024llmosint} has demonstrated the technical feasibility of that component but a legally analyzed, ethically justified, and operationally validated methodology for obtaining, preserving, and presenting dark web intelligence in forms that are admissible in legal proceedings and defensible under the human rights law applicable in the investigator's and subject's jurisdictions. That contribution does not exist in the current corpus, and its absence represents the most significant practical gap between the OSINT-AI literature and operational OSINT reality for law enforcement and national security contexts.

\begin{figure*}[!t]  
  \centering  
  \includegraphics[width=0.8\linewidth]{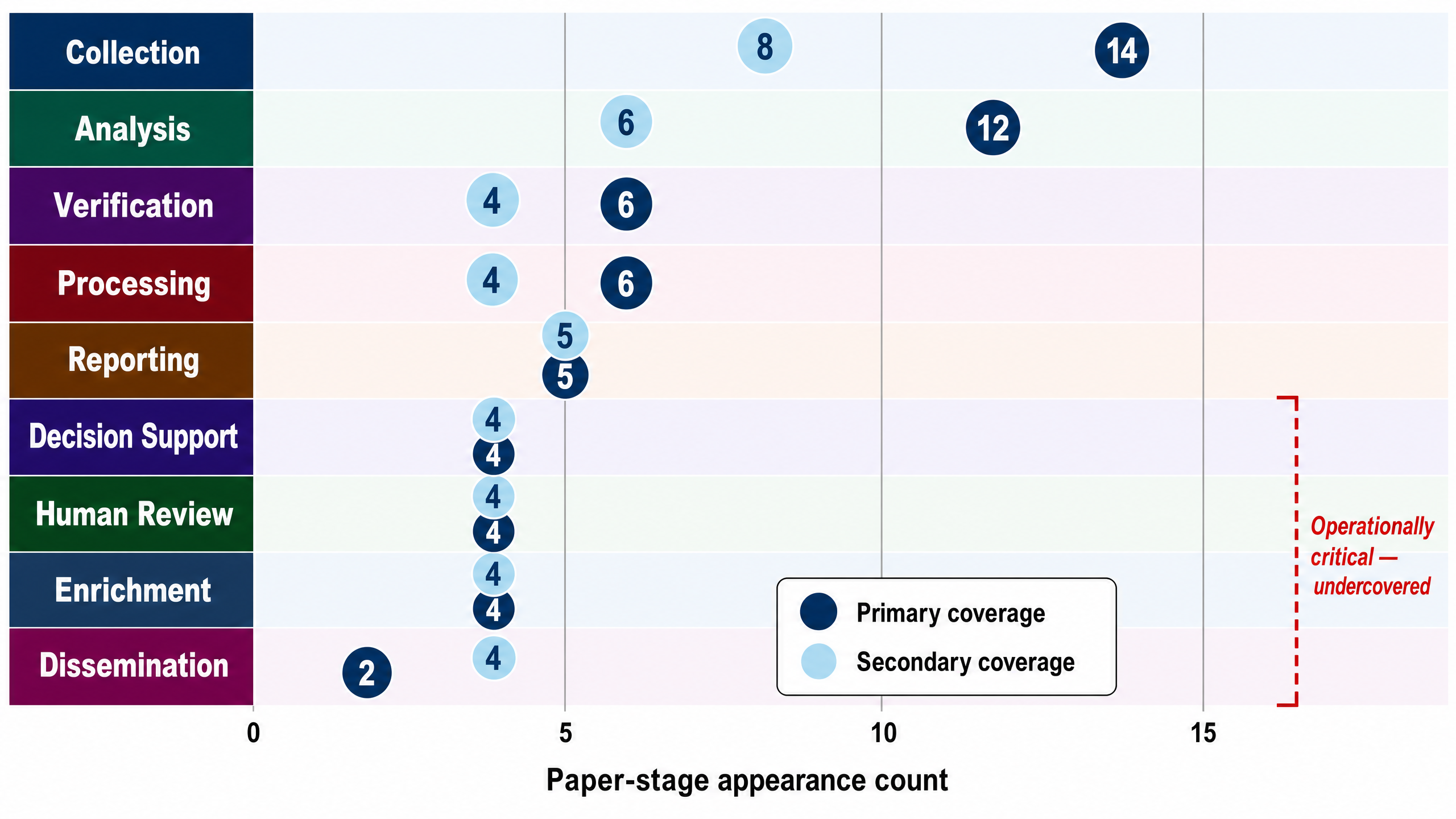}  
  \caption{Bubble Chart of Studies versus OSINT Workflow Stages.   Bubble size reflects the number of papers addressing each stage as primary or secondary coverage (data from Table~\ref{tab:table-6}). A single study may be counted in multiple stages, so values represent paper-stage appearances rather than unique paper counts; the sum across all stages exceeds the corpus total of 74 papers.}
  \label{fig:chart3}  
\end{figure*}


\section{Applications of Agentic and Generative AI in OSINT and Cyber Investigation}
\label{sec:4-applications-of-agentic-and-generative-ai-in-osint-and-dig}

Section 3 presents a taxonomy that defines the technologies and their composition. This section addresses what they are used for in the operational OSINT tasks to which LLMs, generative AI, and agentic architectures have been applied across the 74 studies comprising this corpus. The shift from taxonomy to application is not merely one of organisational level; it is a shift in the evaluative question that drives the analysis. Whereas Section 3 asked which approaches have been developed and with what technical properties, this section asks which OSINT workflow problems these approaches actually address, how effectively they address them, and where the gap between demonstrated capability and operational requirement remains widest.

The answer, across the nine application domains examined in Sections 4.1 through 4.9, is consistently asymmetric. Intelligence collection and analysis, the tasks most amenable to automation, most readily evaluated by computational metrics, and most frequently studied are reasonably well served by the current literature. Verification, dissemination, decision support, and human review the tasks most critical to intelligence reliability, most legally consequential, and most requiring of human judgement are systematically underserved. This asymmetry is not incidental: it reflects the structural preference of academic research for technically tractable problems with quantifiable outputs over the epistemologically complex, legally embedded, and empirically harder problems that constitute the practical core of professional OSINT practice.

\begin{table*}[htbp]
  \centering
  \caption{OSINT workflow coverage by study (stage-level summary). Pri.\ and Sec.\ give primary/secondary study counts. Identifiers index the supplementary Evidence Matrix (Table~S1).}
  \label{tab:table-6}
  \footnotesize
  \setlength{\tabcolsep}{4pt}\renewcommand{\arraystretch}{1.15}
\begin{tabularx}{\textwidth}{@{}l c >{\raggedright\arraybackslash}p{0.26\linewidth} c >{\raggedright\arraybackslash}X@{}}
\toprule
\textbf{Stage} & \textbf{Pri.} & \textbf{Representative Primary} & \textbf{Sec.} & \textbf{Assessment} \\
\midrule
Collection & 14 & \cite{almeidapalmieri2025framework}, \cite{mukhopadhyay2024osintclinic}, \cite{radoi2023ai}, \cite{shamunesh2023cybercheck,nagra2024kia,sermakani2024open,agrahari2025osintct2,pais2014osint,yurtalan2025redefining,rheault2024active,shen2024llmosint}, \cite{yang2024geolocator}, \cite{pervez2023ease}, \cite{amin2025darklens} & 8 & Most densely covered; API automation and surface-web dominant; dark-web (\cite{shen2024llmosint}, \cite{amin2025darklens}) and adversarial risks (\cite{yang2024geolocator}, \cite{niu2024bullshint}) critical \\
\addlinespace[2pt]
Processing & 6 & \cite{karakikes2025aiosint}, \cite{tseng2025disarm}, \cite{hassanin2024llms}, \cite{riad2024finetuning}, \cite{yang2024threatmodeling}, \cite{shafee2024evaluation} & 4 & Adequate; strongest are domain-adapted NER/classification (\cite{hassanin2024llms}, \cite{shafee2024evaluation}) \\
\addlinespace[2pt]
Enrichment & 4 & \cite{shen2024llmosint,su2025opensource,allam2025cybervision,zhao2023verifyandedit} & 4 & Moderate; KG (\cite{shen2024llmosint}, \cite{su2025opensource}) and RAG (\cite{allam2025cybervision}, \cite{zhao2023verifyandedit}) dominate; KG poisoning (\cite{ranade2021fake}) most consequential \\
\addlinespace[2pt]
Analysis & 12 & \cite{rajendran2024comprehensive}, \cite{pastorgalindo2020osint}, \cite{almeidapalmieri2025framework}, \cite{allam2025cybervision}, \cite{hassanin2024llms}, \cite{nila2023llms}, \cite{yigit2024review}, \cite{kobayashi2024subjective}, \cite{ranade2021fake}, \cite{tihanyi2024cybermetric}, \cite{chen2026cyberthreateval}, \cite{shafee2024evaluation} & 6 & Second-most covered; CTI analysis (\cite{tihanyi2024cybermetric}, \cite{chen2026cyberthreateval}, \cite{shafee2024evaluation}) and bias (\cite{kobayashi2024subjective}); \cite{chen2026cyberthreateval} finds no LLM adequate \\
\addlinespace[2pt]
Verification & 6 & \cite{allam2025cybervision}, \cite{zhao2023verifyandedit}, \cite{dekens2023practical}, \cite{kobayashi2024subjective}, \cite{niu2024bullshint}, \cite{sun2025decision} & 4 & Underrepresented; only \cite{allam2025cybervision}, \cite{zhao2023verifyandedit} provide technical mechanisms; gap noted by \cite{dekens2023practical}, \cite{chen2026cyberthreateval} \\
\addlinespace[2pt]
Reporting & 5 & \cite{yuan2024empowering}, \cite{zhao2023verifyandedit}, \cite{yigit2024review}, \cite{cerny2024implications}, \cite{chen2026cyberthreateval} & 5 & Moderate; \cite{cerny2024implications} frames reporting competency; automated generation evaluated only in \cite{chen2026cyberthreateval} \\
\addlinespace[2pt]
Dissemination & 2 & \cite{mukhopadhyay2024osintclinic}, \cite{yigit2024review} & 4 & Sparse; \cite{mukhopadhyay2024osintclinic} checkpoint architecture; \cite{yigit2024review} governance; no empirical automation \\
\addlinespace[2pt]
Decision Support & 4 & \cite{yigit2024review}, \cite{tihanyi2024cybermetric}, \cite{chen2026cyberthreateval}, \cite{srikanth2024usability} & 4 & Sparse; \cite{tihanyi2024cybermetric}/\cite{chen2026cyberthreateval} frame analyst support; no agentic system evaluated \\
\addlinespace[2pt]
Human Review & 4 & \cite{mukhopadhyay2024osintclinic}, \cite{rheault2024active}, \cite{chen2026cyberthreateval}, \cite{srikanth2024usability} & 4 & Most sparse relative to importance; oversight empirically validated in zero agentic papers \\
\midrule
\multicolumn{5}{@{}>{\raggedright\arraybackslash}p{\textwidth}@{}}{\textbf{Coverage gap:} Collection (14) and Analysis (12) dominate, whilst Dissemination (2), Decision Support (4), and Human Review (4) are systematically underrepresented the field has optimised for intelligence \emph{production} over \emph{governance}.} \\
\bottomrule
\end{tabularx}
\end{table*}

\begin{figure*}[!t]  
  \centering  
  \includegraphics[width=0.75\linewidth]{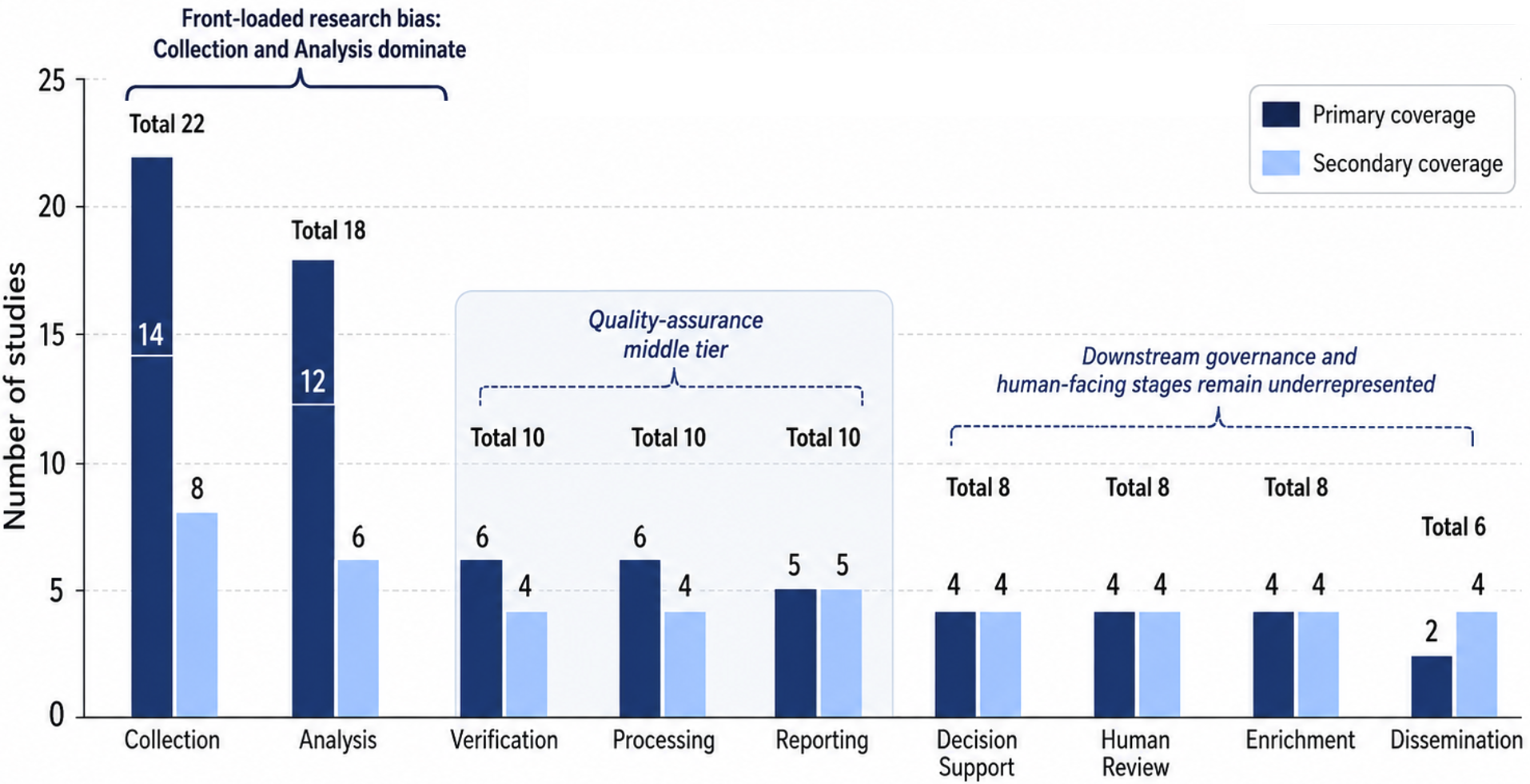}  
  \caption{Distribution of Studies by OSINT Workflow Stage. Bars show the count of paper-stage assignments per stage, disaggregated into primary coverage (darker) and secondary coverage (lighter). As with Figure~\ref{fig:chart3}, a single study may contribute to more than one stage; primary-stage assignments sum to~57 across the 37~primary/core papers (mean 1.5~stages per paper), and total assignments including secondary coverage are given in Table~\ref{tab:table-6}.}
  \label{fig:chart4}  
\end{figure*}


\subsection{Intelligence Collection and Source Discovery}
\label{sec:4-1-intelligence-collection-and-source-discovery}

Intelligence collection is the most extensively addressed domain in the corpus, with fourteen primary papers \cite{almeidapalmieri2025framework,mukhopadhyay2024osintclinic,radoi2023ai,shamunesh2023cybercheck,nagra2024kia,sermakani2024open,agrahari2025osintct2,pais2014osint,yurtalan2025redefining,rheault2024active,shen2024llmosint,pervez2023ease,amin2025darklens,sun2025decision}. Four generations are evident: LLM-augmented conventional pipelines \cite{radoi2023ai}; domain-specific structured-source automation \cite{shamunesh2023cybercheck,nagra2024kia,sermakani2024open,agrahari2025osintct2,pais2014osint}; ethically grounded collection \cite{rheault2024active,sun2025decision}; and agentic collection \cite{almeidapalmieri2025framework,shen2024llmosint,mukhopadhyay2024osintclinic}. The cloud-versus-on-premise decision is sharpest here: API pipelines expose targets to third-party logging \cite{radoi2023ai}, whereas on-premise models achieve sufficient performance for law enforcement \cite{yurtalan2025redefining} and military \cite{nila2023llms} OSINT, and the narrowing performance differential \cite{shafee2024evaluation} strengthens the on-premise case wherever investigation confidentiality is required.

\subsection{Information Extraction, Entity Recognition, and Summarisation}
\label{sec:4-2-information-extraction-entity-recognition-and-summarisat}

Information extraction encompassing NER, relationship extraction, coreference resolution, event detection, and intelligence summarisation is the intellectual operation through which raw open-source information becomes intelligence. The corpus addresses this domain with greater empirical precision than any other, producing the most reproducible performance figures in the review and also its most consequential unresolved deficit: the 25\% NER performance gap that \cite{shafee2024evaluation} measures with precision and that the corpus fails, collectively, to address.

\begin{figure}[htbp]  
  \centering  
  \includegraphics[width=\columnwidth]{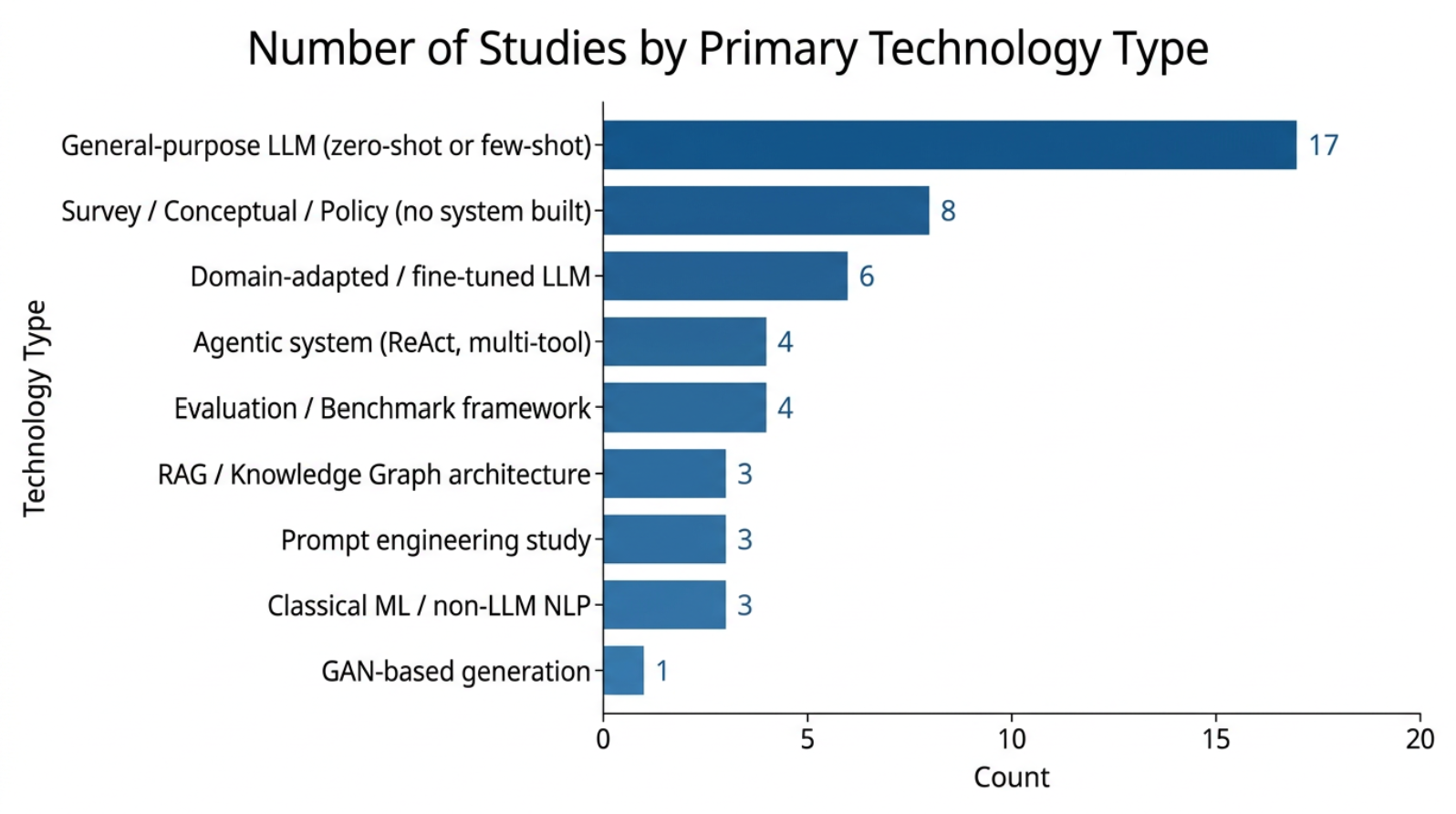}  
  \caption{Number of Studies by Primary Technology Type}  
  \label{fig:chart5}  
\end{figure}

\begin{table*}[htbp]
  \centering
  \caption{Data for Figure 10: studies by primary technology type. Identifiers index the supplementary Evidence Matrix (Table~S1).}
  \label{tab:data-chart-5}
  \footnotesize\setlength{\tabcolsep}{4pt}\renewcommand{\arraystretch}{1.15}
\begin{tabularx}{\textwidth}{@{}>{\raggedright\arraybackslash}p{0.27\linewidth} c >{\raggedright\arraybackslash}X >{\raggedright\arraybackslash}X@{}}
\toprule
\textbf{Technology Type} & \textbf{n} & \textbf{Representative Studies} & \textbf{Notes} \\
\midrule
General-purpose LLM (zero/few-shot; no extension) & 17 & \cite{rajendran2024comprehensive}, \cite{pastorgalindo2020osint}, \cite{radoi2023ai}, \cite{shamunesh2023cybercheck}, \cite{sermakani2024open,agrahari2025osintct2,pais2014osint,yurtalan2025redefining,rheault2024active}, \cite{nila2023llms}, \cite{yigit2024review}, \cite{kobayashi2024subjective}, \cite{yang2024geolocator}, \cite{golda2024privacy}, \cite{cerny2024implications}, \cite{sun2025decision}, \cite{shafee2024evaluation} & Largest category; dominant across application papers \\
\addlinespace[2pt]
Survey / Conceptual / Policy & 8 & \cite{zhou2024application}, \cite{pastorgalindo2020osint}, \cite{nagra2024kia}, \cite{yigit2024review}, \cite{pervez2023ease}, \cite{hwang2022osint}, \cite{amin2025darklens}, \cite{dekens2023practical} & Some overlap with LLM category; no system built \\
\addlinespace[2pt]
Domain-adapted / fine-tuned LLM (LoRA, PEFT) & 6 & \cite{tseng2025disarm}, \cite{hassanin2024llms}, \cite{ranade2021fake}, \cite{riad2024finetuning}, \cite{yang2024threatmodeling}, \cite{anas2024sentiments} & Consistently outperforms zero-shot on domain tasks \\
\addlinespace[2pt]
Agentic system (ReAct, LangChain) & 4 & \cite{yuan2024empowering}, \cite{almeidapalmieri2025framework}, \cite{mukhopadhyay2024osintclinic}, \cite{shen2024llmosint} & Highest complexity; lowest replicability \\
\addlinespace[2pt]
Evaluation / Benchmark framework & 4 & \cite{tihanyi2024cybermetric}, \cite{chen2026cyberthreateval}, \cite{srikanth2024usability}, \cite{sikand2024greencode} & \cite{sikand2024greencode} excluded; most ecologically valid papers \\
\addlinespace[2pt]
RAG / Knowledge Graph & 3 & \cite{su2025opensource}, \cite{allam2025cybervision}, \cite{zhao2023verifyandedit} & \cite{shen2024llmosint} counted under Agentic (KG\,+\,RAG) \\
\addlinespace[2pt]
Prompt engineering study & 3 & \cite{park2024performance}, \cite{cerny2024implications}, \cite{sun2025decision} & \cite{cerny2024implications}, \cite{sun2025decision} also use LLMs; assigned by primary contribution \\
\addlinespace[2pt]
Classical ML / non-LLM NLP & 3 & \cite{karakikes2025aiosint}, \cite{niu2024bullshint}, \cite{mouthami2025political} & Pre-LLM or hybrid; \cite{karakikes2025aiosint} BERT-based NER \\
\addlinespace[2pt]
GAN-based generation & 1 & \cite{saddi2024genai} & Outlier; low replicability \\
\bottomrule
\end{tabularx}
\end{table*}

Shafee et al. \cite{shafee2024evaluation} provide the most reproducible quantitative evidence for LLM extraction performance on OSINT-derived data. Evaluating seven LLM chatbots on Twitter-sourced cybersecurity content, binary classification achieves F1\,=\,0.94 for GPT-4 and F1\,=\,0.90 for open-source GPT4all, sufficient for large-scale OSINT triage under human supervision. The critical boundary emerges at the extraction task: NER performance falls approximately 25\% below specialized pre-trained models in zero-shot and few-shot settings. The entities NER must identify IP addresses, CVE identifiers, malware family names, threat actor attributions, domain names, and infrastructure indicators are precisely the artefacts that constitute the core analytical output of OSINT-derived intelligence. A system achieving excellent triage classification but missing one in four named entities has not solved the extraction problem; it has solved a proximal one while leaving the constitutive one unresolved.

The NER performance gap is largely unaddressed in the corpus. Domain-adapted models consistently outperform general-purpose equivalents on CTI NER tasks \cite{hassanin2024llms}, and parameter-efficient fine-tuning reduces the computational barrier to adaptation \cite{riad2024finetuning}. Yet no paper in the 74-item corpus fine-tunes a model specifically on OSINT-domain entity extraction and measures recovery of the \cite{shafee2024evaluation} deficit. The problem is measured precisely, the solution pathway is identifiable from adjacent fine-tuning work, and no paper connects them, a non-cumulative pattern exemplifying the field's broader methodological fragmentation.

The most structured entity taxonomy for OSINT-relevant extraction is provided by the DISARM disinformation framework, applied in \cite{tseng2025disarm} for Twitter-based election OSINT and referenced in \cite{karakikes2025aiosint} for NER and relationship extraction. DISARM's ontological specificity distinguishing actors, behaviors, channels, and narratives within a structured hierarchy enables supervised extraction training at a level of granularity that general-purpose NER systems do not achieve. Its potential generalisation to terrorist financing, criminal network mapping, and corporate intelligence extraction is not evaluated in the corpus.

Agentic systems integrate extraction as a continuous pipeline component. In \cite{almeidapalmieri2025framework} and \cite{shen2024llmosint}, NER occurs iteratively after each tool invocation: entities extracted from Shodan results are immediately linked in the Neo4j knowledge graph, enabling the agent to direct subsequent collection steps based on partial entity knowledge, a form of closed-loop collection without precedent in non-agentic systems. However, neither paper reports NER precision or recall for entities extracted within the agentic pipeline, leaving extraction quality entirely unevaluated.

Intelligence summarisation is addressed through prompt engineering \cite{cerny2024implications,chen2026cyberthreateval} and output quality evaluation \cite{chen2026cyberthreateval,srikanth2024usability}. The most rigorous assessment is from \cite{chen2026cyberthreateval}, whose analyst-centric evaluation finds that no LLM produces reports of analyst-acceptable quality for complex threat actor attribution. This constrains the optimism implied by \cite{shafee2024evaluation}: even where entity extraction is reasonably accurate, synthesising extracted entities into a coherent, actionable intelligence summary is a qualitatively higher-order task at which current LLMs fall short under realistic workflow conditions.

\subsection{Verification, Fact-Checking, and Misinformation Detection}
\label{sec:4-3-verification-fact-checking-and-misinformation-detection}

Verification is the workflow stage most critical to intelligence reliability and the most underexplored in the corpus. Despite widespread acknowledgement that LLM-generated intelligence is susceptible to hallucination, fabrication, and adversarial contamination, only one paper directly measures hallucination in an OSINT-specific system \cite{allam2025cybervision}, and no paper proposes a complete verification pipeline integrating cross-source fact-checking, provenance tracking, and adversarial content detection. The absence reflects the field's structural preference for building and demonstrating systems over rigorously validating their outputs.

Allam \cite{allam2025cybervision} provides the sole empirical anchor for OSINT AI verification. The RAG-augmented system using Claude 3.7 Sonnet achieves a hallucination rate of 4\%, a factual accuracy rate of 92\%, a citation accuracy rate of 96\%, and a task completion rate of 89\%, establishing that well-designed RAG can substantially reduce incidental hallucination. These figures do not, however, extend to smaller on-premise models, less curated knowledge bases, adversarially contaminated corpora, or domains outside the private evaluation dataset. The 4\% figure is an empirical floor under favourable conditions, a ceiling for current best practices rather than a representative operational rate.

Verify-and-Edit \cite{zhao2023verifyandedit} provides a complementary mechanism addressing propagation of reasoning errors across multi-step chains. By parsing generated reasoning into individual factual claims and revising those that fail external verification, the framework achieves a 4.5\% improvement in Exact Match on AdvHotpotQA and 5.9\% on 2WikiMultiHop over chain-of-thought baselines. Its convergence with \cite{shen2024llmosint}'s self-reflection loop and \cite{sun2025decision}'s iterative revision suggests that staged post-generation verification against accumulated knowledge is emerging as a consensus principle for OSINT AI output quality.

The adversarial dimension is addressed most rigorously by Ranade et al. \cite{ranade2021fake}, who report that professional threat hunters were equally likely to judge generated fake CTI and authentic CTI as true, with the majority of fake samples labelled true in the human evaluation ($n=10$; cautious generalisation is required). The significance is not that it characterises the average case but the adversarial one, which matters most for intelligence reliability: a verification architecture that handles incidental hallucination but fails against deliberately crafted adversarial content provides false assurance. No paper proposes a mechanism designed specifically to detect adversarially generated OSINT content, and \cite{ranade2021fake} demonstrates that standard NLP indicators fluency, coherence, and citation style are insufficient discriminators when the adversarial model has been trained on the same professional lexicon as the content it impersonates.

Niu et al. \cite{niu2024bullshint} address misinformation from the collection environment rather than the generation side. The 9\% misinformation rate within professional OSINT Twitter communities, identified via GraphSAGE community detection across 1.96 million tweets, establishes an environmental contamination rate that precedes any LLM processing. Automated collection pipelines \cite{shamunesh2023cybercheck,nagra2024kia,sermakani2024open,agrahari2025osintct2,pais2014osint} and social media approaches \cite{tseng2025disarm,riad2024finetuning} inherit this rate before any extraction or analysis is performed. The compounded effect of environmental contamination and LLM hallucination operating simultaneously has not been studied in the corpus; no paper attempts to quantify the combined error rate.

Kobayashi and Yamaguchi \cite{kobayashi2024subjective} identify a verification dimension that standard NLP metrics cannot detect: LLM-embedded political and ideological bias is transmitted into intelligence products. Citation accuracy at 96\% \cite{allam2025cybervision} confirms the truthfulness of stated claims but provides no assurance about the analytical framing that contextualizes them, a fundamental gap where evidence selection and weighting are as consequential as accuracy. Dekens's practitioner guidance \cite{dekens2023practical} captures this concern: experienced OSINT analysts report that AI tools produce plausible-sounding outputs requiring sceptical scrutiny and that the primary skill shift is learning to verify LLM outputs, not merely to use the tools.

\subsection{Cyber Threat Intelligence and Threat Awareness}
\label{sec:4-4-cyber-threat-intelligence-and-threat-awareness}

The application of LLMs to cyber threat intelligence production is the most empirically developed application domain in the corpus and the one that most clearly exposes the distinction between capability measured under evaluation conditions and capability required under operational conditions. The central finding of the CTI application literature is not a performance figure but a methodological lesson: what an LLM can do on a structured knowledge benchmark and what it can do in an operational threat intelligence workflow are different enough that the former provides no reliable evidence about the latter.

CyberMetric \cite{tihanyi2024cybermetric} demonstrates the benchmark performance ceiling: GPT-4o achieves 91.25\% accuracy on a 2,000-question multiple-choice cybersecurity knowledge assessment against a mean human expert score of 72.24\%. CyberThreat-Eval \cite{chen2026cyberthreateval} tests this finding against operational reality through a three-stage CTI workflow triage of 488 threat-intelligence articles, a deep search across 55 URLs, and full report drafting derived from real Microsoft security operations. No LLM achieves analyst-acceptable performance on complex threat actor differentiation, and the evaluation explicitly critiques lexical metrics (ROUGE, BLEU, BERTScore) as insufficient, measuring surface similarity rather than factual accuracy, analytical completeness, or actionability.

The contrast between \cite{tihanyi2024cybermetric} and \cite{chen2026cyberthreateval} is not an empirical contradiction but a demonstration of how evaluation design determines performance conclusions. Multiple-choice assessment measures structured recall under zero ambiguity; operational CTI production requires open-ended reasoning from incomplete and potentially adversarially manipulated evidence. These are categorically different competencies, and benchmark superiority on the former provides no assurance about the latter. The lesson for deployment decisions is that benchmark performance must be critically assessed for ecological validity relative to the operational tasks under consideration.

The performance evidence from Shafee et al. \cite{shafee2024evaluation} refines this picture with OSINT-derived data. Binary classification of cybersecurity-relevant content: the triage function that corresponds most closely to \cite{tihanyi2024cybermetric}'s classification task achieves F1 = 0.94 for GPT-4 and F1 = 0.90 for open-source GPT4all, performance levels comparable to human expert classification and sufficient to support large-scale alert triage under analyst oversight. Named entity recognition, the extraction function that \cite{shafee2024evaluation} shows falling 25\% below specialized model performance corresponds most closely to the entity-intensive aspects of \cite{chen2026cyberthreateval}'s deep search and report drafting stages. The practical boundary defined by \cite{shafee2024evaluation} is therefore consistent with the operational finding of \cite{chen2026cyberthreateval}: Current general-purpose LLMs handle triage-class tasks reasonably well and fail or perform below specialist standards on the entity-intensive, context-dependent, multi-source synthesis tasks that constitute operational CTI production.

Domain adaptation partially recovers this deficit. Hassanin and Moustafa \cite{hassanin2024llms} demonstrate that CyberBERT and SecureBERT consistently outperform general-purpose BERT in CTI entity extraction and relationship classification; Yang et al. \cite{yang2024threatmodeling} demonstrate comparable gains in banking-sector threat modelling. The CTI performance boundary is not fixed by the LLM paradigm but by insufficient domain adaptation, a recoverable limitation. No paper, however, evaluates whether domain-adapted models close the \cite{shafee2024evaluation} NER gap, achieve the \cite{chen2026cyberthreateval} operational quality standard, or resist the adversarial attack documented by \cite{ranade2021fake}.

The adversarial dimension transforms the CTI picture in ways the evaluation literature does not address. Ranade et al. \cite{ranade2021fake} establish that the same open channels from which CTI systems collect intelligence are adversarially exploitable sources of convincing fake CTI; reported results suggest professional threat hunters were equally likely to judge fake and authentic CTI as true ($n=10$), with the majority of fake samples labelled true. When applied against KG-based architectures via deliberate NER pipeline poisoning, the knowledge infrastructure that provides the grounding benefit of \cite{allam2025cybervision} becomes an amplification mechanism for adversarial error. This gap between the benign conditions of \cite{chen2026cyberthreateval} and the adversarial conditions of \cite{ranade2021fake} is the most critical ecological validity failure in the CTI application literature.

Srikanth et al. \cite{srikanth2024usability} add a sociotechnical constraint: experienced CTI professionals (median experience 4/5) have a median LLM familiarity of only 2/5, revealing that CTI AI capability and deployability are distinct conditions that the evaluation literature conflates. Technical performance under uncritical acceptance is precisely the condition under which \cite{ranade2021fake}'s adversarial content is most dangerous.

\subsection{SOCMINT and Social Media Analysis}
\label{sec:4-5-socmint-and-social-media-analysis}

Social media intelligence (SOCMINT) is the most frequently utilized source environment in the corpus, with Twitter/X serving as the primary platform for studies spanning disinformation classification \cite{tseng2025disarm}, cybersecurity content analysis \cite{shafee2024evaluation}, sentiment analysis \cite{mouthami2025political,riad2024finetuning}, misinformation detection \cite{niu2024bullshint}, and automated OSINT pipelines \cite{radoi2023ai,pais2014osint}. This concentration reflects Twitter's historical API openness and the genuine intellectual relevance of social media, but also a research convenience bias: the availability of Twitter data has shaped which problems the field has studied, not which are most operationally significant.

Niu et al. \cite{niu2024bullshint} provide the most important empirical characterisation of the SOCMINT source environment. Applying GraphSAGE-based community detection to 1.96 million tweets from 48 professional OSINT and threat-intelligence Twitter communities (with the Russo-Ukrainian War as a focal event), the study finds that approximately 9\% of the content circulating within professional communities constitutes misinformation. This is epistemologically significant: professional community membership does not provide reliable signal quality. An automated collection system that applies no community-quality filter inherits this 9\% contamination rate at scale, and even during periods of acute geopolitical scrutiny, the quantity and urgency of information production actively facilitate misinformation propagation by overwhelming verification capacity.

Almutairi et al. \cite{tseng2025disarm} demonstrate the application of fine-tuned LLMs to election-related OSINT on Twitter using the DISARM disinformation framework for annotation and classification. The consistent superiority of fine-tuned models over zero-shot and few-shot approaches for OSINT-relevant disinformation classification confirms the domain adaptation consensus established in Section 3.8. The DISARM framework's structured entity taxonomy distinguishing actors, behaviors, channels, and narratives enables classification at a level of analytical granularity that general-purpose classification approaches do not achieve, and the framework's reusability across disinformation contexts beyond election campaigns is not exploited in the corpus.

Sentiment analysis is addressed by Patel et al. \cite{mouthami2025political} and Riad et al. \cite{riad2024finetuning}. The 98\% accuracy claimed by \cite{mouthami2025political} for a Bi-LSTM enhanced with LLaMA 3 is acknowledged within the paper as likely inflated by dataset simplicity and overfitting; the LLaMA 3 contribution is not demonstrably separable from the Bi-LSTM baseline. \cite{riad2024finetuning}'s application of LoRA and PEFT to tweet sentiment classification provides methodologically more reliable evidence: parameter-efficient fine-tuning outperforms full fine-tuning at lower computational cost, with a more clearly specified evaluation protocol.

Political and ideological bias embedded in LLM outputs \cite{kobayashi2024subjective} is a particular concern for SOCMINT, where subject matter systematically intersects with training-data biases, causing systematic over- or under-classification depending on political alignment. Sun et al.'s \cite{sun2025decision} self-reflection approach partially addresses bias in report generation, but whether it is sufficient for social media classification where bias affects entity extraction and event labelling, not only report synthesis, remains untested.

The reviewed SOCMINT sub-corpus is English-language and Twitter-centric: no paper evaluates LLM-based SOCMINT on Facebook, Telegram, Reddit, WeChat, or any other platform, and no paper addresses multilingual OSINT at scale despite the fact that operationally significant events routinely involve communication across multiple languages and platforms. The Twitter API restrictions introduced in 2023 are not addressed in any corpus paper, despite affecting the reproducibility of every study that used the platform. The SOCMINT literature provides a detailed picture of a single platform under conditions that no longer fully obtain, leaving the operationally critical multilingual, multi-platform challenge entirely unaddressed.

\subsection{Dark Web and Specialized Source Monitoring}
\label{sec:4-6-dark-web-and-specialized-source-monitoring}

The intelligence value of the dark web, encrypted messaging platforms, invitation-only criminal forums, and non-indexed repositories is not proportional to their coverage in the literature. These environments host communications from threat actors, criminal market infrastructure, and leaked credential databases that are unavailable through surface-web collection. The corpus dedicates only two papers to this domain: one working implementation and one conceptual proposal, a disproportion that is itself a finding.

The only working dark web OSINT capability is within the Tsinghua system \cite{shen2024llmosint}: a Tor-based crawling module with IP rotation and dynamic .onion domain handling enables dark web intelligence to be ingested and stored within the same Neo4j knowledge graph as surface-web collection. This integration is architecturally significant, demonstrating that a general OSINT pipeline can extend to dark-web sources without a parallel system, but the capability is unvalidated for dark-web NLP, where content is frequently multilingual, idiomatic, and deliberately obfuscated. The empirical question of whether surface-web extraction performance transfers to dark-web text is not addressed by \cite{shen2024llmosint}.

The DarkLens framework \cite{amin2025darklens} proposes the design space for a dedicated dark web OSINT tool through a six-module taxonomy: Search, Alerts, Rules, Profiling, Dashboard, and Severity Classification. The Severity Classification module, which proposes to map dark web content automatically to country-specific legal codes, raises jurisdictional and due process concerns that the paper does not engage with. The paper is useful for establishing what a comprehensive dark web OSINT tool would need to accomplish; it provides no evidence that any component of the proposed architecture has been implemented or evaluated. Its citation practices, which include `\texttt{Google Scholar}' as a listed reference and several entries with placeholder URLs, fall below the quality expected of peer-reviewed research. DarkLens should be cited only as a design proposal.

Beyond these two papers, the operational challenges of dark-web monitoring are almost entirely unaddressed: how .onion domains are discovered when URLs change frequently; how investigator anonymity is maintained against honeypot detection; how Telegram and encrypted messaging platforms are systematically monitored; and how collected intelligence is preserved with chain-of-custody integrity for legal proceedings. The last omission is most consequential: dark web intelligence is operationally valuable only if its collection methodology can withstand legal scrutiny. No paper proposes a methodology explicitly designed to produce legally admissible dark web intelligence.

\begin{figure*}[!t]  
  \centering  
  \includegraphics[width=\linewidth]{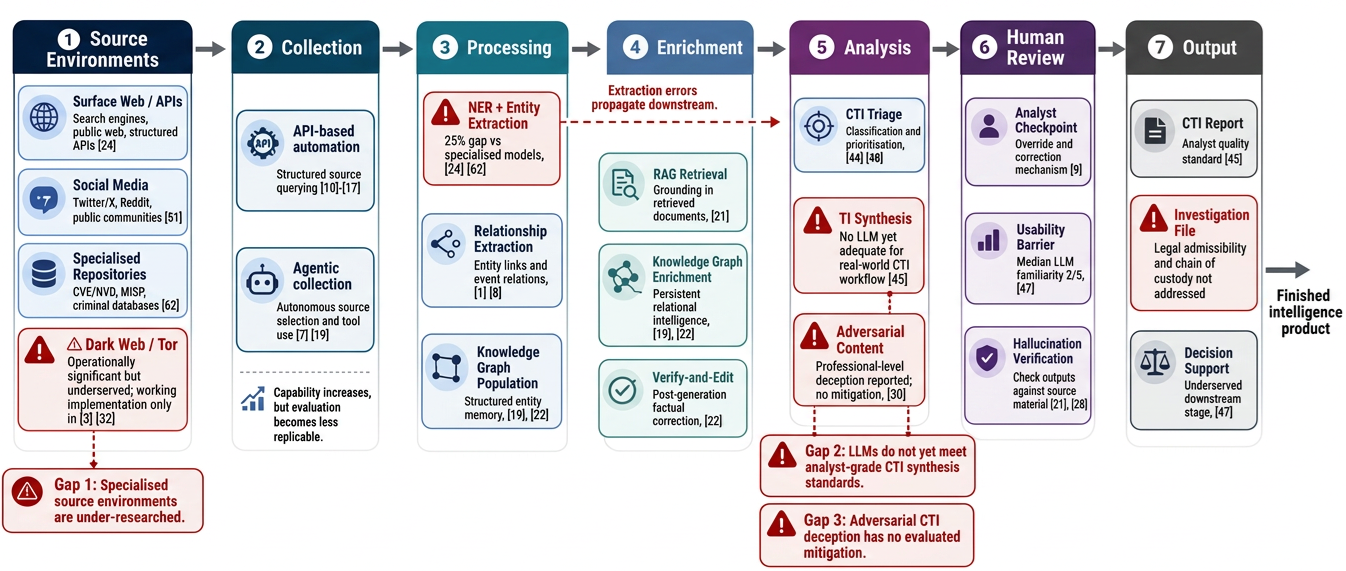}  
  \caption{OSINT-to-CTI and Cyber Investigation pipeline: a left-to-right flow across six stages from source environments to final output, with study identifiers shown where corpus evidence exists. The figure highlights three critical capability gaps.}  
  \label{fig:ctipipeline}  
\end{figure*}

\subsection{Report Generation and Analyst Decision Support}
\label{sec:4-7-report-generation-and-analyst-decision-support}

Report generation, which translates collected and analyzed OSINT into actionable products for decision-makers, is among the most analytically demanding tasks in the lifecycle. The corpus addresses it as an output category in several studies \cite{almeidapalmieri2025framework,mukhopadhyay2024osintclinic,allam2025cybervision,cerny2024implications,sun2025decision,chen2026cyberthreateval} but evaluates report quality rigorously in only one \cite{chen2026cyberthreateval}. The distance between producing a report and producing a verified, legally defensible intelligence product is precisely the distance most papers in this domain do not cross.

The corpus consistently elides the distinction between report \emph{drafting} and verified intelligence \emph{production}. Report drafting, which involves generating a structured text organised into an intelligence-shaped document, is within the demonstrated capabilities of LLMs: \cite{almeidapalmieri2025framework} produces structured collection summaries, \cite{allam2025cybervision} generates cited products with 96\% citation accuracy, and \cite{cerny2024implications} produces structured outputs from targeted prompts. Verified intelligence production in which every claim is source traceable, confidence levels are calibrated, source disagreements are surfaced, and an analyst reviews for errors and adversarial content is a fundamentally more demanding standard that no corpus paper demonstrates an LLM system meeting in full.

CyberThreat-Eval \cite{chen2026cyberthreateval} provides the most rigorous evidence for this distinction. Its three-stage analyst workflow evaluation, triage, deep search, and threat intelligence report drafting finds that, whilst LLMs can generate structured drafts at the triage and search stages, no model produces TI reports of analyst-acceptable quality for complex threat actor attribution and technical infrastructure analysis. The deficit manifests specifically in the stages that require synthesis under uncertainty: distinguishing between threat actors with overlapping TTPs, contextualising infrastructure indicators within the evolving tactical landscape, and communicating confidence levels in ways that support rather than mislead downstream decision-making. These are precisely the analytical capabilities that the multiple-choice benchmark performance of \cite{tihanyi2024cybermetric} does not measure and cannot predict.

The iterative report refinement approach of Sun et al. \cite{sun2025decision} represents the most developed attempt in the corpus to improve report quality through structured revision. By applying a self-reflection loop in which the model evaluates its own output for political balance and factual consistency before finalising, \cite{sun2025decision} demonstrates measurable quality improvements for bias-prone OSINT topics. The approach converges with \cite{zhao2023verifyandedit}'s Verify-and-Edit correction and \cite{shen2024llmosint}'s self-reflection mechanism, reinforcing the consensus that report generation in OSINT AI should be treated as a multi-stage process with explicit revision steps, not as a single-pass generation task. {\v{C}}ern{\'y} \cite{cerny2024implications}'s prompt engineering framework provides structural guidance for generating collection-appropriate reports across targeted, bulk, and monitored collection modes, though without evaluating whether the reports generated meet operational quality standards comparable to those applied in \cite{chen2026cyberthreateval}.

Decision support receives less direct attention than report generation, despite being the ultimate purpose of the intelligence lifecycle. Mukhopadhyay and Luther \cite{mukhopadhyay2024osintclinic} argue that human oversight checkpoints are required before AI-generated intelligence reaches decision-makers; Yigit et al. \cite{yigit2024review}'s MetaAID framework addresses the dual-use risk of products reaching decision-makers without adequate epistemic provenance. No paper evaluates decision quality outcomes, whether decisions made with AI-assisted intelligence support are better calibrated, faster, or more accurate. The corpus can document what LLMs produce but cannot provide evidence that what they produce improves the decisions they are intended to support.

\subsection{Human-AI Collaboration in OSINT Workflows}
\label{sec:4-8-human-ai-collaboration-in-osint-workflows}

The human analyst cannot be optimised out of the OSINT workflow as automation advances, not from professional conservatism but because of the specific failure modes documented here: adversarial content judged genuine by professional threat hunters \cite{ranade2021fake} ($n=10$); hallucination measured in only one system \cite{allam2025cybervision}; NER gaps automated verification cannot self-detect \cite{shafee2024evaluation}; embedded bias citation metrics cannot reveal \cite{kobayashi2024subjective}; and benchmark performance leaving operational capability undemonstrated \cite{tihanyi2024cybermetric,chen2026cyberthreateval}. Mukhopadhyay and Luther\ \cite{mukhopadhyay2024osintclinic} make the most principled architectural case: agentic autonomy \cite{almeidapalmieri2025framework,shen2024llmosint} is inappropriate where errors propagate, and oversight checkpoints reflect the legal accountability structure of intelligence practice. Rheault et al.\ \cite{rheault2024active} show this accountability must be embedded at the design stage, a standard neither agentic system approaches.

The practical obstacle is the adoption gap \cite{srikanth2024usability}: professionals with high CTI experience but low LLM familiarity (median 2/5) cite hallucinations and verification difficulties as the primary barriers. Effective oversight requires not only mechanisms but also the LLM literacy to use them, neither of which is systematically addressed in the corpus. The practitioner literature \cite{dekens2023practical} reinforces this point: the key skill shift is treating the LLM as a source requiring validation, not a producer of validated outputs. Framing prompt engineering as a trainable analyst competency \cite{cerny2024implications} identifies a further gap the absence of curricula, proficiency standards, and certification pathways that technical design alone cannot close. Table~\ref{tab:table-7} summarises the oversight provisions across application domains.

\begin{table*}[htbp]
  \centering
  \caption{Human oversight and safeguards across OSINT application domains. Status codes: \textbf{O}=oversight mechanism proposed, \textbf{E}=empirically evaluated, \textbf{A}=adversarial robustness addressed, \textbf{L}=legal/ethical safeguard, \textbf{I}=IRB or equivalent; ${\checkmark}$ full, ${\sim}$ partial, ${\times}$ absent. Identifiers index Table~S1.}
  \label{tab:table-7}
  \footnotesize
  \setlength{\tabcolsep}{4pt}\renewcommand{\arraystretch}{1.15}
\begin{tabularx}{\textwidth}{@{}>{\raggedright\arraybackslash}p{0.20\linewidth} >{\raggedright\arraybackslash}p{0.15\linewidth} >{\raggedright\arraybackslash}p{0.12\linewidth} >{\raggedright\arraybackslash}X@{}}
\toprule
\textbf{Application Domain} & \textbf{Status (O\,E\,A\,L\,I)} & \textbf{Studies} & \textbf{Assessment} \\
\midrule
Collection \& Source Discovery & $\sim\;\times\;\times\;\sim\;\checkmark$ & \cite{rheault2024active}, \cite{radoi2023ai}, \cite{yurtalan2025redefining} & Only IRB-approved collection paper (\cite{rheault2024active}); no agentic paper evaluates oversight; legal/privacy risks documented, not mitigated \\
\addlinespace[2pt]
Extraction, NER \& Summarisation & $\times\;\times\;\times\;\times\;\times$ & \cite{shafee2024evaluation}, \cite{hassanin2024llms}, \cite{karakikes2025aiosint} & No oversight in any extraction paper; systems autonomous; errors propagate unchecked \\
\addlinespace[2pt]
Verification \& Fact-Checking & $\sim\;\times\;\times\;\times\;\times$ & \cite{allam2025cybervision}, \cite{zhao2023verifyandedit}, \cite{niu2024bullshint} & RAG (\cite{allam2025cybervision}) and Verify-and-Edit (\cite{zhao2023verifyandedit}) partial; misinformation detection (\cite{niu2024bullshint}) detection-only; no adversarial evaluation \\
\addlinespace[2pt]
CTI \& Threat Awareness & $\checkmark\;\sim\;\times\;\sim\;\times$ & \cite{chen2026cyberthreateval}, \cite{srikanth2024usability} & \cite{chen2026cyberthreateval} analyst review at reporting; \cite{srikanth2024usability} usability ($n=10$); no adversarial CTI tested; cloud legal exposure (\cite{golda2024privacy}) \\
\addlinespace[2pt]
SOCMINT \& Social Media & $\times\;\times\;\times\;\times\;\times$ & \cite{tseng2025disarm}, \cite{niu2024bullshint}, \cite{kobayashi2024subjective} & No oversight; bias (\cite{kobayashi2024subjective}) and misinformation (\cite{niu2024bullshint}) identified, not mitigated; fine-tuning lacks consent framework \\
\addlinespace[2pt]
Dark Web \& Specialized Sources & $\times\;\times\;\times\;\times\;\times$ & \cite{shen2024llmosint}, \cite{amin2025darklens} & No oversight, safeguard, or chain-of-custody; \cite{shen2024llmosint} Tor integration unevaluated; legal risks unaddressed \\
\addlinespace[2pt]
Reporting \& Decision Support & $\sim\;\sim\;\times\;\sim\;\times$ & \cite{chen2026cyberthreateval}, \cite{mukhopadhyay2024osintclinic}, \cite{yigit2024review} & \cite{chen2026cyberthreateval} analyst quality review; \cite{mukhopadhyay2024osintclinic} checkpoint architecture (unevaluated); \cite{yigit2024review} MetaAID governance \\
\addlinespace[2pt]
Human--AI Collaboration & $\checkmark\;\sim\;\times\;\times\;\checkmark$ & \cite{mukhopadhyay2024osintclinic}, \cite{rheault2024active}, \cite{srikanth2024usability} & \cite{mukhopadhyay2024osintclinic} checkpoint design (unevaluated); \cite{srikanth2024usability} adoption gap (familiarity 2/5); \cite{rheault2024active} IRB benchmark; operationally unvalidated \\
\addlinespace[2pt]
Cyber Investigation \& Forensics & $\checkmark\;\times\;\times\;\sim\;\checkmark$ & \cite{yigit2024review}, \cite{rheault2024active} & \cite{yigit2024review} MetaAID most complete governance; \cite{rheault2024active} ethical validation; neither evaluated in operational forensic conditions \\
\midrule
\multicolumn{4}{@{}>{\raggedright\arraybackslash}p{\textwidth}@{}}{\textit{Overall:} oversight proposed in 5/9 domains, validated in $<$1/9, adversarial robustness in 0/9, IRB in 1/9 (\cite{rheault2024active}). The gap between architectural proposal and operational validation is the review's second central finding.} \\
\bottomrule
\end{tabularx}
\end{table*}

\subsection{Cyber Investigation and Forensic Support}
\label{sec:4-9-digital-investigation-and-forensic-support}

LLM-assisted Cyber Investigation occupies a position of genuine promise and significant caution. The promise lies in the alignment between LLM capabilities, extraction from heterogeneous text, relationship mapping, cross-source pattern recognition, report generation and investigative demands. The caution lies in the near-total absence of engagement with evidentiary standards, chain-of-custody requirements, and admissibility frameworks. Yigit et al.'s MetaAID \cite{yigit2024review} is the most relevant contribution, a governance model addressing dual-use, ethical obligations, and accountability, but it establishes governance questions without showing how provenance and interpretability would be operationalized in a legal evidence framework. Rheault et al.\ \cite{rheault2024active} provide the most forensically rigorous methodology, with OSINT-V being the closest the corpus comes to a chain-of-custody standard, though it was designed for human-led collection and its extension to agentic systems is unaddressed.

The agentic and RAG systems produce outputs that could in principle support investigation citation-bearing summaries \cite{almeidapalmieri2025framework}, 96\% citation accuracy \cite{allam2025cybervision}, a structured entity graph \cite{shen2024llmosint} yet none was designed for legal admissibility. The 4\% hallucination rate \cite{allam2025cybervision}, manageable under intelligence review, is not acceptable for evidence subject to adversarial challenge. Chain-of-thought \cite{wei2022cot} provides partial reasoning transparency and citation metrics for limited source traceability, but neither meets the legal standard of full provenance, and no paper addresses whether AI-generated intelligence can serve as primary evidence or under what conditions it can be authenticated. The domain is best characterised as one of demonstrated technical relevance and underdeveloped legal infrastructure, whose responsible development requires engagement with legal scholars and forensic specialists that is absent from every relevant paper.

\section{Case Studies}
\label{sec:5-case-studies}

The four case studies presented in this section are not selected as exemplars of best practice in isolation. They are selected because, collectively, they expose the four most consequential tensions in the OSINT-AI corpus: the gap between architectural ambition and empirical validation; the near-total absence of hallucination measurement against the sole empirical data point that fills it; the distinction between benchmark performance and operational readiness; and the challenge of designing evaluation frameworks that are simultaneously rigorous, ecologically valid, and reproducible. Each case study is examined across eight dimensions problem addressed, proposed system or method, data sources and workflow, evaluation approach and results, strengths, weaknesses and limitations, relevance to the wider review, and what the case reveals about the maturity of the field to provide the deepest analytical engagement that the primary sources support.

\subsection{Case Study 1: Palmieri \cite{almeidapalmieri2025framework} --- An Agentic and Generative AI OSINT Framework}
\label{sec:5-1-case-study-1-palmieri}

Palmieri et al. \cite{almeidapalmieri2025framework} is the corpus's central paper and its most complete agentic OSINT proof-of-concept. It asks whether a general-purpose LLM, operating in a structured reasoning loop with access to real OSINT tools, can autonomously plan and execute multi-step collection without task-specific engineering. The system is a ReAct agent that alternates thought steps (natural-language reasoning about the task state) with Act steps (invocations of Shodan, Maltego, VirusTotal, Censys, and theHarvester), whose observations update subsequent reasoning. A natural-language objective is decomposed into a tool-call sequence spanning social, web, and technical OSINT within a single loop. The evaluation is a proof-of-concept comparison reporting task-level figures across multi-tool tasks; it is the first OSINT-agent evaluation to address completion, tool selection, attribution, and verification in one protocol.

Its principal contribution is that ReAct reasoning generalizes across heterogeneous tool types, is independent of the specific, non-reproducible figures, and provides a reusable evaluation template the field has not built upon. The limitations are equally instructive: the evaluation tasks are undisclosed, so results are non-replicable; testing is exclusively benign, assuming genuine tool outputs, whereas Ranade et al.\ \cite{ranade2021fake} show that the open channels these tools draw on routinely carry adversarial content; and no human-oversight mechanism is assessed. Palmieri thus sets the state of the art against which architectural alternatives \cite{shen2024llmosint} and evaluation advances \cite{chen2026cyberthreateval} are measured throughout this review, while illustrating the field's position: capable of demonstrating agentic OSINT under favourable conditions, but not yet of validating it under deployment conditions.

\subsection{Case Study 2: Shen et al. \cite{shen2024llmosint} LLM-Based OSINT Agent with Memory, Knowledge Integration, Tool Use, and Self-Reflection}
\label{sec:5-2-case-study-2-tsinghua}

Where Palmieri addresses single-session orchestration, the Tsinghua system \cite{shen2024llmosint} asks whether one architecture can also accumulate intelligence across sessions, persist entities and relationships, integrate dark-web sources, and validate outputs against prior intelligence. It integrates four components: LangChain orchestration (the agentic planning layer, with multi-step session support); a Neo4j knowledge graph persisting extracted entities, persons, organisations, domains, CVEs, malware families, threat actors, and locations as queryable triples across sessions; session memory maintaining continuity across tool calls; and a self-reflection mechanism that checks outputs against the accumulated graph before finalising, converging independently with the Verify-and-Edit approach of \cite{zhao2023verifyandedit} and the iterative revision of \cite{sun2025decision}. Tor-based crawling with IP rotation brings \texttt{.onion} sources into the same pipeline as surface-web collection.

This is the most architecturally complete system in the corpus and, simultaneously, the least evaluated: no metrics are reported for extraction quality, graph-population accuracy, self-reflection effectiveness, or dark-web reliability; the paper demonstrates function rather than measuring performance. Consequently, its operational reliability is unknown, dark-web NLP performance is uncharacterised, and the legal, operational-security, and chain-of-custody dimensions of automated Tor crawling receive no analysis; its status as a non--peer-reviewed preprint further limits independent verification of its claims. Its knowledge-graph design also inherits the poisoning vulnerability demonstrated by \cite{ranade2021fake}, defining the central tension in OSINT knowledge infrastructure: the mechanisms that reduce hallucination also amplify adversarial error when the knowledge base is compromised. The inversion of design ambition over validation rigour is the signature of a field in its early maturity.

\subsection{Case Study 3: Allam \cite{allam2025cybervision} AI Impact on OSINT Technologies and Hallucination Measurement}
\label{sec:5-3-case-study-3-fernandes}

Hallucination is named the primary reliability threat in more than twenty corpus papers, yet empirical measurement in OSINT-specific systems is otherwise absent. Allam \cite{allam2025cybervision} supplies the only such measurement across the 48 primary and methodological support papers subjected to full systematic extraction; none of the remaining 26 background or excluded studies provides an equivalent OSINT-specific rate. The system augments Claude 3.7 Sonnet with a purpose-built OSINT retrieval-augmented generation (RAG) knowledge base, grounding outputs in sources retrieved at inference time so that each claim traces to a retrieved document rather than to parametric knowledge. Evaluation uses a private, curated knowledge base a favourable condition across six metrics: precision-at-5 (0.83), mean reciprocal rank (0.76), factual accuracy (92\%), citation accuracy (96\%), task completion rate (89\%), and a hallucination rate of 4\%, the single most significant reliability data point in the corpus and the most comprehensive RAG-OSINT protocol within it.

The 96\% citation accuracy establishes traceability as a precondition for analytical accountability, and the result reinforces RAG alongside the self-reflection of \cite{shen2024llmosint} and the correction of \cite{zhao2023verifyandedit} as the best-supported hallucination-reduction architecture. Its limitations bound the claim: the private dataset precludes replication and comparison; the curated corpus makes 4\% an empirical \emph{floor} under favourable conditions rather than an operational rate; the proprietary, cloud-deployed model leaves on-premise performance unknown; the 11\% task-completion failure (one query in nine) is operationally significant yet underemphasised; and the rate under adversarial knowledge-base contamination, which \cite{ranade2021fake} shows corrupts outputs, is unmeasured. The contrast between near-universal identification of hallucination and a single favourable-condition measurement is the starkest evidence of the field's evaluation immaturity.

\subsection{Case Study 4: Chen et al. \cite{chen2026cyberthreateval} --- CyberThreat-Eval: Evaluating LLMs for Real-World CTI Tasks}
\label{sec:5-4-case-study-4-cyberthreateval}

CyberThreat-Eval \cite{chen2026cyberthreateval} tests the implicit assumption that performance on structured CTI benchmarks predicts performance on open-ended, multi-source analytical work and finds it fails. Rather than a new system, it proposes an analyst-centric, end-to-end evaluation derived from a real CTI workflow in three stages: triage (relevance assessment of 488 threat-intelligence articles), targeted deep search (multi-hop retrieval over 55 identified URLs), and report drafting (synthesis into a structured report with attribution, indicator extraction, confidence assignment, and recommendations). The framework explicitly rejects lexical metrics (ROUGE, BLEU, and BERTScore) in favour of analyst-judged criteria: factual accuracy, attribution completeness, infrastructure-analysis quality, confidence appropriateness, and operational actionability. Under these criteria no evaluated LLM, including GPT-4-class models, reaches analyst-acceptable performance on complex attribution at the drafting stage; triage performance is adequate (consistent with \cite{shafee2024evaluation}'s F1\,=\,0.94), then degrades through deep search and drafting as synthesis under uncertainty increasingly diverges from structured recall.

This is the most ecologically valid evaluation in the corpus, testing models against an operationally meaningful standard; its paper, code, and benchmark are public, though the single-organisation origin and incompletely published expert criteria limit full replication, and only the LLM component, not agentic architectures \cite{almeidapalmieri2025framework, shen2024llmosint}, is assessed. Its decisive contribution is methodological: the contrast between \cite{tihanyi2024cybermetric}'s finding that GPT-4o surpasses human experts on structured knowledge and CyberThreat-Eval's finding that no LLM meets operational CTI quality demonstrates that evaluation design determines the conclusions drawn. The lesson that benchmark performance does not establish operational readiness is the field's most urgent methodological correction, and this paper is the only one in the corpus to have made it.

\begin{table*}[htbp]
  \centering
  \caption{Case study comparison matrix across the four flagship systems.}
  \label{tab:table-9}
  \footnotesize
  \setlength{\tabcolsep}{4pt}\renewcommand{\arraystretch}{1.15}
\begin{tabularx}{\textwidth}{@{}>{\raggedright\arraybackslash}p{0.135\linewidth} >{\raggedright\arraybackslash}X >{\raggedright\arraybackslash}X >{\raggedright\arraybackslash}X >{\raggedright\arraybackslash}X@{}}
\toprule
\textbf{Attribute} & \textbf{Palmieri et al. } \cite{almeidapalmieri2025framework} & \textbf{Shen et al. } \cite{shen2024llmosint} & \textbf{Allam} \cite{allam2025cybervision} & \textbf{Chen et al.} \cite{chen2026cyberthreateval} \\
\midrule
Architecture & Agentic / ReAct multi-tool & Agentic + KG + memory + Tor integration & RAG-augmented retrieval and generation & Analyst-centric three-stage CTI evaluation framework \\
\addlinespace[2pt]
Primary task & Multi-tool collection and target analysis & Multi-source collection, KG enrichment, dark-web monitoring, self-reflection & OSINT retrieval and question-answering & CTI triage, targeted search, report generation \\
\addlinespace[2pt]
Data sources & Shodan, Maltego, VirusTotal, Censys, theHarvester & Surface web; Tor .onion nodes & Private OSINT knowledge base & Real-world CTI workflow (public) \\
\addlinespace[2pt]
Evaluation & Proof-of-concept; no open benchmark & None (qualitative architecture only) & Six-metric RAG evaluation & Three-stage analyst-judged workflow \\
\addlinespace[2pt]
Key metrics & Task-level figures (unverified) & None & P@5, MRR, factual/citation accuracy, TCR, hallucination & Factual accuracy, completeness, actionability \\
\addlinespace[2pt]
Hallucination measured & No & No & \textbf{Yes --- 4\%} & No \\
\addlinespace[2pt]
Adversarial eval. & No & No & No & No \\
\addlinespace[2pt]
Human oversight & No & No & No & Yes (analyst review at reporting) \\
\addlinespace[2pt]
Data availability & Private & Private & Private & Public (paper, code, benchmark) \\
\addlinespace[2pt]
Key finding & Most complete agentic PoC; multi-tool ReAct & Architecturally richest agent; KG + memory + dark web in one system & Only OSINT hallucination measurement (4\%) & No LLM adequate across three-stage CTI workflow \\
\addlinespace[2pt]
Key limitation & Benign-only; private set; no oversight & Quantitatively unevaluated; non--peer-reviewed preprint & Private data; single favourable measurement & Operational context limits replication; annotation not fully standardised \\
\addlinespace[2pt]
Corpus role & Case Study 1; central agentic reference & Case Study 2; cited \S3.3, \S3.5, \S3.11, \S4.6 & Case Study 3; hallucination anchor & Case Study 4; CTI benchmark; foil to \cite{tihanyi2024cybermetric} \\
\bottomrule
\end{tabularx}
\end{table*}

\begin{figure*}[!t]  
  \centering  
  \includegraphics[width=\linewidth]{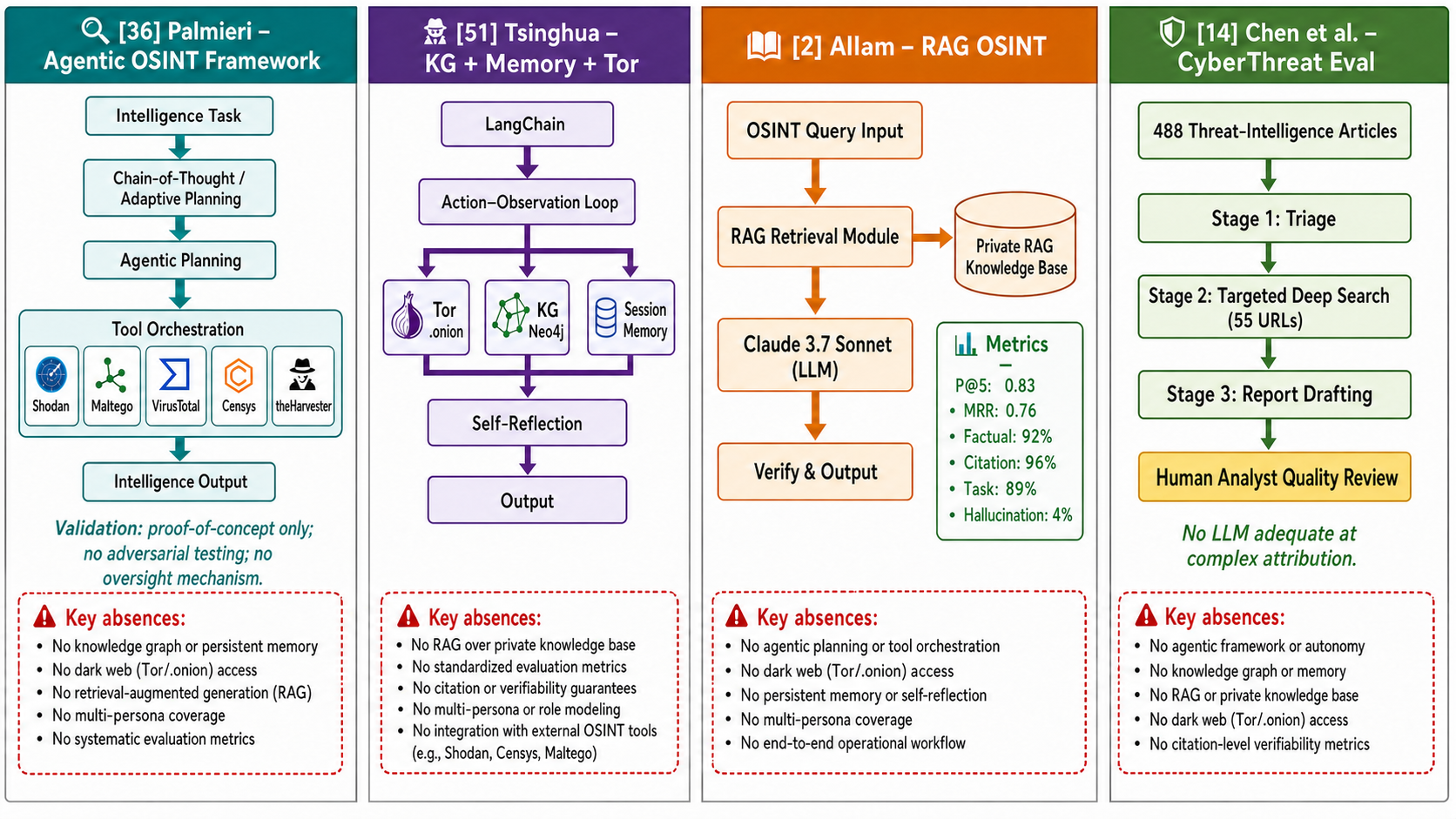}  
  \caption{Comparative Architecture of the Four Case Studies. Each panel shows the internal architecture, tool integrations, key metrics (where reported), and principal limitations of one flagship system: Palmieri~\cite{almeidapalmieri2025framework} (agentic ReAct framework), the Tsinghua system~\cite{shen2024llmosint} (KG, memory, and Tor integration), Allam~\cite{allam2025cybervision} (RAG-augmented OSINT with hallucination measurement), and Chen et al.~\cite{chen2026cyberthreateval} (analyst-centric CTI evaluation). No single system implements all components simultaneously.}  
  \label{fig:casestudy}  
\end{figure*}

\subsection{Cross-Case Synthesis}
\label{sec:5-5-cross-case-synthesis}

The four case studies do not represent four independent research trajectories. They represent four different attempts to answer the same foundational question how far LLMs and agentic AI can advance the capability, reliability, and trustworthiness of OSINT practice from four different analytical positions: architectural capability (\cite{almeidapalmieri2025framework,shen2024llmosint}), reliability measurement (\cite{allam2025cybervision}), and operational evaluation realism (\cite{chen2026cyberthreateval}). What they collectively demonstrate, and what they collectively fail to validate, defines both the current state of the field and the conditions under which operational deployment can be responsibly considered.

\textbf{What the four case studies collectively demonstrate.} Technical feasibility of agentic OSINT collection is established by \cite{almeidapalmieri2025framework} and \cite{shen2024llmosint}: an LLM in a structured reasoning loop can orchestrate real OSINT tools, accumulate intelligence across cycles, persist extracted entities in a queryable graph, and access dark-web sources through Tor. Under favourable conditions, the hallucination rate of RAG-grounded OSINT generation is 4\% \cite{allam2025cybervision} low enough to be operationally tolerable with appropriate verification, but not without it. The gap between structured benchmark performance and operational CTI workflow performance is wide and consequential \cite{chen2026cyberthreateval}: evaluation design determines performance conclusions, and the dominant benchmark paradigm systematically overestimates operational readiness.

\textbf{What the four case studies fail to validate.} None has been evaluated under adversarial conditions against fake CTI \cite{ranade2021fake}, adversarially poisoned knowledge graphs, or the misinformation-contaminated environments that \cite{niu2024bullshint} characterises as the operational baseline. None incorporates or evaluates a human oversight mechanism shown to function effectively. None uses an open, independently replicable evaluation dataset. No pair of systems has been evaluated using the same metrics. The hallucination rate of the architecturally most complete system \cite{shen2024llmosint} is entirely unknown, and the operational quality of \cite{almeidapalmieri2025framework}'s outputs has not been assessed by trained intelligence analysts.

\textbf{Why the field is not yet ready for fully autonomous OSINT deployment.} The case studies jointly reveal a field in which technical demonstration has outpaced empirical validation of adversarial robustness, human oversight, reproducibility, and legal defensibility. Systems are evaluated under conditions simpler than operational reality, without adversarial inputs, without human analyst quality review, and without engagement with legal, ethical, and evidentiary requirements. The most architecturally complete system \cite{shen2024llmosint} has no quantitative evaluation. The only empirical hallucination measurement \cite{allam2025cybervision} rests on a private dataset and proprietary model, whose conditions cannot be reproduced on-premises. The only operationally realistic evaluation \cite{chen2026cyberthreateval}, while now publicly available, still limits full ecological replication. Under these conditions private tasks, no adversarial testing, and unknown operational quality, the claim that agentic OSINT systems are ready for autonomous deployment has no empirical foundation.

\textbf{Why a human-in-the-loop co-pilot model is the most defensible current architecture.} Convergent evidence from \cite{mukhopadhyay2024osintclinic} (oversight checkpoint design), \cite{rheault2024active} (IRB-grounded collection), \cite{srikanth2024usability} (adoption gap), \cite{dekens2023practical} (practitioner scepticism), and \cite{chen2026cyberthreateval} (operational quality failure) points towards a co-pilot rather than an autonomous model. In this model, the LLM executes collection, extraction, and preliminary analysis to the extent that demonstrated capability supports; human analysts review at defined checkpoints; and the pipeline maintains provenance documentation sufficient for legal and analytical accountability. As evaluation frameworks mature and adversarial robustness testing becomes routine, the balance between autonomous and supervised operation will shift, but that shift must be driven by evidence, not architectural ambition.

\textbf{From four flagship systems to a field-level agenda.} The four case studies are not isolated artefacts; each localises, in a single concrete system, a deficit that recurs across the full corpus. The absence of adversarial evaluation in \cite{almeidapalmieri2025framework} and \cite{shen2024llmosint} mirrors the benign-only evaluation regime documented throughout the corpus's risk literature; the unknown hallucination rate of the architecturally most complete system \cite{shen2024llmosint} generalizes to the corpus-wide scarcity of OSINT-specific hallucination measurement; the non-reproducibility of the single 4\% datapoint \cite{allam2025cybervision} reflects the field's broader reliance on private datasets and proprietary models; and the benchmark-versus-workflow divergence isolated by \cite{chen2026cyberthreateval} exposes the absence of shared, operationally valid evaluation protocols. The transition from these four diagnoses to a field-level research programme is therefore direct: the agenda that follows is organised around exactly these gaps, with each direction tied to a specific corpus deficit and naming the methodological, institutional, or normative investment whose absence most immediately obstructs the move from demonstrated capability to validated operational deployment.

\section{Research Challenges and Future Directions}
\label{sec:6-research-challenges-and-future-directions}

Across every layer of the corpus, the technical capability of LLMs and agentic AI for OSINT has advanced well ahead of the field's capacity to validate, govern, and responsibly deploy it. The challenges below are therefore structural deficits in evaluation, legal engagement, human factors, reproducibility, and governance that better models alone cannot resolve. Each future direction is grounded in a specific corpus gap and identifies the methodological, institutional, or normative investment whose absence most immediately obstructs progress.

\subsection{Reliability, Hallucination, and Source Verification}
\label{sec:6-1-reliability-hallucination-and-source-verification}

The hallucination-validation gap is this review's most consequential finding: hallucination is named the primary reliability concern in more than twenty papers, yet OSINT-specific empirical measurement rests on a single favourable-condition datapoint, Allam's 4\% RAG rate \cite{allam2025cybervision}, that does not transfer across models, knowledge-base quality, source domains, or adversarial contamination. The deeper problem is the absence of standardised, adversarial, and reproducible protocols that would make such measurement cumulative. Verify-and-Edit \cite{zhao2023verifyandedit} is the best-evaluated post-generation mechanism (+4.5\%/+5.9\% exact-match over CoT), and its independent convergence with self-reflection \cite{shen2024llmosint} and iterative revision \cite{sun2025decision} supports staged verification as effective against \emph{incidental} error; none of these is shown to defend against the \emph{adversarial} content that Ranade et al.\ \cite{ranade2021fake} demonstrate professional analysts cannot reliably detect.

\textbf{Future directions.} The field requires standardised hallucination-validation protocols specifying minimum dataset size, the distinction between incidental and adversarial hallucination, per-claim provenance tracing, and the conditions under which cross-system rate comparison is valid; a dedicated adversarial-hallucination detection capability targeting content crafted with knowledge of the system's collection channels; and the provenance chain standards (Section~6.6) needed for legal admissibility.

\subsection{Evaluation and Benchmarking Gaps}
\label{sec:6-2-evaluation-and-benchmarking-gaps}

The absence of shared, open, cumulative evaluation infrastructure is the principal obstacle to scientific progress. The corpus exhibits a non-cumulative metric pattern: custom OSINT-agent metrics \cite{yuan2024empowering} unused by \cite{almeidapalmieri2025framework}; proof-of-concept figures \cite{almeidapalmieri2025framework} adopted by no later paper; six private RAG metrics \cite{allam2025cybervision} replicated by none; analyst-centric metrics \cite{chen2026cyberthreateval} whose operational origin limits replication; and reproducible but unadopted F1 evaluation \cite{shafee2024evaluation}. Every significant paper builds its own evaluation on its own private data, making cross-study comparison structurally impossible. The CyberMetric and CyberThreat-Eval contrast quantifies the cost: GPT-4o reaches 91.25\% on multiple-choice cybersecurity knowledge, surpassing the 72.24\% expert mean \cite{tihanyi2024cybermetric}, while no LLM meets analyst-acceptable quality on real CTI workflows \cite{chen2026cyberthreateval}. The results are not contradictory they measure different competencies, but without a commensurable framework, procurement decisions are made on benchmark figures that do not predict operational performance.

A distinct but related failure concerns agentic system evaluation specifically. The four agentic OSINT papers \cite{yuan2024empowering,almeidapalmieri2025framework,mukhopadhyay2024osintclinic,shen2024llmosint} each propose agent-level performance criteria for task completion, tool-call accuracy, and output coherence that are neither standardised nor adopted by any subsequent paper (Section~4.8). This non-adoption means that agentic OSINT capability, arguably the field's most consequential advance, remains the least cumulatively validated component of the literature. The metrics used at the agent level must be designed to feed back into the broader benchmark suite, not to exist as isolated proof-of-concept indicators: agent task-completion rates are only interpretable when grounded in the same ground-truth datasets and NER taxonomies against which RAG and extraction performance are measured. Without this integration, benchmark improvements and agentic improvements develop on incommensurable tracks, and the combined system, the practical deployment target, has no unified evaluation standard.

\textbf{Future directions.} The community's most urgent investment is an open, adopted OSINT-AI benchmark suite covering the following: collection accuracy against verified ground truth; NER precision/recall on OSINT cybersecurity text using an established taxonomy (DISARM, MITRE ATT\&CK); expert-reviewed product factual accuracy; hallucination rate under the Section~6.1 protocol; analyst task-completion with and without AI; agentic pipeline metrics (tool-call accuracy, graceful degradation, and completion under adversarial conditions) aligned to the same ground truth; and adversarial robustness at the capability level of \cite{ranade2021fake}. A community working group modelled on SuperGLUE or ARC, with practitioners and legal advisors, should oversee it.

\subsection{Tool Use, API Reliability, and Automation Risks}
\label{sec:6-3-tool-use-api-reliability-and-automation-risks}

The feasibility of agentic OSINT depends on the reliability, security, and legal compliance of the external tools invoked by the model, as well as the layer of the corpus that is least analyzed. No reviewed study evaluates agent behavior under API failure (rate limiting, authentication expiry, or unavailability), the cost of sustained high-volume monitoring, or the security of orchestration frameworks such as LangChain \cite{shen2024llmosint} as dependencies in a security-critical pipeline. The cloud-versus-on-premise tension of R\u{a}doi's API approach \cite{radoi2023ai} versus Yurtalan's local deployment \cite{yurtalan2025redefining} remains the most consequential unresolved deployment question: a four-point F1 gap \cite{shafee2024evaluation} is narrow enough to justify on-premise serving where operational security or data sovereignty preclude cloud use, yet the corpus offers no systematic cost-benefit framework, and the dual-use exposure of API-accessible tooling \cite{pervez2023ease} compounds the concern.

\textbf{Future directions.} Research should address graceful degradation under tool failure, cost modelling for operational-scale monitoring, orchestration-security attack surfaces, and API-access governance for dual-use tools such as Shodan and Censys. The deployment choice should be supported by a single decision framework integrating performance \cite{shafee2024evaluation}, data sovereignty \cite{yurtalan2025redefining}, legal compliance \cite{golda2024privacy}, and dual-use risk \cite{pervez2023ease}.

\subsection{Data Quality, Bias, and Coverage}
\label{sec:6-4-data-quality-bias-and-coverage}

OSINT data is not a neutral substrate, and the corpus reveals three distinct challenges. Environmental contamination: Niu et al.\ \cite{niu2024bullshint} establish a 9\% misinformation base rate across 1.96M tweets from 48 professional communities, an input condition inherited before any LLM processing, whose compounding with incidental hallucination is unmodelled. Embedded bias: Kobayashi and Yamaguchi \cite{kobayashi2024subjective} show that LLMs transmit training-corpus political bias into analyses, a validity threat where neutrality is a legal requirement; self-reflection \cite{sun2025decision} partially mitigates report-stage bias but not the upstream extraction and classification stages. Coverage: The corpus is effectively English-language and Twitter-centric, with at-scale evaluation lacking on non-English sources and on Facebook, Telegram, or Reddit platforms, which are an operational baseline.

\textbf{Future directions.} Annotated multilingual OSINT datasets (at minimum Arabic, Mandarin, Russian, Farsi, and Spanish), incorporating adversarial content at a representative contamination rate \cite{niu2024bullshint}, would be the highest-impact data investment. OSINT-specific bias-measurement protocols addressing entity selection and event classification, not only report framing, should accompany the Section~6.2 benchmark.

\subsection{Dark Web, Multimodal, and Multilingual OSINT Challenges}
\label{sec:6-5-dark-web-multimodal-and-multilingual-osint-challenges}

Three operationally critical environments are effectively absent from the literature. Dark-web OSINT rests on one working demonstration \cite{shen2024llmosint} and one conceptual proposal \cite{amin2025darklens}; the NLP challenges of coded, multilingual, obfuscated content, the legality of automated \texttt{.onion} access under computer-misuse statutes, and chain-of-custody for evidential use are all unaddressed. Multimodal OSINT is represented by a single paper: GeoLocator \cite{yang2024geolocator} infers precise location from one photograph yet offers no countermeasure, and no paper addresses image, video, or audio OSINT despite their operational centrality. Multilingual coverage is minimal even though intelligence environments are multilingual by default.

\textbf{Future directions.} Dark-web research should prioritise a legally validated, evidentially admissible collection methodology designed from the outset for chain-of-custody and disclosure, developed jointly by NLP researchers, legal scholars, and forensic specialists. Multimodal research should prioritise defensive capabilities (metadata sanitisation, counter-inference) against attacks such as \cite{yang2024geolocator}. Multilingual work should target entity recognition and relationship extraction, the tasks at which LLMs are most deficient even in English \cite{shafee2024evaluation}.

\subsection{Security, Privacy, Ethics, and Legal Constraints}
\label{sec:6-6-security-privacy-ethics-and-legal-constraints}

The risk landscape is extensive, empirically grounded, and inadequately mitigated, resting on six papers: adversarial CTI generation and KG poisoning \cite{ranade2021fake}; geo-privacy inference \cite{yang2024geolocator}; community misinformation \cite{niu2024bullshint}; cloud-API GDPR exposure \cite{golda2024privacy}; critical-infrastructure exposure via tooling \cite{pervez2023ease}; and IRB-grounded ethical collection as the positive benchmark \cite{rheault2024active}. The most immediate is adversarial: a modestly fine-tuned model deceived professional threat hunters at the rate of authentic CTI ($n=10$) and poisoned the knowledge graphs that give RAG its grounding benefit, turning the mitigation into an amplification mechanism; no corpus paper proposes a detection architecture, and disclosure is unconfirmed. The geo-privacy and infrastructure attacks both demonstrate a failure in responsible disclosure, as they involve a complete attack methodology that was published without confirmed remediation, highlighting the lack of established disclosure norms that have long been standard in cybersecurity. GDPR exposure \cite{golda2024privacy} generalises in principle to any jurisdiction with data protection law; the on-premise \cite{yurtalan2025redefining,shafee2024evaluation,nila2023llms} and PEFT \cite{riad2024finetuning} evidence supports a compliant alternative in which no paper converts into a decision framework. Against this, \cite{rheault2024active} alone treats compliance as a primary design requirement, a standard roughly 60\% of primary papers do not meet.

\textbf{Future directions.} The field needs a responsible-disclosure standard for OSINT-AI covering both offensive demonstrations \cite{ranade2021fake,yang2024geolocator,pervez2023ease} and misuse-enabling defensive proposals; a comprehensive legal-compliance framework spanning data protection, intelligence, computer access, and evidentiary law, built with legal scholars and civil-liberties organisations; and an IRB-analogue for agentic collection systems processing personal data at scale, with oversight proportionate to privacy-intrusion potential.

\subsection{Human-AI Teaming and Analyst Oversight}
\label{sec:6-7-human-ai-teaming-and-analyst-oversight}

Human oversight is the most proposed and least evaluated component of responsible OSINT AI. The consensus that analysts must remain in the loop is near-universal, yet the only directly relevant evidence \cite{ranade2021fake}'s finding that expert reviewers judged adversarial CTI as genuine at the authentic rate ($n=10$) suggests review is ineffective against adversarial content. Mukhopadhyay and Luther\ \cite{mukhopadhyay2024osintclinic} provide the most principled checkpoint architecture but do not evaluate whether analysts actually catch agent errors at those checkpoints. The adoption gap \cite{srikanth2024usability} faced by professionals with low familiarity with LLMs, who cite hallucination and verification difficulties as primary barriers, can be addressed by enhancing the analysts' ability to work with the model rather than by changing the model itself, which \cite{cerny2024implications} describes as a trainable prompt-engineering competency. Rheault et al.\ \cite{rheault2024active} show that accountability must be embedded at the design stage rather than applied retrospectively, and \cite{dekens2023practical} characterises the key skill shift as principled scepticism grounded in domain expertise.

\textbf{Future directions.} Three empirical questions are unaddressed: when does human review catch AI errors rather than merely approve them (effectiveness, not presence); what minimum LLM familiarity enables meaningful oversight of a given system (qualification standards); and how do oversight quality, completion time, and accuracy change across co-pilot, supervisory, and autonomous modes (autonomy calibration). Structured LLM-literacy training with measurable proficiency outcomes should be developed alongside the systems it oversees and the \cite{mukhopadhyay2024osintclinic} checkpoints evaluated using the analyst metrics \cite{srikanth2024usability} that show are currently absent.

\subsection{Reproducibility and Open-Source Tooling}
\label{sec:6-8-reproducibility-and-open-source-tooling}

The corpus fails the reproducibility standards of its publication venues. Private datasets \cite{allam2025cybervision,almeidapalmieri2025framework,chen2026cyberthreateval}, unnamed models \cite{saddi2024genai}, absent code, and idiosyncratic, unadopted metrics \cite{yuan2024empowering,almeidapalmieri2025framework,allam2025cybervision,chen2026cyberthreateval} mean that most papers cannot be replicated from the published text, and the strongest empirical findings are, by virtue of private data and proprietary infrastructure, the least replicable. The causes are partly inherent (OSINT data often involves personal, sensitive, or commercial content) and partly cultural; the non-cumulative metric pattern is both cause and consequence. Open-source tooling is equally immature: no paper releases a complete, deployable system, and the OSINT-specific intelligence layer prompt designs, entity ontologies, KG schemas, and evaluation protocols remain proprietary or under-documented, even though the invoked tools and orchestration libraries are themselves well-documented third-party components.

\textbf{Future directions.} The community should adopt NLP/ML-style reproducibility norms: open code as a publication condition (or explicit justification otherwise); open or synthetic/anonymised datasets where privacy law permits; pre-registration of evaluation protocols; and replication studies as valued contributions. An open OSINT-AI agent framework analogous to a model hub providing standardised tool interfaces, a documented prompt-template library, and a shared evaluation harness implementing the Section~6.2 suite would supply the infrastructure for cumulative development.

\subsection{Ten-Point Research Agenda for Responsible Agentic and Generative AI in OSINT and Cyber Investigation}
\label{sec:6-9-ten-point-research-agenda-for-responsible-agentic-and-ge}

\textbf{Agenda Point 1: Establish a standardised OSINT-AI hallucination measurement protocol.} Develop and publish a protocol distinguishing incidental from adversarial hallucination, specifying minimum dataset size, provenance-tracing mechanism, and comparison conditions. Apply it as a mandatory evaluation component for all new OSINT AI system publications. \textit{Motivated by: the hallucination-validation gap (\cite{allam2025cybervision} provides the only measurement; over twenty papers acknowledge the problem without measuring it).}

\textbf{Agenda Point 2: Develop an open, community-adopted OSINT-AI evaluation benchmark suite.} Establish a community working group to develop an open benchmark covering: collection accuracy; NER precision and recall on OSINT-derived text; intelligence product factual accuracy by expert review; hallucination rate; analyst task completion with and without AI; and adversarial robustness at the capability level of \cite{ranade2021fake}. \textit{Motivated by: benchmark fragmentation across \cite{yuan2024empowering,almeidapalmieri2025framework,allam2025cybervision,tihanyi2024cybermetric,chen2026cyberthreateval,srikanth2024usability,shafee2024evaluation}.}

\textbf{Agenda Point 3: Evaluate API reliability, cost, and orchestration security under operational conditions.} Systematically evaluate agentic OSINT pipelines under API failure, rate limiting, authentication expiry, and orchestration framework vulnerability conditions absent from all current evaluations. Develop a cost-modelling framework for sustained OSINT monitoring, enabling cloud versus on-premise comparisons across performance, security, legal compliance, and economic dimensions. \textit{Motivated by: the API reliability gap and the \cite{radoi2023ai} vs.\ \cite{yurtalan2025redefining} deployment tension.}

\textbf{Agenda Point 4: Build multilingual, multi-platform OSINT evaluation datasets.} Develop annotated datasets covering at minimum Arabic, Mandarin, Russian, Farsi, and Spanish across social media, forum, and dark web source types, including adversarially crafted content at the \cite{niu2024bullshint} environmental contamination rate. \textit{Motivated by: the complete absence of non-English OSINT evaluation in the corpus.}

\textbf{Agenda Point 5: Develop a legally admissible dark web OSINT collection methodology.} In collaboration with legal scholars, law enforcement, and forensic evidence specialists, develop a dark-web collection methodology designed from the outset to satisfy chain-of-custody, authentication, and disclosure requirements, including investigator anonymity and the legal status of automated Tor access. \textit{Motivated by: the dark-web legal void in \cite{shen2024llmosint,amin2025darklens}.}

\textbf{Agenda Point 6: Establish responsible disclosure norms for OSINT-AI research. } Develop and adopt a disclosure standard for both offensive capability demonstrations and defensive architecture publications, specifying minimum disclosure obligations, coordination timelines, and conditions under which demonstrations should be withheld pending remediation. \textit{Motivated by: the disclosure failures of \cite{ranade2021fake,yang2024geolocator}, and \cite{pervez2023ease}.}

\textbf{Agenda Point 7: Create structured LLM literacy training for OSINT practitioners.} Develop LLM literacy training for OSINT and CTI analysts, with measurable proficiency outcomes integrated into professional certification pathways, addressing, at a minimum: LLM failure modes in OSINT contexts (hallucination, bias, prompt brittleness); structured prompt engineering for OSINT objectives; and conditions under which outputs require verification. \textit{Motivated by: \cite{srikanth2024usability}'s adoption gap (median LLM familiarity 2/5) and \cite{cerny2024implications}'s competency framing.}

\textbf{Agenda Point 8: Evaluate human oversight effectiveness empirically.} Conduct user studies ($n \geq 30$) measuring not only whether oversight mechanisms are architecturally present but also whether analysts successfully identify hallucinated claims, adversarially generated content, and biased framing in LLM-assisted OSINT outputs. \textit{Motivated by: \cite{srikanth2024usability}'s small-sample usability study and \cite{ranade2021fake}'s finding that professional threat hunters were equally likely to judge fake CTI as true ($n=10$).}

\textbf{Agenda Point 9: Mandate open datasets, code repositories, and pre-registered evaluation protocols.} Adopt reproducibility standards for OSINT-AI research: open code repositories (or documented justification); open or synthetic evaluation datasets; pre-registered protocols preventing post-hoc metric selection; and replication studies as a valued contribution. {Motivated by reproducibility failures across \cite{su2024aiagent,su2025opensource,allam2025cybervision,saddi2024genai,amin2025darklens}, and \cite{chen2026cyberthreateval}.}

\textbf{Agenda Point 10: Develop a comprehensive legal and governance framework for OSINT AI.} In collaboration with legal scholars, intelligence practitioners, civil liberties organizations, and data protection regulators, develop a jurisdiction-sensitive framework addressing GDPR and equivalent data protection regimes \cite{golda2024privacy}; computer access law as applied to automated dark web collection \cite{shen2024llmosint}; evidentiary admissibility of AI-generated intelligence \cite{rheault2024active,yigit2024review}; and the ethical governance obligations of OSINT AI developers and operators. \textit{Motivated by: the near-total absence of legal engagement in the primary corpus and governance failures in \cite{ranade2021fake,yang2024geolocator,golda2024privacy,pervez2023ease}.}

\section{Limitations and Threats to Validity}
\label{sec:7-limitations}

The findings reported above should be read in light of several limitations arising from the scope, construction, and methodology of this review. We state them explicitly so that readers can calibrate the strength of each conclusion and so that subsequent work can address them directly.

\textbf{Temporal coverage and the moving research frontier.} The corpus was assembled from searches conducted in the first half of 2026 and spans publications from 2014 \cite{pais2014osint} to 2026 \cite{chen2026cyberthreateval}. Agentic and generative AI is among the fastest-moving areas of contemporary computer science, and a review of it is necessarily a snapshot: systems, benchmarks, and model capabilities released after the search window are not represented, and approximately eleven grey-literature items carry no formal publication date, complicating precise temporal placement. The capability claims surveyed here should therefore be treated as a baseline that newer models may already have surpassed. The \emph{structural} findings concerning evaluation and validation, by contrast, are properties of the field's methodology rather than of any single model generation, and are correspondingly more robust to the passage of time.

\textbf{Publication and venue bias.} The corpus is drawn predominantly from work reporting successful capability demonstrations. Null results, failed integrations, and negative operational findings are systematically underrepresented in the publishable literature, a well-documented bias that this review cannot correct and that likely inflates the apparent maturity of the field. The most consequential finding of this review---the scarcity of hallucination measurement and adversarial evaluation---is itself partly a manifestation of this bias: studies that would have reported poor reliability under adversarial conditions were, for the most part, never designed, run, or published.

\textbf{Language bias.} Screening was conducted in English, and the corpus is overwhelmingly English-language. Relevant work published in other languages---including the Korean-language evaluation by Park et al.\ \cite{park2024performance}, which entered the corpus only through an English-language route---is likely underrepresented. OSINT is a globally practised discipline, and a monolingual review risks omitting regionally significant systems, datasets, and threat models.

\textbf{Database and source coverage.} Because the corpus was assembled by expert curation within a defined conceptual scope rather than by an exhaustive automated Boolean export, several major indexes---including Scopus, Web of Science, DBLP, and SSRN---were not queried to exhaustion, and the grey-literature component (preprints, theses, and technical reports) was identified opportunistically rather than systematically. Relevant studies indexed only in those sources may have been missed; this is the principal coverage threat to the review's completeness.

\textbf{Expert curation and reproducibility of corpus construction.} The corpus was constructed through expert judgement applied to the conceptual scope defined in Section~\ref{sec:2-2-search-strategy-and-corpus-construction}, not through a fully mechanised query that a third party could replay to reproduce the exact study set. This approach was adopted deliberately, because the boundary between OSINT-relevant and OSINT-adjacent AI research is conceptual rather than keyword-separable, but it limits the bit-for-bit reproducibility a purely automated protocol would offer. To mitigate this, the inclusion and exclusion criteria (Section~\ref{sec:2-1-study-selection-criteria}), the screening flow (Figure~\ref{fig:prisma_flow}), and the disposition of every item (Table~S1) are documented in full, so that the curation can be independently audited even where it cannot be mechanically re-executed.

\textbf{Single-screener selection and coding.} Study selection, tier assignment, and taxonomic coding were performed by a single screener rather than by two or more independent reviewers with a measured inter-rater agreement statistic (for example, Cohen's~$\kappa$). This introduces a risk of selection and classification subjectivity that a multi-reviewer protocol would reduce. The risk is mitigated, though not eliminated, by the explicit, documented criteria applied uniformly across the corpus and by the per-item evidence matrix (Table~S1), which exposes every coding decision to external scrutiny.

\textbf{Reliance on preprints and non-peer-reviewed sources.} Several pivotal items in the corpus---including the architecturally most complete agentic system \cite{shen2024llmosint} and the only OSINT-specific hallucination measurement \cite{allam2025cybervision}---are preprints or theses that have not completed peer review. Conclusions resting on them, in particular the 4\% hallucination figure, are correspondingly provisional and should be revisited as peer-reviewed versions appear. The review status of these sources is flagged at each point of substantive reliance.

\textbf{Private datasets and unreproducible evaluations.} A substantial share of the quantitative evidence summarised in Table~\ref{tab:table-4} rests on private datasets, proprietary models, or operational deployments whose conditions cannot be reproduced on-premises. The 4\% hallucination rate \cite{allam2025cybervision}, for instance, was obtained on a private corpus using a proprietary cloud model. Such results are informative but not independently verifiable, and the review treats them as indicative rather than settled.

\textbf{Generalisability.} The reviewed systems are evaluated predominantly on surface-web, English-language, benign, proof-of-concept tasks. Their behavior on dark-web sources, non-English content, adversarially contaminated inputs, and sustained operational workloads is largely unmeasured, so capability claims should not be extrapolated to those settings. The review's central recommendation---the analyst-in-the-loop co-pilot model---is an evidence-based inference from the convergence of oversight, adoption, and evaluation findings rather than a directly tested operational result; it represents the most defensible reading of the current evidence, not an empirically validated deployment outcome.

\textbf{Net effect on the review's conclusions.} These limitations bound the precision of the review but do not undermine its three central findings. The capability-versus-validation imbalance, the hallucination--validation gap, and the divergence between benchmark and operational performance each rest not on a single study but on convergent evidence from many independent groups, and are therefore robust to the omission, misdating, or later supersession of any individual item. What the limitations qualify is the exactness of specific figures and the breadth of safe extrapolation---not the direction or the strength of the field-level conclusions.

\section{Conclusion}
\label{sec:8-conclusion}

This systematic literature review has surveyed 74 research and grey-literature items to characterize the current state of agentic and generative AI as applied to open-source intelligence. The review was motivated by the convergence of two developments: the emergence of sufficiently capable LLMs to make autonomous multi-step reasoning over heterogeneous open-source data technically feasible, and the growth in OSINT collection volume to the point where human-only analytical workflows can no longer achieve the speed, scale, or consistency that operational requirements demand. The central question is not whether LLMs and agentic AI can improve OSINT; they demonstrably can, but the key issues are under what conditions this improvement can be achieved, at what level of risk, and with what governance architecture to ensure it is done responsibly.

The 74 items span twelve thematic folders plus one excluded/peripheral category, classified into four analytical tiers: 37 primary or core papers, 11 methodological support papers, 22 background or context papers, and 4 excluded or minimal-use papers. The corpus is unevenly distributed across the OSINT workflow lifecycle: collection and analysis are well served by the primary literature; verification, reporting, dissemination, and decision support remain systematically underexplored, reflecting a structural research preference for technically tractable problems over the epistemologically complex, legally embedded tasks that constitute professional OSINT practice.

The review's principal findings may be summarised as follows. First, the basic feasibility of LLM-assisted OSINT collection, entity extraction, and preliminary analysis is established across multiple studies \cite{almeidapalmieri2025framework,radoi2023ai,shen2024llmosint,shafee2024evaluation}. Second, agentic OSINT systems represent the most promising architectural direction for scalable OSINT but are evaluated exclusively under benign conditions that do not reflect the adversarial environments the corpus's own risk literature confirms are operationally normal \cite{ranade2021fake,yang2024geolocator,niu2024bullshint}. Third, RAG and knowledge graph integration achieve a 4\% hallucination rate in the only OSINT-specific empirical measurement \cite{allam2025cybervision}, yet these same grounding mechanisms amplify adversarial knowledge-base poisoning \cite{ranade2021fake}, and no paper addresses both failure modes within a single architecture. Fourth, the critical finding from the CTI evaluation literature is not a performance figure but a methodological lesson: benchmark performance \cite{tihanyi2024cybermetric} and operational workflow failure \cite{chen2026cyberthreateval} measure different competencies, and the gap between them is consequential for every deployment decision. Fifth, hallucination is identified as the primary reliability threat across more than twenty papers and is empirically measured in only one \cite{allam2025cybervision}, the most significant failure of the field's collective commitment to the reliability problem it universally identifies. Sixth, the risk landscape is empirically grounded and inadequately mitigated: adversarial CTI generation was reported to deceive professional threat hunters at rates equal to authentic CTI \cite{ranade2021fake} ($n=10$; cautious generalisation required), geo-privacy attacks operate from single photographs \cite{yang2024geolocator}, misinformation contaminates professional communities at 9\% \cite{niu2024bullshint}, and cloud API deployment creates GDPR exposure \cite{golda2024privacy}, yet no paper proposes a governance architecture that addresses this landscape in combination. Seventh, human oversight is the consensus position, but its effectiveness under operational conditions is unevaluated \cite{ranade2021fake}.

The practical implication is clear and consistent. The most defensible current deployment model for agentic and generative AI in OSINT and Cyber Investigation is an analyst-in-the-loop co-pilot architecture. In this model, LLM-based systems execute collection, extraction, and preliminary analysis only to the extent supported by demonstrated capability and measured reliability, while human analysts review outputs at defined oversight checkpoints with the domain expertise and LLM literacy required to identify and correct the specific failure modes documented across the corpus. This is not a concession to technical limitation but an accurate calibration of demonstrated capability against operational risk under current evidence conditions. As open benchmarks, adversarial evaluation frameworks, and larger-scale human oversight studies develop, the appropriate boundary between autonomous and supervised operation may evolve; the present recommendation reflects the state of evidence as of this review, not a permanent claim about the limits of agentic AI's capabilities.

The research implication is equally clear. The field requires, as a priority investment, standardised hallucination validation protocols and open benchmarks; adversarial robustness evaluation as a mandatory validation component; legal compliance frameworks for cloud deployment; structured LLM literacy training; reproducibility standards; and multi-disciplinary engagement with legal scholars and forensic evidence specialists to address the dark web collection and evidence admissibility gaps that the current corpus leaves entirely unaddressed.

Agentic and generative AI for OSINT is not a future capability waiting to be realized. It is a present capability waiting to be responsibly validated. The ten-point research agenda proposed in Section 6 charts the path from technical demonstration to operational trustworthiness. The distance between the two is not a measure of the technology's potential but of the field's commitment to earning the trust that operational intelligence deployment requires.

\section*{Acknowledgments}
This research received no external funding.

\section*{Conflict of Interest}
The authors declare that they have no known competing financial interests or personal relationships that could have appeared to influence the work reported in this paper.

\section*{Data Availability Statement}
\label{sec:appendix-pointer}
The complete corpus reference apparatus is provided as supplementary material to keep the main manuscript within length. It comprises the full \emph{Evidence Matrix} Summarising all 74 studies by stable identifier (01--74), analytical tier, OSINT relevance, principal contribution, principal limitation, and recommended citation locus; and the \emph{Folder-to-Theme Mapping}, relating the twelve source folders to the taxonomy of section~\ref{sec:3-taxonomy-of-agentic-and-generative-ai-in-osint-and-digital}. The study identifiers used throughout the methodology and corpus-inventory tables of this paper are defined in the summary Table. \href{https://github.com/eduardo-palmieri/Agentic-AI-OSINT-Survey-Supplementary/tree/master}{https://github.com/eduardo-palmieri/Agentic-AI-OSINT-Survey-Supplementary/tree/master}

\bibliographystyle{IEEEtran}
\bibliography{references}

\end{document}